 \documentclass[11pt]{article}

 \pdfoutput=1

\usepackage{graphicx} 
\usepackage{amsmath}

\usepackage{enumitem}

\usepackage[francais,english]{babel}

\usepackage{float}
\usepackage{ifthen}
\usepackage{hyperref}

\begin{document}


\title{On the computation of moist-air specific thermal enthalpy.}

\author{by Pascal Marquet {\it M\'et\'eo-France}}


\date{19th of January, 2014}

\maketitle


\begin{center}
{\em Paper accepted for publication in the \underline{Quarterly Journal of the Royal Meteorological Society}. }\\
{\em Last revised version in January 2014.} \\
{\em \underline{Corresponding address}: pascal.marquet@meteo.fr}
\end{center}
\vspace{1mm}


\begin{abstract}
The specific thermal enthalpy of a moist-air parcel is defined analytically following a method in which specific moist entropy is derived from the Third Law of thermodynamics.
Specific thermal enthalpy is computed by integrating specific heat content with respect to absolute temperature and including the impacts of various latent heats (i.e., solid condensation, sublimation, melting, and evaporation).
It is assumed that thermal enthalpies can be set to zero at $0$~K for the solid form of the main chemically inactive components of the atmosphere (solid-$\alpha$ oxygen and  nitrogen, hexagonal ice).
The moist thermal enthalpy is compared to already existing formulations of moist static energy (MSE).
It is shown that the differences between thermal enthalpy and the thermal part of MSE may be quite large.
This prevents the use of MSE to evaluate the enthalpy budget of a moist atmosphere accurately, a situation that is particularly true when dry-air and cloud parcels mix because of entrainment/detrainment processes along the edges of cloud.
Other differences are observed when MSE or moist-air thermal enthalpy is plotted on a psychrometric diagram or when vertical profiles of surface deficit are plotted.
\end{abstract}


 \section{Introduction.} 
 \label{section_intro}

This paper shows that local values of internal energy and enthalpy within a moist atmosphere can be better computed so as to study the energy and enthalpy directly for global/local domains.
Difficulties encountered in the past were due to a paradox: the budget equation of temperature was easier to compute than the local values of enthalpy, whereas it was easier to compute the local values of entropy than to solve the budget equation.

One example of this paradox is given by the conservation or imbalance properties of internal energy 
$e_i = h - R\: T$ (see Appendix~A) or for the enthalpy $h \equiv e_i+p/\rho$. 
It is important to evaluate observed changes in radiative forcing and imbalance properties for energy fluxes, in order i) to properly assess observed climate change impacts and ii) to improve our ability to understand energy changes in the climate system at regional scales.
In studying the conservation of energy in NWP models and  GCMs, it is common to assess global energy fluxes from the surface to the top of the atmosphere using re-analysis products.
Conservation and/or imbalance properties are thus currently monitored via the computation of  fluxes of energy at the surface and  the top of the atmosphere rather than by directly computing  regional or global integrals of energy throughout the atmosphere.

\begin{figure}[t]
\centering
 \includegraphics[width=0.85\linewidth,angle=0,clip=true]{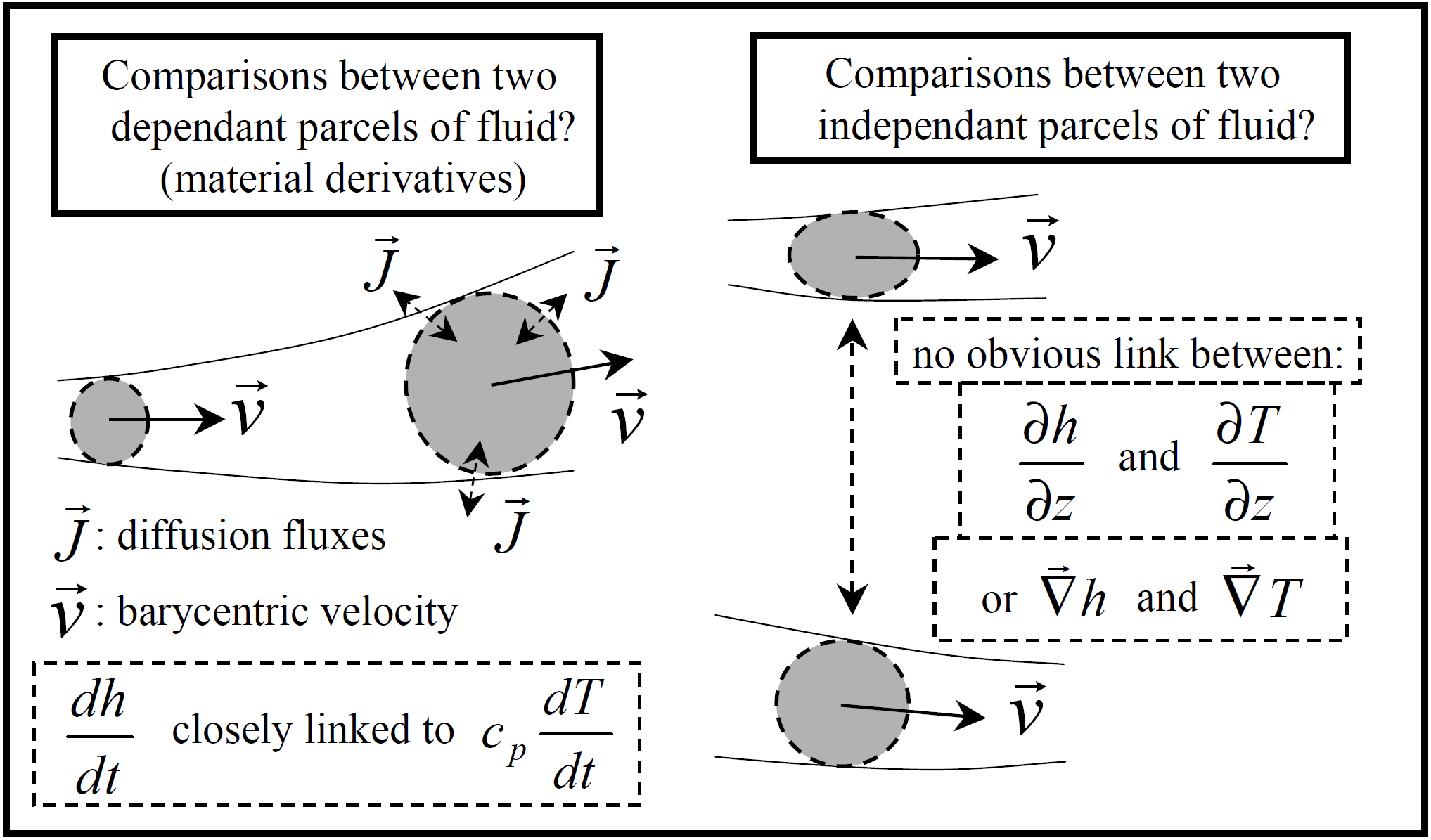}
\caption{
{\it \small
Comparison of enthalpy of moist-air parcels.
Left panel: for parcels on a given streamline.
Right panel: for parcels with different trajectories and histories.
}
\label{Fig_Barycentric}}
\end{figure}
Another example of this paradox appears when the enthalpy equation is used in NWP models or in GCMs.
The enthalpy equation is never written in terms of $d\,h/dt$, with the specific enthalpy $h$ to be evaluated at each grid-point.
It is recalled on the left side of Fig.(\ref{Fig_Barycentric}) that the enthalpy equation can be written in terms of $c_p \: d\,T/dt$ or $d\,(c_p\:T)/dt$, where $T$ is the temperature and $c_p$ the moist-air specific heat.
These properties are consequences of material derivatives applied to barycentric motions of open parcels of fluid described by De Groot and Mazur (1962) or Glansdorff et Prigogine (1971) and applied by Marquet (1993),  Zdunkowski and Bott  (2004) and  Catry \textit{et al.} (2007), among others.

It is recalled below that most of the present moist-air formulae for $h$ and $s$ were derived using the hypothesis of closed parcels of fluid.
The problem concerning the knowledge of reference values for $h$ or $s$ can be briefly illustrated by starting with the dry air enthalpy written as $h_d = c_{pd} \: T + [ \: (h_d)_r - c_{pd}\:T_r \: ]$.
This is a valid expression provided that the specific heat $c_{pd}$ is assumed to be a constant for $T$ close to $T_r$ and $(h_d)_r$ is associated with $T_r$.
The enthalpy of a clear-air parcel of moist air is computed as the sum of $q_d \: h_d + q_v \: h_v$, where $h_v$ is computed with $c_{pd}$ and $(h_d)_r$ replaced by $c_{pv}$ and $(h_v)_r$.
Since the term of $h_d$ in brackets is multiplied by $q_d=1-q_t$, it can be neglected either if $q_t$ is a constant or if the bracketed term is equal to zero, i.e. if $(h_d)_r = c_{pd}\:T_r$.
But, in fact, this is not valid.
It is only if $q_t$ is a constant that changes in enthalpy or entropy can be easily computed, with no need to determine reference values for enthalpies and entropies.

The above properties are only valid if the parcels of fluid are closed.
There are, however, circumstances in which it is necessary to compare absolute values of enthalpy or entropy.
This is true particularly when open systems are considered and if parcels of fluid are not connected by barycentric motion.
This is illustrated on the right side of Fig.(\ref{Fig_Barycentric}).
It is not possible to transform $dh$ into $c_p \: dT$ for these independent parcels since the reference values of $h$ and $s$ are multiplied by varying terms, all of which depend on $q_t$.
In consequence, spatial integrals, vertical fluxes and 3D gradients must be computed by using absolute definitions for $h$ and $s$.
Accordingly, Richardson (1922) and Businger (1982), hereafter referred to as R22 and B82,  searched for relevant choices for zero entropy and enthalpy for both dry air and water species.

A way to compute the specific moist-air enthalpy can be found by examining the other aspect of the paradox, concerning the entropy.
It is shown in the following sections that it is impossible to compute all the terms forming the entropy budget equation.
Nevertheless, it is possible to use the Third Law of thermodynamics to compute the moist air entropy in an absolute way, as explained in Hauf and H\"{o}ller (1987) and Marquet (2011).
This is done by setting the zero of entropies for the most stable crystalline form of all atmospheric species at $0$~K.

This paper explores new possibilities for computing and analyzing isenthalpic or isentropic processes in open systems, defined by $h=\:$Constant or $s=\:$Constant.
This is equivalent to setting $dh/dt=0$ and $ds/dt=0$ (using barycentric formalism for the enthalpy and entropy equations).
Direct local definitions for $h$ and $s$ may allow the study of new classes of processes associated with open systems.
They include the impact of diffusion of water on $h$ or $s$  (entrainment and detrainment at cloud edges), and/or the impact of evaporation of water from the surface.
For instance, it is reported in M11 that moist-air entropy is constant within the whole boundary layer of marine stratocumulus, with almost no jump in $s$ at the top of the boundary layer.
These properties could only be observed with the open-system formulation derived in M11, since large vertical gradients and jumps are observed for $q_t$, $\theta_l$ and $\theta_e$  with marine stratocumulus.

Other examples of new applications can be found in Marquet and Geleyn (2013) and Marquet (2013), where the moist-air Brunt-V\"{a}is\"{a}l\"{a} frequency and  potential vorticity are defined in terms of absolute values of the specific moist-air entropy derived in M11.
These moist-air quantities are computed with the help of vertical and 3D gradients of $s$, and thus depend on relevant absolute definitions for the reference entropies of dry air and water vapour.

It is worth noting that the search for absolute and local definitions for $h$ and $s$ does not correspond to new definitions for the well-known potential temperatures $\theta'_w$, $\theta_e$ or $\theta_l$.
It is shown in following sections that the absolute formulation of M11 must be understood as being a generalization to open systems of the two specific entropies derived in Pauluis \textit{et al.} (2010, hereafter referred to as P10), which correspond to the (pseudo-) adiabatic potential temperatures $\theta_e$ or $\theta_l$.

The important issue in this paper is to determine which kind of enthalpy might be defined in an absolute way.
It is assumed that this is possible for the ``thermal'' enthalpies, which correspond to those enthalpies for N${}_2$, O${}_2$ and H${}_2$O substances that are generated by the variations of $c_p(T)$ generated by progressive excitations of the translational, rotational and vibrational states of the molecules, and by the possible changes of phase represented by the latent heats (with negligible impact of changes of pressure).

Both entropy and thermal enthalpy are thermodynamic state functions.
Therefore, both of them can be computed following any possible reversible path that connects the dead state at $0$~K and the actual atmospheric states.
The Third Law of thermodynamics can be used to set the entropy to zero for the dead states corresponding to the most stable cryogenic substances at $0$~K.
It is assumed that the same hypothesis can be applied for thermal enthalpies at $0$~K provided that the fluid is free of chemical or nuclear reactions.

The paper is organized as follows.
Section~\ref{subsection_dhdt_dTdt} explains in some detail why the enthalpy equation can be transformed into the temperature equation, with no impact of the reference values for enthalpies.
The importance of these reference values for computing the turbulent enthalpy fluxes is then illustrated in Section~\ref{subsection_enthalpy_fluxes} using results from B82.
Some impacts of reference values  on the computations of moist-air entropy are described in Section~\ref{subsection_specific_entropy}, where the essential results of M11 are recalled.
Comparisons  between the moist-air entropy derived in M11 and those described in P10 are presented in Section~\ref{subsection_entropy_P10}.
Following R22, it is explained, in Section~\ref{subsection_Richardson}, why it is possible to apply Nernst's result (namely the Third Law of thermodynamics) to compute the water entropies of moist air. 
In Section~\ref{subsection_unit_mass_dry_air}, it is shown that the method of computing moist-air enthalpy or entropy ``per unit mass of dry air'' does not obviate the need to determine the reference values of moist-air species in the case of open systems.
A review of different moist static energies is  briefly presented in Section~\ref{subsection_MSE} in order to list these well-known quantities, which are often considered to be similar to the sum of the moist-air enthalpy and the potential energy.

Section~\ref{section_data_method} corresponds to the data and method part of the paper.
The general formula for the specific moist-air thermal enthalpy is derived in Section~\ref{subsection_enthalpy}.
Computations of the standard and reference values are then presented in Section~\ref{subsection_STD_h} (details are given in Appendix~B).

Numerical applications are considered in Section~\ref{section_results}.
An analysis of the validity of the approximation $(h_d)_r \approx (h_l)_r$ is shown in  Section~\ref{subsection_Trouton} and provides new insights into the questions raised in B82.
Comparisons of the thermal enthalpy with different moist static energies are presented in Section~\ref{subsection_comp_h_MSE}.
The properties of the moist-air thermal enthalpy diagrams are analysed in Section~\ref{subsection_s_h_diagrams}.
Numerical values of the thermal enthalpy are computed and plotted in Section~\ref{subsection_h_TMSE_profiles} for several  stratocumulus and cumulus case studies (see Appendix~C).
The link between the thermal enthalpy diagram and the computation of the wet-bulb temperature are analysed in Section~\ref{subsection_Tw_psychro}, where several psychrometric equations are used and compared.

In Section~\ref{section_conclusion}, other applications are suggested for the various findings presented in this paper.

 \section{Review of results of moist-air thermodynamics.} 
 \label{section_Review}

    \subsection{Enthalpy and temperature equations.} 
    \label{subsection_dhdt_dTdt}

The enthalpy and temperature equations are described in some detail because reference will be made to them in later sections.
The enthalpy equation can be written as 
\begin{align}
  \frac{dh}{dt}  & = \:
           q_k \: \frac{dh_k}{dt} 
   \: + \: h_k \: \frac{dq_k}{dt} 
  \:  . \label{def_barycentric_1} 
\end{align}
Equation~(\ref{def_barycentric_1}) is the development of the material derivative of the implicit weighted sum $h=q_k\:h_k=\sum_k q_k\:h_k$, where the index $k$ applies to for dry air ($d$), water vapour ($v$), liquid water ($l$) and ice ($i$).
It is assumed that it is possible to write $h_k = c_{pk}\:T + b_k$ for each species, where $c_{pk}$ and $b_k$ are constant terms attributed to each gas.
Equation~(\ref{def_barycentric_2}) 
\begin{align}
  \frac{dh}{dt}  & = \:
           c_p \: \frac{dT}{dt} 
   \: + \: h_k \: \frac{d_i\: q_k}{dt} 
   \: + \: h_k \: \frac{d_e\: q_k}{dt} 
  \:   \label{def_barycentric_2}
\end{align}
is obtained from the properties $dh_k/dt = c_{pk}\: dT$ (no impact of  $b_k$), $c_p = q_k\:c_{pk}$ and $d q_k/dt = d_e\:q_k/dt + d_i\:q_k/dt$.
External changes $d_e\:q_k/dt = - \:{\rho}^{-1}\: \boldmath{\nabla} \, . \, \boldmath{J}_k$ are created by the divergence of the diffusion fluxes $\boldmath{J}_k$, where $\boldmath{J}_k = \rho_k \: (\boldmath{v}_k - \boldmath{v})$ represent departures from the barycentric mean $ \rho \: \boldmath{v}$.
Internal changes $d_i\:q_k/dt$ are created by phase changes of water.

The First Law equation (\ref{def_barycentric_3bis}) is expressed in terms of barycentric motion of an open parcel of fluid.
\begin{align}
  \frac{dh}{dt}  & = \:
   \frac{1}{\rho} \frac{dp}{dt} 
   \: - \:\frac{1}{\rho} \: 
   \boldmath{\nabla} \, . \left( h_k \: \boldmath{J}_k \, \right)
   \: + \: \dot{Q}
  \:  , \label{def_barycentric_3bis} 
 \\
  \frac{dh}{dt}  & = \:
   \frac{1}{\rho} \frac{dp}{dt} 
   \: - \:\frac{1}{\rho} \: \boldmath{J}_k \, . \boldmath{\nabla} h_k
   \: + \: h_k \: \frac{d_e\: q_k}{dt} 
   \: + \: \dot{Q}
  \:  . \label{def_barycentric_3} 
 \end{align}
The term involving change in pressure represents the expansion work.
The divergence of the enthalpy flux $h_k \: \boldmath{J}_k$ is separated into two parts in (\ref{def_barycentric_3}), with $- \:{\rho}^{-1}\: \boldmath{\nabla} \, . \, \boldmath{J}_k$ equal to the external changes $d_e\:q_k/dt$.
The heating rate $\dot{Q}$ corresponds to other diabatic processes (mainly radiation).

The temperature equation $dT/dt$ is then obtained by subtracting (\ref{def_barycentric_3}) from (\ref{def_barycentric_2}) and using the result $\boldmath{\nabla} h_k = c_{pk} . \boldmath{\nabla} T$ (no impact of  $b_k$), which leads to (\ref{def_barycentric_4}).
\begin{align}
  c_p \: \frac{dT}{dt} 
  & = \:
   \frac{1}{\rho} \frac{dp}{dt} 
   \: - \frac{1}{\rho} \: c_{pk} \: \boldmath{J}_k \,  . \boldmath{\nabla} T
   \: -  h_k \: \frac{d_i\: q_k}{dt} 
   \: + \: Q
  \:  . \label{def_barycentric_4} 
 \end{align}
The terms depending on external changes cancel out in the difference (\ref{def_barycentric_2}) minus (\ref{def_barycentric_3}).
The implicit sum depending on the diffusion fluxes $\boldmath{J}_k$ acts in regions where gradients of temperature exist.
The implicit sum depending on internal changes represents the impact of phase changes of water.
It can be written as $L_{vap}\: \dot{q}_{vap} + L_{sub}\: \dot{q}_{sub} + L_{fus}\: \dot{q}_{fus}$, in terms of the vaporization, sublimation and fusion rates.

Therefore all the terms in (\ref{def_barycentric_4}) can be easily computed and there is no need to determine the absolute values for enthalpies to express the First Law in terms of the temperature equation.
It is also possible to compute $d(c_p \:T)/dt$, since it is equal to $c_p \: dT/dt + T \: dc_p/dt$ and since the sum $c_p = \sum_k q_k \: c_{pk}$ is known.
These results correspond to the first part of the paradox.

    \subsection{Turbulent enthalpy fluxes: Businger (1982).} 
    \label{subsection_enthalpy_fluxes}

The problem of finding relevant values for the reference enthalpies $(h_k)_r$ is explicitly addressed  in B82, where the specific enthalpies are written as $h_k = c_{pk}\:T + b_k$ (as in the previous section), corresponding to the definitions $b_k = (h_k)_r - c_{pk}\:T_r$ and $h_k - (h_k)_r = c_{pk}\: (T - T_r)$.

It is a usual practice in atmospheric science to set reference enthalpies of dry air equal to zero for a given reference temperature, typically for $T_r = 0^{\circ}$C.
The choices of zero-enthalpy for water species are more variable.
It is set to zero either for the water vapour or for the liquid water enthalpies, for the same reference temperature $0^{\circ}$C and with the latent heats obviously connecting the other water-enthalpies through $L_{vap} = h_v - h_l$ or $L_{fus} = h_l - h_i$.
It is shown in B82 that the choices of zero-enthalpies for both dry air and liquid water are is in agreement with well-established procedures for computing surface turbulent fluxes.
Otherwise, additional fluxes of $q_t$ would have to be added, leading to other definitions of the moist-air enthalpy flux.
The same hypothesis $(h_d)_r = (h_l)_r = 0$ is used in the review by Fuehrer and Friehe (2002) and for $T_r = 0^{\circ}$C.

It is, however, unlikely that such arbitrary fluxes of $q_t$ may be added to or subtracted from the enthalpy flux, leading to arbitrary closure for the computation of turbulent fluxes of moist air.
The same is true for the vertical integral and the horizontal or vertical gradients of $h$, for which terms depending on $q_t$ could be of real importance if $q_t$ is not constant.
Accordingly, one of the aims of the coming sections will be to better justify these assumptions concerning the reference values of partial enthalpies, and possibly to define them more accurately for any value of the reference temperature $T_r$.

Another motivation for the search for accurate definitions of moist-air enthalpy (or entropy) is to try to answer the following question illustrated on the right side of Fig.(\ref{Fig_Barycentric}): Is it possible to determine whether two parcels of fluid have the same local values of $h$ or $s$?

It is easy to answer to this question for the motion of closed parcels of moist-air, for which $q_t$ is a constant and the $b_k$ may be cancelled, or for pseudo-adiabatic motion, for which it is possible to rely on the differential equations given by Saunders (1957) or Iribarne and Godson (1973).
These differential equations are established without the need for precise definitions of partial entropies.
These adiabatic or pseudo-adiabatic processes correspond to the definitions of the well-known liquid-water, equivalent and wet-bulb potential temperatures $\theta_l$, $\theta_e$ and $\theta'_w$, respectively.

Accordingly, the search for absolute and local definitions for $h$ and $s$ does not correspond to a change in the definitions of $\theta_l$, $\theta_e$ or $\theta'_w$ and the corresponding real temperatures.
A new way to answer these questions and to compute the specific moist-air ``thermal enthalpy'' will be found by examining the other aspect of the paradox, concerning the entropy, with the term ``thermal'' explained in the following sections.

    \subsection{Reference values of specific entropy.} 
    \label{subsection_specific_entropy}

The other aspect of the paradox concerns the entropy: it is easier to compute the local values of entropy than to manage its budget equation.
Like the enthalpy equation (\ref{def_barycentric_3}), the entropy equation derived by De Groot and Mazur (1962) is suitable for open systems.
As in Section~\ref{subsection_enthalpy_fluxes}, the Gibbs equation (\ref{def_Gibbs_1}) is expressed for barycentric motion with diffusion fluxes $\boldmath{J}_k$ and with the local Gibbs function defined by ${\mu}_k = h_k - T \: s_k$.
\begin{align}
  T \: \frac{ds}{dt}  & = \:
   \frac{dh}{dt} 
   \: - \: \frac{1}{\rho} \frac{dp}{dt} 
   \: - \: {\mu}_k \: \frac{d q_k}{dt} 
  \:  , \label{def_Gibbs_1} \\
  \frac{ds}{dt}  & =
   \: - \:\frac{1}{\rho} \: c_{pk} \: \frac{\boldmath{J}_k \, . \,  \boldmath{\nabla} T}{T}
   \: + \: s_k \: \frac{d_e\: q_k}{dt} 
   \: + \: \frac{\dot{Q}}{T}
  \:  . \label{def_Gibbs_2} 
 \end{align}
The First Law expressed by (\ref{def_barycentric_3}) can be introduced in (\ref{def_Gibbs_1}), leading to the entropy equation (\ref{def_Gibbs_2}).
A term $- \: ({\mu}_k/T )\: d_i\:q_k/ dt$ should appear on the right hand side of (\ref{def_Gibbs_2}) but it is equal to zero if it is assumed that changes of phases of water species are reversible and occur at equal Gibbs potential ${\mu}_k$ between the two phases.
This excludes the possibility of supercooled water for instance.

The first reason why it is difficult to evaluate the entropy equation accurately is due to another term, $\dot{S}_{irr}$, which should be added into the right hand side of (\ref{def_Gibbs_2}) in order to take account of other sources of irreversibility that exist in the real atmosphere.
An entropy equation similar to (\ref{def_Gibbs_2})  is derived in P10 for the simplified case of moist air composed of dry air, water vapour and liquid water (no ice).
This equation can be rewritten as
\begin{align}
  \frac{ds}{dt}  & =
   \:      \dot{S}_{irr}
   \: + \: \frac{\dot{Q}}{T}
   \: + \: (s_v-s) \: \dot{q}_{v} 
   \: + \: (s_l-s) \: \dot{q}_{l} 
 \:  , \label{def_P10_dsdt} 
 \end{align}
where the implicit sum $s_k \: d_e\:q_k/dt$ in (\ref{def_Gibbs_2}) is expressed in  (\ref{def_P10_dsdt}) in terms of the rate of change of $q_v$ and $q_l$ due to diffusion of water vapour and precipitation processes, denoted $\dot{q}_{v}$ and $\dot{q}_{l}$, respectively.
The impact of the diffusion term depending on $\boldmath{\nabla} T$ is not taken into account in (\ref{def_P10_dsdt}).

Another  reason why it is difficult to evaluate the entropy equations (\ref{def_Gibbs_2}) or (\ref{def_P10_dsdt}) accurately is due to the barycentric view, which corresponds to a unit mass of moist air.
The consequence is that external input or output of $q_v$ or $q_l$ must be balanced at the expense of opposite changes in the moist-air value $s$.
This explains the two terms that multiply $\dot{q}_{v}$ and $\dot{q}_{l}$ in (\ref{def_P10_dsdt}), while $s$ is approximated by $s_d$ in P10.

It is thus important to determine the absolute value of $s$ in order to compute, from (\ref{def_P10_dsdt}), the impact of diffusion of water (entrainment and detrainment at the edges of clouds), or the evaporation of water from the surface.
The same is true if the aim is to determine the impact of these processes on $h$ starting from the enthalpy equation (\ref{def_barycentric_3}).

It is possible to by-pass these difficulties.
The aim is to compute $s$ directly, without trying to compute $ds/dt$ via the associated source-sink terms.
Following HH87, it is explained in M11 that it is indeed possible to compute the specific moist-air entropy by determining the reference values from the Third Law, which states that the entropy of any substance is equal to zero for the most stable crystalline form of the substance and at absolute zero temperature (the Third Law cannot be applied to liquids or gases).

A Third-Law-based formulation for the specific moist-air entropy $s$ is written in M11 as
\begin{align}
  s \,(\theta_s)  & =  \: s_{ref} + c_{pd} \: \ln(\theta_s)
 \:  , \label{def_M11_s} \\
\theta_s   
   & = 
        \: \theta \;
         \exp \left( - \:
                     \frac{L_{vap}\:q_l + L_{sub}\:q_i}{{c}_{pd}\:T}
                \right)
        \exp \left( {\Lambda}_r\:q_t \right)
\nonumber \\
       &  \quad  \times \;
        \left( \frac{T}{T_r}\right)^{{\lambda} \:q_t}
        \left( \frac{p}{p_r}\right)^{-\kappa \:\delta \:q_t}
     \left(
      \frac{r_r}{r_v}
     \right)^{\gamma\:q_t}
  \;\;\;
      \frac{(1+\eta\:r_v)^{\:\kappa \: (1+\:\delta \:q_t)}}
           {(1+\eta\:r_r)^{\:\kappa \:\delta \:q_t}}
  \: . \label{def_M11_Thetas}
 \end{align}
Both $s_{ref}$ and $c_{pd}$ are constant terms and the entropy potential temperature $\theta_s$ depends on the adimensional parameter ${\Lambda}_r = [\: (s_v)_r - (s_d)_r \: ] \:/ c_{pd}$.
This parameter must be evaluated from the reference values of the partial entropies of dry-air and water-vapour components.

It is important to notice that the two notions of ``reference'' and ``standard'' entropies correspond to different numerical values.
The reference values $(s_v)_r$ and $(s_d)_r$ are associated with the reference temperature $T_r$, the total pressure $p_r$, the liquid-water saturation pressure $e_r$ and the water-vapour mixing ratio $r_r$.
These reference values are determined according to the standard values $s^0_v$ and $s^0_d$ which correspond to the same standard values for temperature and pressure for all species, with for instance $T_0=0^{\circ}$~C (or $T_0=25^{\circ}$~C) and $p_0=1000$~hPa, even for water vapour.
The Third-Law standard values of specific entropies are available in thermodynamic tables for all atmospheric species.
The values recalled in Appendix~A and computed with the datasets of Appendix~B are the same as those used in HH87 and M11 (see the end of Section~\ref{subsection_STD_h}).

The aim of the study is thus to arrive at a formulation for the moist-air enthalpy, $h$, that would be similar to results (\ref{def_M11_s}) and (\ref{def_M11_Thetas}) valid for the moist air entropy, $s$.

    \subsection{Moist entropy formulations: Pauluis \textit{et al.} (2010).} 
    \label{subsection_entropy_P10}

Since the aim of the article is to compute the moist-air enthalpy by determining the reference and standard enthalpies for each species, it is important to explain why the same method is relevant for the moist-air entropy and how it is possible to recover the well-known formulae used in atmospheric science and based on the equivalent and liquid-water potential temperatures $\theta_e$ and $\theta_l$, respectively.

A synthetic view of existing formulations of moist-air entropy derived in Iribarne and Godson, 1973, Betts, 1973 (hereafter referred to as B73); or Emanuel, 1994 (hereafter referred to as E94), is given in P10.
Two moist-air entropies are defined.
The first one is called  ``moist entropy'' and is denoted $S_m$ in P10.
It is written as $S_e$ in (\ref{def_P10_Se}), since it is associated with $\theta_e$.
\begin{align}
  S_e & = \: \left[ \: c_{pd} + q_t \: (c_l-c_{pd} )\: \right] 
         \ln\left( \frac{T}{T_r}\right)
       + \: q_v \:\frac{L_{vap}}{T}
       - \: q_d \: R_d \: \ln\left( \frac{p-e}{p_r-e_r}\right)
       - \: q_v  \: R_v \: \ln\left( \frac{e}{e_{sw}}\right)
   . \label{def_P10_Se}
\end{align}
The second one is given by (\ref{def_P10_Sl}).
It is called  ``dry entropy'' and denoted $S_l$ in P10.
It is associated with the B73' value $\theta_l$.
\begin{align}
  S_l & = \: \left[ \: c_{pd} + q_t \: (c_{pv}-c_{pd}) \: \right] 
         \ln\left( \frac{T}{T_r}\right)
       - \: q_l \:\frac{L_{vap}}{T}
       - \: q_d \: R_d \: \ln\left( \frac{p-e}{p_r-e_r}\right)
       - \: q_t  \: R_v \: \ln\left( \frac{e}{e_r}\right)
  . \label{def_P10_Sl}
\end{align}
It is possible to show that the difference between (\ref{def_P10_Sl}) and (\ref{def_P10_Se}) is equal to $S_e - S_l = q_t \: L_{vap}(T_r) / T_r$.
This is slightly different from the value given in (A5) of P10 because the term ${e}/{e_{sw}}$ was written as ${e}/{e_{sw}(T_r)}$ in (A3) of P10.
The important feature is that  $S_e - S_l$ depends on $q_t$ (both here and in P10).
Therefore, since there is only one physical definition for the moist-air entropy and since $q_t$ is not a constant in the real atmosphere, $S_e$ and $S_l$ cannot represent the more general form of moist-air entropy at the same time.

More precisely, the comparisons between the Third Law-based formulation $s(\theta_s)$ derived in M11 and the two formulations $S_e(\theta_e)$ or $S_l(\theta_l)$ can be written as
\begin{align}
  s \, (\theta_s) & = S_e (\theta_e) + q_t  \left[ \: (s_l)_r - (s_d)_r \: \right] + (s_d)_r
 \:  , \label{def_P10_s_Se} \\
  s  \, (\theta_s) & = S_l (\theta_l) + q_t  \left[ \: (s_v)_r - (s_d)_r \: \right] + (s_d)_r
 \:  . \label{def_P10_s_Sl}
\end{align}

The first result is that, if $q_t$ is a constant, then  $S_e$ and $S_l$ can become specialized versions of the moist-air entropy, associated with the use of the conservative variables $\theta_e$ and $\theta_l$, respectively.
However, even if $S_e$ and $S_l$ are equal to $s$ up to true constant terms, the constant terms are not equal to zero and they are not the same for $S_e$ and $S_l$.
Moreover, they depend on the value of $q_t$.
Therefore, even if $q_t$ is a constant (for instance for a given vertical ascent of moist air), it is not possible to compare values of $S_e$ or $S_l$ with those for other columns, since the values of $q_t$ and the ``constant'' terms in (\ref{def_P10_s_Se}) and  (\ref{def_P10_s_Sl}) differ from one column to another.
This means that it is not possible to compute relevant spatial averages, fluxes or gradients of $S_e(\theta_e)$ and $S_l(\theta_l)$, because the link between $\theta_e$ or $\theta_l$ and the moist-air entropy must change in space and in time.

The other result is that, if $q_t$ is not a constant, then $s = S_e + (s_d)_r$ only if $(s_l)_r = (s_d)_r$.
Similarly, $s = S_l + (s_d)_r$ only if $(s_v)_r = (s_d)_r$.
The fact that $S_e(\theta_e)$ or $S_l(\theta_l)$ may represent the moist air entropy for an open system and varying $q_t$ is thus dependent on these arbitrary choices for the reference entropies.

Since the Third Law is a general thermodynamic property, this study considers that only the Third-Law-based formulation $s (\theta_s)$ is general enough to allow relevant computations of spatial average, fluxes or gradients of moist air entropy, and thus of $\theta_s$.
This can be taken in relation with the problem studied in B82 for computing the moist-air surface turbulent fluxes, which are performed with arbitrary choices of the reference enthalpies.

It is suggested in Appendix~C of  P10 that the weighted average $S_a = (1-a)\:S_e + a \: S_l$, where $a$ is an arbitrary constant, is a valid definition of the entropy of moist air.
The adiabatic formulation $S_l$ corresponds to $a=1$ and $\theta_l$, whereas the pseudo-adiabatic formulation $S_e$ corresponds to  $a=0$ and $\theta_e$.
The weighted sum  $S_a$ applied to (\ref{def_P10_s_Se}) and (\ref{def_P10_s_Sl}) leads to
\begin{align}
  s  \, (\theta_s) & \; = \: S_a \:+\: (s_d)_r 
    \:+\: q_t \: \left[ \: (s_l)_r - (s_d)_r \:+\: a \; \frac{L_{vap}(T_r)}{T_r} \: \right] 
  . \label{def_P10_Sa} 
\end{align}
This result (\ref{def_P10_Sa}) shows that, if the term by which $q_t$ is multiplied is not equal to zero, $S_a$ will be different from the Third Law formulation $s$.
If the value $\Lambda_r \approx 5.87$ of M11 is considered, this term is equal to zero for $a = [ \: (s_d)_r - (s_l)_r \: ]  \: T_r / L_{vap}(T_r) \approx 0.356$, which represents (and allows the measurement of) the specific entropy of moist air in all circumstances since no other hypothesis is made concerning the values of the reference entropies, or on adiabatic or pseudo-adiabatic properties, or on constant values for $q_t$.
This provides another explanation for the result derived in M11: the Third-Law potential temperature $\theta_s$ is in a position about ``$(1-a)$'' $\approx 2/3$ versus ``$a$'' $\approx 1/3$  between $\theta_l$ and $\theta_e$. 

The Third Law cannot be by-passed when evaluating the general formula of moist air entropy. 
Reference values must be set to the standard ones obtained with zero-entropy for the most stable crystalline form at $T=0$~K.
If $q_t$ is not a constant, the formulations for $s(\theta_s)$, $S_e$, $S_l$ and $S_a$ are thus different.
In M11, it is claimed that the Third-Law formulation $s(\theta_s)$ is the more general one and, in this work, we present results to further support this claim.

    \subsection{Richardson's view (1922).} 
    \label{subsection_Richardson}

The issue of whether the Third Law can be applied to atmospheric studies or not, and how this can be managed practically, is an old question.
Richardson (R22, pp.159-160) already wondered if it could be possible to ascribe a value to energy and entropy for a unit mass of (water) substance.
He first proposed taking absolute zero temperature as the zero origin of entropies.
He recognized that the entropy varied as $c_p \:dT/T$ and could cause the integral to take infinite values when $T=0$~K.
But he mentioned that Nernst had shown that the specific heats tended to zero at $T=0$~K in such a way that the entropy remained finite there.
This was due to Debye's Law, which is valid for all solids and for which $c_p(T)$ is proportional to $T^3$.
The entropy defined by $ds=c_p\:dT/T$ is thus proportional to $T^3$, which is not singular at $T = 0$~K.

However, Richardson did not use the Third Law and he suggested considering the lowest temperature occurring in the atmosphere ($180$~K) as the most practical value for the zero origin of entropies.
This is in contradiction with the above conclusions of the comparisons of $s$, $S_e$ and $S_l$, and it is shown in HH87 and M11 that it is indeed possible to use the Third Law in atmospheric science.

The Third Law is not used in P10.
It is explained that the entropy of an ideal gas is fundamentally incompatible with Nernst's theorem as it is singular for $T$ approaching $0$~K.
This statement is true for ideal gases, but it does not invalidate the application of the Third Law in atmospheric science since only the most stable solid states and Debye's Law should be considered to apply Nernst's theorem.

According to the advice of Richardson and to the conclusion of comparisons between the Third Law entropy with previous results recalled in $S_a$ of P10, the aim of the article will be to mimic what is done in M11 for entropy in order to compute the thermal enthalpy (equal to the internal energy plus $R\:T$) for a unit mass of moist air.

    \subsection{Views ``per unit mass of dry air''.} 
    \label{subsection_unit_mass_dry_air}

There is another possibility that could avoid the need to use absolute values for entropies or enthalpies, and thus the Third Law.
The method is to assume that $q_d=1-q_t$ is a constant and to express moist-air entropy and enthalpy ``per unit mass of dry air'', and not as specific values expressed ``per unit mass of moist air''.
This is the choice made, in particular, in Normand, 1921 (hereafter referred to as N21), B73 and E94, where $s/q_d$ is expressed as any of
\begin{align}
  s/q_d  & = \: s_d + r_v \: s_v + r_l \: s_l
 \:  , \label{def_s_qd_1}  \\
  s/q_d  & = \: s_d + r_t\: s_l + r_v \: (s_v - s_l) 
 \:  , \label{def_s_qd_2}  \\
  s/q_d  & = \: s_d + r_t\: s_v - r_l \: (s_v - s_l) 
 \:  , \label{def_s_qd_3} 
\end{align}
with $r_t = r_v+r_l$ the total water mixing ratio.
Similar formulae are valid for the enthalpy expressed per unit of dry air if $s$ is replaced by $h$.
These formulae are valid for a mixture of dry air, water vapour and liquid water, with (\ref{def_s_qd_2}) and (\ref{def_s_qd_3}) corresponding to $S_e(\theta_e)$ and $S_l(\theta_l)$ respectively.

Clearly, the term $s_d$ can be transformed into the sum of $[ \: s_d - (s_d)_r \: ]$ plus $(s_d)_r$ on the right hand sides of (\ref{def_s_qd_2}) and (\ref{def_s_qd_3}).
The reference value $(s_d)_r$ is a global offset having no physical meaning.
The bracketed terms can be easily computed, together with $s_v - s_l$ and  $s_v - s_i$ in (\ref{def_s_qd_2}) and (\ref{def_s_qd_3}), which are equal to $L_{vap}/T$ and $L_{sub}/T$.

However, the reference values involved in $s_l = (s_l)_r + [ \: s_l - (s_l)_r \: ]$ in (\ref{def_s_qd_2}), or in $s_v = (s_v)_r + [ \: s_v - (s_v)_r \: ]$, in (\ref{def_s_qd_3}) are multiplied by $r_t$ and they can be neglected if and only if  $r_t$ is a constant.
Conversely,  they would acquire physical meanings in the regions and  for the processes where $r_t$ is not a constant.
Moreover, the study of the term $s/q_d$ would become irrelevant for the case of open systems  and varying $q_d$, for which it would be necessary to multiply the right hand sides by $q_d=1-q_t$ in order to properly compute $s$, with the terms $q_d \: (s_d)_r$ taking on a physical meaning.
There is thus no real improvement with regard to the study of specific values as in (\ref{def_P10_Se}) and (\ref{def_P10_Sl}).

The aim of the article is thus to apply the method suggested in R22 and demonstrated in M11 for the entropy, in order to derive a formulation of the specific moist-air thermal enthalpy $h$  with a minimum of hypotheses.
The result is intended to be valid for varying $q_t$ and with the reference values of enthalpies defined from physical properties, and not prescribed arbitrarily.

    \subsection{A review of various MSE quantities.} 
    \label{subsection_MSE}

Specific values of moist-air enthalpy are often computed by using the well-known MSEs.
There are, however, several  formulations for MSEs, each of them corresponding to different assumptions for the zero-enthalpies of dry air, water vapour or liquid-water, and to different ways of deriving MSE from either the First or Second Laws.

On the one hand, MSE formulations are derived from the Second Law and are often presented as being similar to the equivalent potential temperature $\theta_e$, which is  conserved during pseudo-adiabatic ascent or descent of moist-air parcels (N21; Madden and Robitaille, 1970; Betts, 1974; Arakawa and Schubert, 1974; E94; and Ambaum, 2010, hereafter referred to as MR70, B74, AS74 and A10).

The reason why it is possible to associate the moist-air enthalpy, $h$, with MSE quantities (and thus with the Second Law) is to be found in the Gibbs equation (\ref{def_Gibbs_1}).
If steady state vertical motions are considered, the material derivative reduces to $d/dt = w \: {\partial}/{\partial z}$, with all the terms in (\ref{def_Gibbs_1}) being multiplied by the vertical velocity $w$, which can be omitted hereafter.
For vertical hydrostatic motion $- \: {\rho}^{-1}\:{\partial p}/{\partial z} = {\partial \phi}/{\partial z}$.
If the parcel is closed and undergoes reversible adiabatic processes, then ${\partial s}/{\partial z} = 0$ and ${\mu}_k \:{\partial q_k}/{\partial z} = 0$.
The stationary Gibbs equation can then be written as
\begin{align}
  T \: \frac{\partial s}{\partial z}  
  & = \: 
   \frac{\partial ( h + \phi)}{\partial z} 
  = \: 0 
  \:  . \label{def_Gibbs_MSE} 
 \end{align}
The quantity $h + \phi$ is called the generalized enthalpy in A10.
It is thus a quantity that is conserved for vertical motion and provided that all the previous assumptions are valid.

On the other hand, the MSEs are often interpreted as generalized forms of energy or enthalpy (First Law).
For instance, the MSE function is interpreted as the non-kinetic part of the total energy in Emanuel (2004) and Peterson \textit{et al.} (2011), with the thermal part representing the specific moist enthalpy in E94, or the fluctuating part of the total specific enthalpy of a parcel in A10.

The Third-Law-based thermal enthalpy, $h$, derived in the next section will be compared with a selection of existing MSE formulae, which can be written as
\begin{align}
\mbox{MSE}_d
  & = \:
  c_{pd}\:T + L_{vap}\:q_v \:+ \phi
  \: , \label{def_MSEd} \\
\mbox{MSE}_l
 & = \:c_{pd}\:T - L_{vap} \:q_l
       \:+ \phi
  \: , \label{def_MSEl} \\
\mbox{LIMSE}
 & = \:c_{pd}\:T - L_{vap} \:q_l - L_{sub} \:q_i
       \:+ \phi
  \: , \label{def_LIMSE} \\
h^{\star}_v
  & = \:
  \left[ \:
     c_{pd} + (c_l- c_{pd}) \: q_t \:
 \right] T + L_{vap}\:q_v \:+ \phi
  \: , \label{def_MSE_A10} \\
\mbox{MSE}_m
  & = \:
  c_p\:T + L_{vap}\:q_v \:+ \phi
  \: . \label{def_MSEm}
\end{align}
These MSE quantities are made up of three parts.
The first part is the product of the local temperature by a moist-air specific heat, the second part is the product(s) of latent heats by specific contents, and the third part is the potential energy $\phi = g \: z$.
The ``thermal'' counterparts of (\ref{def_MSEd}), (\ref{def_MSEl}), (\ref{def_MSE_A10}) and (\ref{def_MSEm}), where the potential energy is removed, are noted  TMSE$_d$, TMSE$_l$, $h^{\star}_v-\phi$ and TMSE$_m$, respectively.

The quantity MSE$_d$ given by (\ref{def_MSEd}) is the most popular.
It is considered in AS74 as ``approximately conserved by individual air parcels during moist adiabatic processes''.
Betts, 1975 (hereafter referred to as  B75), mentions that it is  an approximate analogue of the equivalent potential temperature $\theta_e$.
It is used as a conserved variable for defining saturated updraughts in some deep-convection schemes (Bougeault, 1985).

A liquid-water version MSE$_l$ is defined in B75 by removing the quantity $L_{vap}\:q_t$ (assumed to be a constant) from (\ref{def_MSEd}), leading to (\ref{def_MSEl}).
It is considered in B75 that MSE$_l$ is an analogue of the liquid-water potential temperature $\theta_l$.
The formulation MSE$_l$ is generalized in Khairoutdinov and Randall (2003) and in Bretherton \textit{et al.} (2005) by removing a term $L_{sub} \:q_i$ from (\ref{def_MSEl}), for the sake of symmetry, which leads to the liquid-ice static energy given by (\ref{def_LIMSE}).
LIMSE is used as a conserved variable for defining saturated updraughts in some shallow-convection schemes (Bechtold \textit{et al.}, 2001).

The generalized enthalpy $h^{\star}_v$ defined by Eq.(5.37) in A10 is given by (\ref{def_MSE_A10}).
The ``moist enthalpy'' (per unit of dry air) defined by Eq.(4.5.4) in E94 is equivalent to $k=(h^{\star}_v-\phi)/q_d$.
Both $h^{\star}_v$ and $k$ are computed in A10 and E94 with the assumptions $h^0_d=h^0_l=0$.
MSE is sometimes defined with  $c_{pd}$ replaced in (\ref{def_MSEd}) by the moist value $c_p$, leading to MSE$_m$ given by (\ref{def_MSEm}).
This version, MSE$_m$, is used in some convective schemes (e.g. Gerard \textit{et al.}, 2009), because $d(c_p\:T)/dt$ can be easily computed from $c_p \: dT/dt$.

The concept of MSE was  not explicitly introduced in Dufour et van Mieghem (1975,  hereafter referred to as DVM75). 
These authors were mainly concerned with extensive versions for enthalpy and entropy, computed for a mass $m$ of moist air, i.e. with $H = m \: h$ and $S = m \: s$.

 \section{Data and method.} 
 \label{section_data_method}

    \subsection{Specific moist-air thermal enthalpy function.} 
    \label{subsection_enthalpy}

The same method already used in M11 is followed in this section to derive the formulation of the specific moist-air thermal enthalpy $h$ in terms of a ``moist enthalpy temperature'' $T_h$, similar to the moist-static-energy temperature MSE$_d/c_{pd}$ introduced in Derbyshire \textit{et al.} (2004).

The moist atmosphere is considered as a mixture of different ideal gases, mainly composed of N${}_2$, O${}_2$, Ar and CO${}_2$ for dry air, plus the three phases of the water species H${}_2$O (vapour, liquid or solid).
It is assumed that the volume of the condensed water species can be neglected, although their impact on the moist definitions of $c_p$ is taken into account.
The dry air is a mixture of 79~\%  N${}_2$ plus 20~\%  O${}_2$, with less than $1$~\% of Ar, CO${}_2$ and other gases.
If the chemical reactions (like the ozone-oxygen cycle in the stratosphere) are neglected , the thermodynamic properties of dry air are thus determined at $99$~\% by the observed and constant concentrations of the gases N${}_2$ and O${}_2$, with no tropospheric sources or sinks for these two gases.

In contrast, the specific contents for the three phases of H${}_2$O are  highly variable in time and space, with the evaporation and the precipitation processes acting as tropospheric sources and sinks, respectively.
The moist atmosphere is thus a multi-component mixture of different ideal gases, i.e. dry air plus vapour, and liquid and solid water species.

Since specific enthalpy is an additive function (Dalton's Law), the moist-air formulation is equal to the weighted average of the individual values for the partial specific enthalpies of the dry air, water vapour, liquid water and ice species, leading to
\begin{align}
  h  & = \; q_d \: h_d  \: + \: q_v \: h_v  \: + \: q_l \: h_l  \: + \: q_i \: h_i 
  \:  . \label{def_h0} 
 \end{align}
The terms can be rearranged (with the result $q_t=q_v+q_l+q_i$), yielding
\begin{align}
  h  & = \; q_d \: h_d  \: + \: q_t \: h_v  \: - \left( \:  q_l \: L_{vap}  \: + \: q_i \: L_{sub} \: \right) 
  , \label{def_h1}
 \end{align}
where the latent heats of vaporization and sublimation are equal to $L_{vap} = h_v - h_l$ and $L_{sub} = h_v - h_i$.
As in M11, the assumption $q_d = 1 - q_t$ is made, with precipitations considered as equivalent to cloud contents and taken to be at the same temperature as the rest of the parcel. 
The result is
\begin{align}
  h  & = \; h_d  
         \: + \: q_t \: \left(  h_v  - h_d \right)  
         \: - \left( \, q_l \: L_{vap} + \: q_i \: L_{sub} \, \right) 
  . \label{def_h2}
 \end{align}
The next step is to express the enthalpies by linearizing around some reference value $T_r$, with the hypotheses of constant values for all the specific heats ${c}_{pd}$ to ${c}_i$ in the atmospheric range of temperature (i.e. from $180$ to $320$~K).
This leads to 
\begin{align}
  h_d  & = \; (h_d)_r \: + \: {c}_{pd} \: ( \: T \: - \: T_r \: ) 
  \: , \label{def_hdr} \\
  h_v  & = \; (h_v)_r \: + \: {c}_{pv} \: ( \: T \: - \: T_r \: ) 
  \: , \label{def_hvr} \\
  h_l  & = \; (h_l)_r \: + \: {c}_{l} \: ( \: T \: - \: T_r \: ) 
  \: , \label{def_hlr} \\
  h_i  & = \; (h_i)_r \: + \: {c}_{i} \: ( \: T \: - \: T_r \: ) 
  \: .  \label{def_hir}
\end{align}
The aim of the next section will be to compute the reference values $(h_d)_r$ and $(h_i)_r$, with $(h_v)_r$ and $(h_l)_r$ determined from $(h_i)_r$ by the latent heats $L_{sub}(T_r)$ and $L_{fus}(T_r)$.

If (\ref{def_hdr}) and (\ref{def_hvr}) are inserted into (\ref{def_h2}), and after rearrangement of the terms, the moist enthalpy can be written as
\begin{align}
  h  & = \; \left[ \: (h_d)_r \: - \: {c}_{pd} \: T_r \: \right]  
   \nonumber \\
     & \quad
      + \:  {c}_{pd} \: T \: \left[ 
              \: 1 \: + \: \lambda \:q_t 
                   \: - \: \left( \:
                   \frac{L_{vap}\:q_l + L_{sub}\:q_i}{{c}_{pd}\: T}
                   \: \right)
           \: \right]
   \nonumber \\
     & \quad   \: + \: q_t \: \left[ \: (h_v)_r \: - \: (h_d)_r \:
         \: - \: ( {c}_{pv} \: - \:{c}_{pd} \: ) \: T_r \: \right] 
  . \label{def_h3}
\end{align}
The formulation (\ref{def_h3}) is equivalently written by the system (\ref{def_h})-(\ref{def_Upsilon}), in a similar way to the moist entropy formulation (\ref{def_M11_s}) derived in M11 in terms of $\theta_s$, but this time in terms of a moist enthalpy temperature $T_h$, yielding
\begin{align}
 h  & = \: h_{ref}  \: + \: {c}_{pd} \: T_h 
   \: . \label{def_h}
\end{align}
The first line of (\ref{def_h3}) defines $h_{ref}$ by (\ref{def_Href}).
The term ${c}_{pd} \: T_h$  and the dimensionless upsilon-term ${\Upsilon\!}_r$ correspond to the last two  lines of (\ref{def_h3}), leading to (\ref{def_Th})-(\ref{def_Upsilon}).
\begin{align}
  h_{ref}  & = \; (h_{d})_r  \: - \: {c}_{pd} \: T_r 
   \: , \label{def_Href} \\
   T_h  & = \: T_{il}
           \: + \: 
     \lambda \; T \: q_t 
           \; + \: 
     \left[ \: 
       T_r \: \left( {\Upsilon\!}_r - \lambda \right)
      \: \right] \: q_t
  \: , \label{def_Th} \\
   T_{il}  & = \: T 
           \: \left[ 
                   \: 1 
                    - \left( \:
                   \frac{L_{vap}(T)\:q_l + L_{sub}(T)\:q_i}{{c}_{pd}\: T}
                   \: \right)
           \: \right]
   , \label{def_T_l}  \\
{\Upsilon\!}_r & = \; \frac{ (h_{v})_r - (h_{d})_r }{ c_{pd} \; T_r}
  \: . \label{def_Upsilon}
\end{align}

It is worth noting that the ice-liquid temperature $T_{il}$ corresponds to a generalization for non-zero $q_i$ of the liquid-water (potential) temperature defined in B73 and involved in the first line of $\theta_s$ recalled in (\ref{def_M11_Thetas}), in the limit where $q_i=0$ and $\exp(-x)\approx 1 - x $ for small $x$.
It is more general than the ice-liquid water (potential) temperature defined in Tripoli and Cotton (1981), in that the constant values $L^0_{vap}(T_0)$ and $L^0_{sub}(T_0)$  are replaced by the varying values $L_{vap}(T)$ and $L_{sub}(T)$ in (\ref{def_T_l}).

\begin{figure}[t]
\centering
\includegraphics[width=0.7\linewidth,angle=0,clip=true]{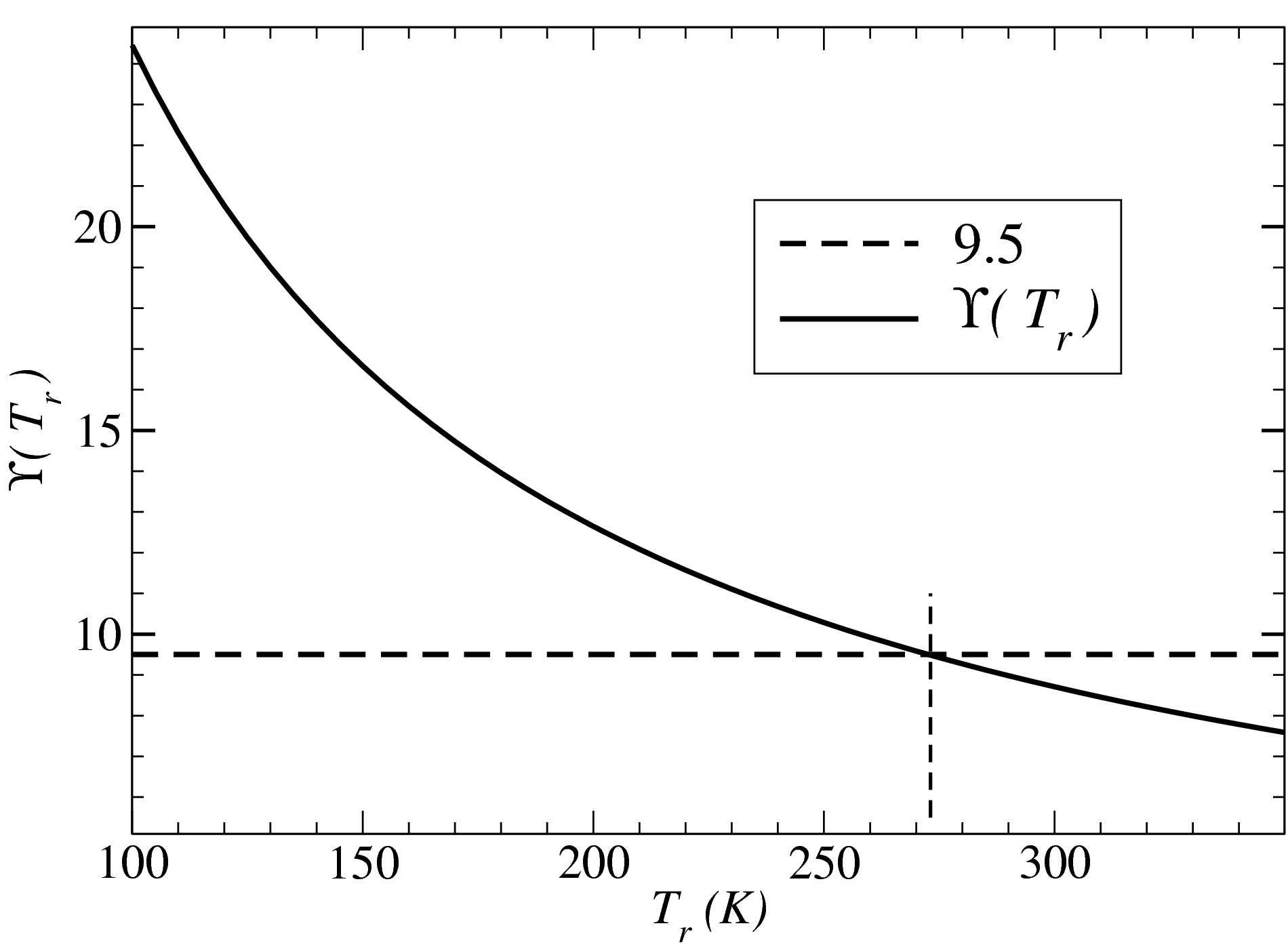}
\caption{
{\it \small
An analysis of the change with $T_r$ of the upsilon term ${\Upsilon\!}_r = {\Upsilon} (T_r)$ (solid curve), in comparison with the constant value ${\Upsilon\!}_0 = {\Upsilon} (T_0) = 9.5$ (dashed straight line).
}
\label{Fig_Upsilon_Tr}}
\end{figure}

In (\ref{def_Upsilon}), the Upsilon-term ${\Upsilon\!}_r$ depends on the absolute values for water-vapour and dry-air enthalpies, evaluated at the temperature $T_r$.
It acts as the dimensionless term $\Lambda_r= [\: (s_{v})_r - (s_{d})_r \:]/c_{pd}$ appearing in the entropy formulation (\ref{def_M11_Thetas}), and it varies with $T_r$ as illustrated in Figure~\ref{Fig_Upsilon_Tr}.

It is important to demonstrate that choosing another value of $T_r$ has no impact on $h$, $h_{ref}$ and $T_h$ as defined by (\ref{def_h}) to (\ref{def_Upsilon}) respectively.
Otherwise, if $h_{ref}$ could be dependent on $T_r$, the enthalpy temperature $T_h$ would not be equivalent to the moist enthalpy $h$, as is expected, and the moist thermal enthalpy itself would not have a clear physical meaning as a state function of the moist atmosphere.

This result is true for $h_{ref}$ because, from (\ref{def_Href}) and for constant values of ${c}_{pd}$, the variations of $(h_{d})_r$ with $T_r$ are exactly balanced by the term $- \:{c}_{pd} \: T_r$, leading to $(h_{d})_r  \: - \: {c}_{pd} \: T_r \: = \: h^0_{d}  \: - \: {c}_{pd} \: T_0$ which corresponds to the definition (\ref{def_hdr}) applied between $ T_r$ and $ T_0$.
The reference enthalpy (\ref{def_Href}) can thus be written as
\begin{align}
  h_{ref}  & = \;  h^0_{d}  \: - \: {c}_{pd} \: T_0 \; \approx \: 256  \mbox{~kJ} \mbox{~kg}^{-1}
   \: . \label{def_Href0}
\end{align}
The numerical value $256 $~kJ~kg${}^{-1}$   is obtained  from the standard value $h^0_{d}$ determined in the next section at $T_0$.

\begin{figure}[t]
\centering
\includegraphics[width=0.7\linewidth,angle=0,clip=true]{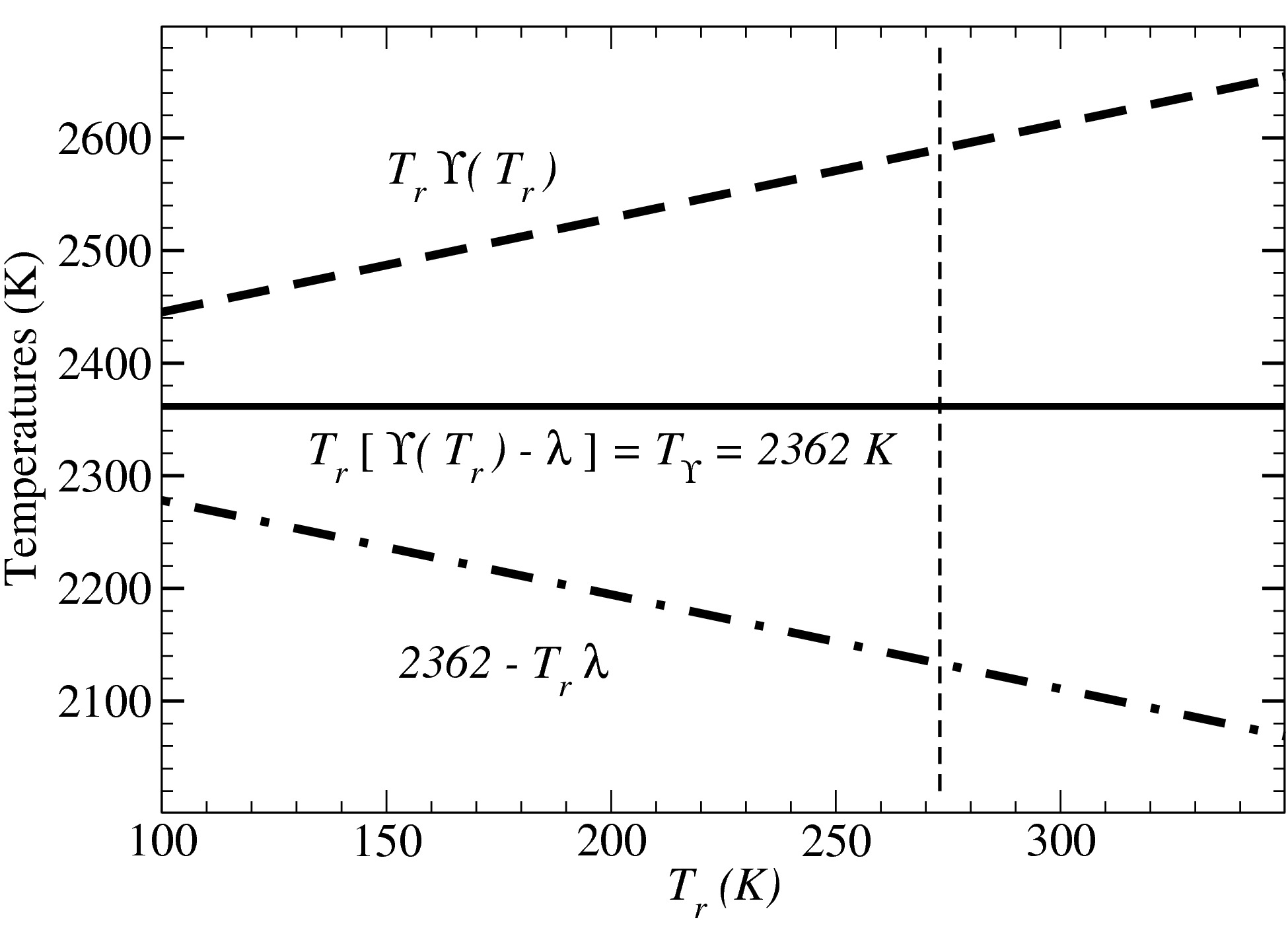}
\caption{
{\it \small
An analysis of the change with $T_r$ of the three terms:
``$\:T_r \: {\Upsilon} (T_r)$'' (dashed, increasing curve) ;
``$\:2362 \: - \:  T_r \: \lambda $''  (dotted-dashed, decreasing curve)
and 
``$\:T_r \: [ \: {\Upsilon} (T_r) \: - \: \lambda \: ]$''$ \: \equiv T_{\Upsilon}  \approx 2362$~K (solid, constant straight line).
}
\label{Fig_H_Cste_Tr}}
\end{figure}

It is shown in Figure~\ref{Fig_Upsilon_Tr} that ${\Upsilon\!}_r $ is not a constant and  decreases strongly with $T_r$. 
However, the bracketed  term in (\ref{def_Th}) is a constant term independent of $T_r$.
From the definitions $c_{pd} \: \lambda = c_{pv}- c_{pd}$ and (\ref{def_Upsilon}), the term $T_r \: ( {\Upsilon\!}_r - \lambda )\: q_t$ is equal to $q_t$ times $[\:(h_{v})_r - {c}_{pv} \: T_r \:] - [\:(h_{d})_r - {c}_{pd} \: T_r\:]$.
Moreover, from (\ref{def_hdr}) and (\ref{def_hvr}), neither $(h_{d})_r  \: - \: {c}_{pd} \: T_r$ nor $(h_{v})_r - {c}_{pv} \: T_r$ varies with $T_r$, as long as ${c}_{pd}$  and ${c}_{pv}$ are  assumed to be constant terms.
The result is that the bracketed  term in (\ref{def_Th}) is equal to
\begin{equation}
   T_r \: [ \: \Upsilon (T_r) - \lambda \: ] 
 \, = \,
   T_0 \: ( {\Upsilon\!}_0 - \lambda ) 
 \, \equiv \, 
   T_{\Upsilon} 
  \approx \: 2362  \mbox{~K}
  \, .  \label{def_Tr_Upsilon_cste} 
\end{equation}
$T_{\Upsilon}$ is a constant, due to compensating variations with $T_r$ of the two terms $T_r \:{\Upsilon}(T_r)$ and $- \: T_r \: \lambda$, as illustrated in Figure~\ref{Fig_H_Cste_Tr}.
The numerical value $T_{\Upsilon} = 2362$~K  is obtained from the standard values given in the next section, leading to ${\Upsilon\!}_0 = \Upsilon (T_0) = (h^0_v-h^0_d)/(c_{pd}\:T_0)\approx 9.5$.

An alternative, synthetic and compact expression for $h$ defined by (\ref{def_h})-(\ref{def_Tr_Upsilon_cste}) is given by
\begin{equation}
\boxed{\; \; 
 h  \;  = \: h_{ref}  
\: + \: 
{c}_{pd} \: T
 \: - \:
   L_{vap}\:q_l  
\: - \:  L_{sub}\:q_i
 \: +\: 
    {c}_{pd}  
\left( \:
  \lambda \; T
 \: + \: 
  T_{\Upsilon} 
 \: \right) \: q_t
 \; \; }
  \: . \label{def_h_bis}
\end{equation}
The constant terms $h_{ref}$ and $T_{\Upsilon}$ are given by (\ref{def_Href0}) and  (\ref{def_Tr_Upsilon_cste}).

It is worth noting that the formula (\ref{def_h_bis}) is symmetric with respect to the condensed water species: it is left unchanged if $q_l$ is replaced by $q_i$ provided that $L_{vap}$ is replaced by $L_{sub}$, where $L_{vap}$ and $L_{sub}$ connect the water vapour with the condensed phases ($q_l$ or $q_i$).

\subsection{Computations of standard partial enthalpies.} 
\label{subsection_STD_h}

One of the goals of this section is to determine the numerical values of the constants $h_{ref}$ and $T_{\Upsilon}$ appearing in (\ref{def_h_bis}).
They both depend, from (\ref{def_Href0})  and (\ref{def_Tr_Upsilon_cste}), on the standard values of the partial enthalpies $h^0_d$ and $h^0_v$.
The term ``standard'' means that the values are expressed at the zero Celsius temperature $T_0=273.15$~K and the conventional pressure $p_0=1000$~hPa.
The aim of this section is thus to provide a general method for computing the partial enthalpies for several components of moist air (O$_2$, N$_2$, H$_2$O) and at any temperature, in particular for the standard conditions.

The partial enthalpies must be understood as ``thermal enthalpies'', i.e. the enthalpies generated by the variations of $c_p(T)$ with $T$.
This corresponds to progressive excitations of the translational, rotational and vibrational states of the molecules, and by the possible changes of phase represented by the latent heats (with negligible impact of changes of pressure).
The consequence is that the concept of thermal enthalpy does not correspond to the concept of ``standard enthalpies of formation or reaction'' which are denoted by $\Delta H^0_f$ and  $\Delta H^0_r$ and are available in chemical tables for most species.

The concepts of standard enthalpies of formation  or reaction are useful in case of chemically reacting species, with the consequence that the relative concentrations of these species are determined by the equilibrium values depending on the law of mass action, depending on the reaction considered.
In that case, the concentrations depend on local pressure and temperature (as in the O${}_3$ chemistry regions).
But, except in the stratosphere, the concentrations of the species are free parameters with, for instance, tropospheric water contents that are modified by diffusion, convection or precipitation processes.
The standard enthalpies of formation or reaction thus cannot be relevant starting points for the determination of the tropospheric values of standard thermal enthalpies $h^0_d$ and $h^0_v$, and thus ${\Upsilon\!}_0(T_0)$.

Moreover, in chemical thermodynamics, it is assumed that the hypothesis $\Delta H^0_f=0$ is valid for all gaseous forms of pure species at the conventional pressure of $1000$~hPa and the standard temperature of $298$~K.
This is true, for instance, for O${}_2$ and N${}_2$ and thus for $99$~\% of dry air.
However, since the thermal enthalpies must depend on the variations of $c_p(T)$ with $T$ and on the latent heats, and since these quantities are different for each gas, the thermal enthalpies of O${}_2$ and N${}_2$ must be different in the standard conditions and $h\neq \Delta H^0_f$.

The  method for computing the thermal enthalpies of N${}_2$, O${}_2$ and H${}_2$O makes use of the important result that enthalpy is a state function: it can be evaluated by following any of the reversible paths that connect the dead state at $0$~K and the actual atmospheric state at temperature $T$.

The challenge is thus to compute the integral of $c_p(T)$ and the sum of all latent heats for a given substance from $0$~K to the standard temperature $T_0$, by following a reversible path involving the most stable forms of this substance at each temperature, first for the solid state(s), then for the liquid state, and finally for the gaseous form of the substance.
All these processes are represented by the various terms of the mathematical formula 
\begin{align}
  h^{0}(T_0)  & = \: h(T=0) 
  + \int^{T_0}_{0} \! c_p(T)\:  dT
  + \sum_k L_k
  \: . \label{def_abs_h}
\end{align}
Other forms of enthalpy are associated with the potential ($\phi = g \: z + \phi_0$), chemical or nuclear energies.

\begin{figure}[t]
\centering
\includegraphics[width=0.9\linewidth,angle=0,clip=true]{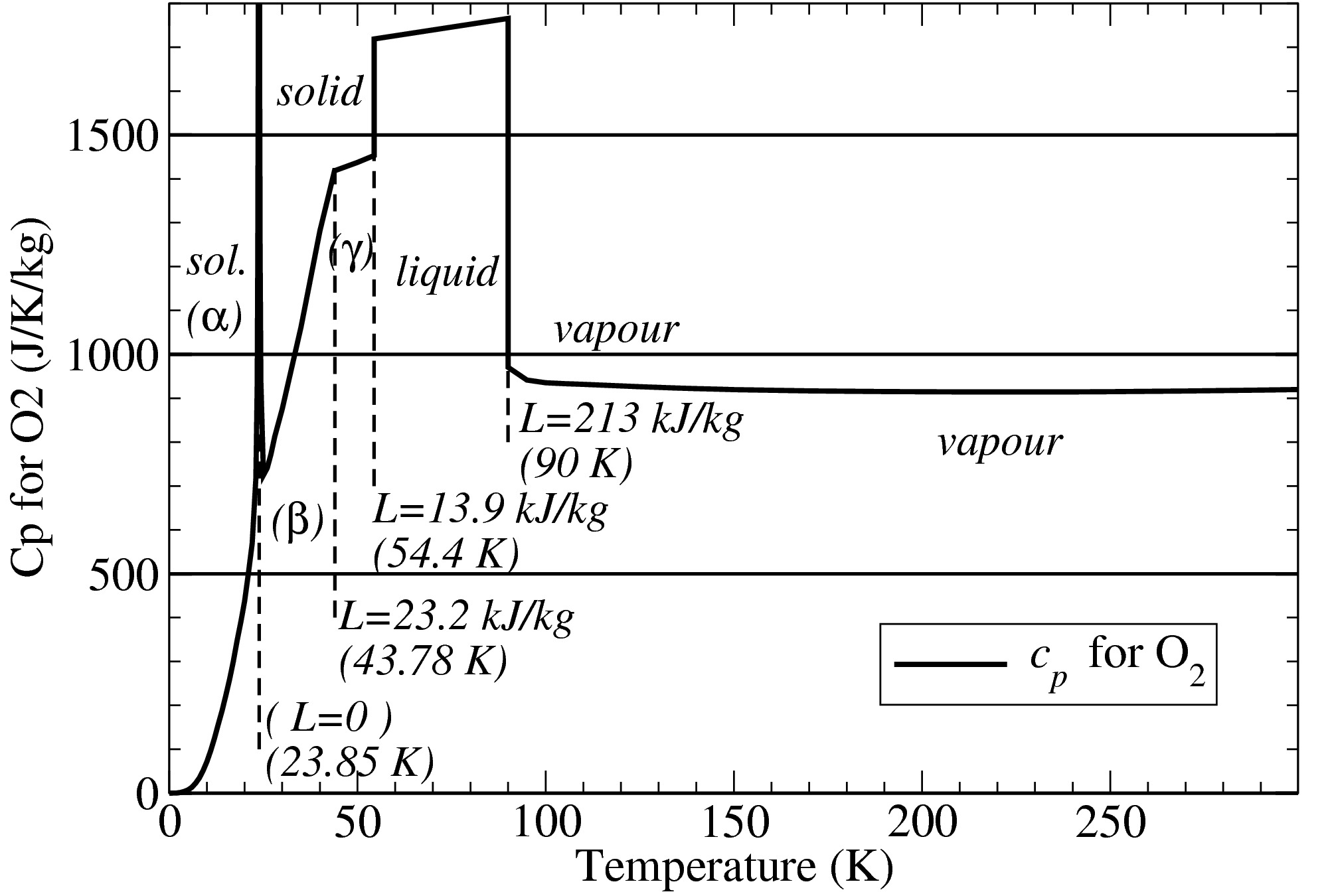}
\caption{
{\it \small
Specific heat capacity at constant pressure for  O${}_2$ corresponding to Table~\ref{Table_cp_O2} in Appendix~B.
Units of $c_p$ are J~K${}^{-1}$~kg${}^{-1}$.
The latent heats are in units of kJ~kg${}^{-1}$.
}
\label{fig_Cp_O2}}
\end{figure}

\begin{figure}[t]
\centering
\includegraphics[width=0.9\linewidth,angle=0,clip=true]{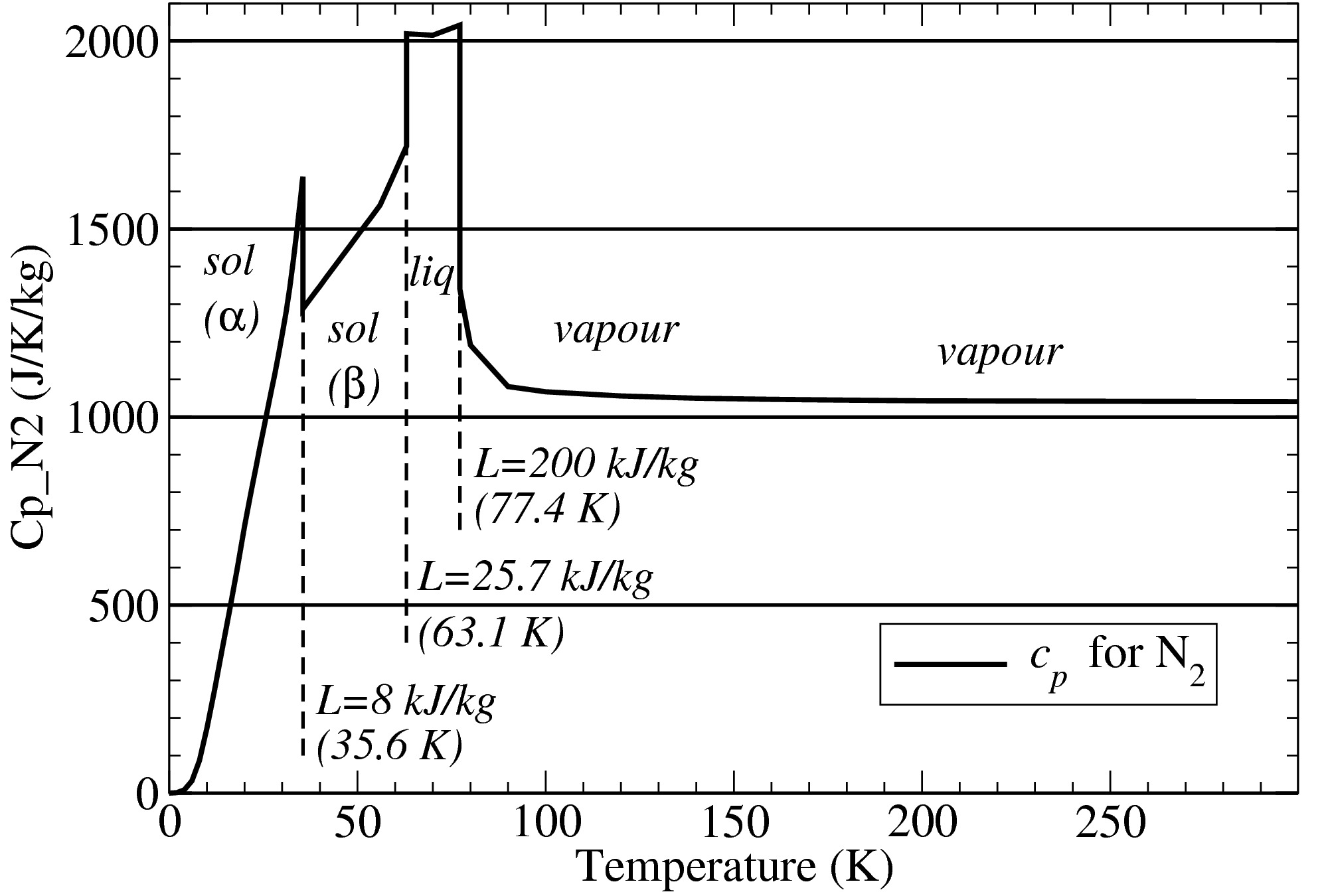}
\caption{
{\it \small
Specific heat capacity at constant pressure for  N${}_2$ corresponding to Table~\ref{Table_cp_N2} in Appendix~B.
Units of $c_p$ are J~K${}^{-1}$~kg${}^{-1}$.
The latent heats are in units of kJ~kg${}^{-1}$.
}
\label{fig_Cp_N2}}
\end{figure}

\begin{figure}[t] 
\centering
\includegraphics[width=0.8\linewidth,angle=0,clip=true]{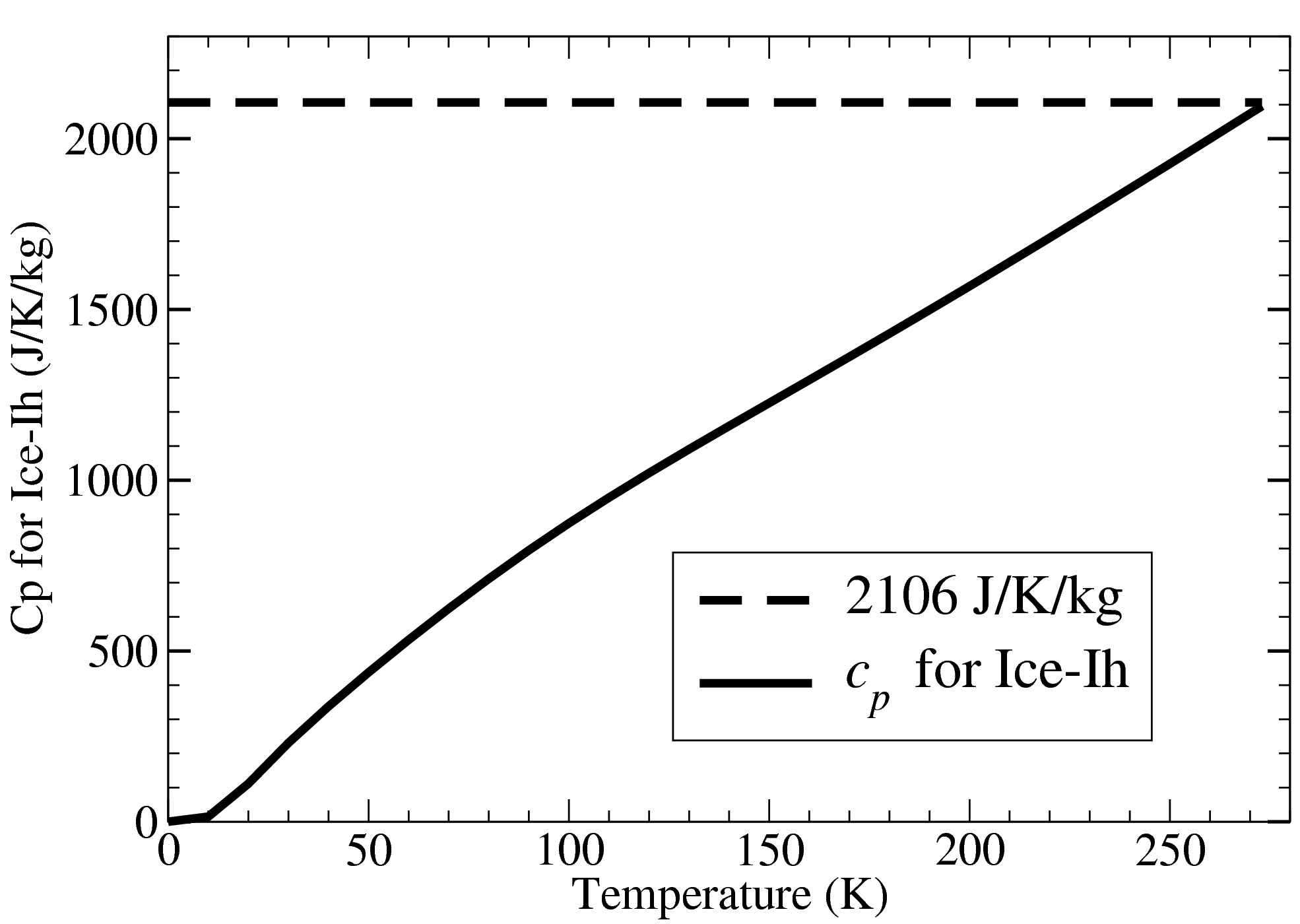}
\caption{
{\it \small
Specific heat capacity at constant pressure for ice-Ih (solid line) corresponding to Table~\ref{Table_H2O_ice} in Appendix~B, and the usual constant value $c_i = 2106$ (dashed line).
Units are J~K${}^{-1}$~kg${}^{-1}$.}
\label{fig_Cp_Ice}}
\end{figure}

The arbitrary reference level ${\phi}_0$ needed for defining $\phi$ is the same for all the molecules at a given point.
The impact of ${\phi}_0$ is thus to add or to retrieve a true constant term independent of $q_v$, $q_l$ and $q_i$.
Therefore ${\phi}_0$  has no physical impact on the energy or on the enthalpy functions.
This is true even from the barycentric standpoint of fluid dynamics, where differential fluxes of matter occur.

Nuclear reactions and the formation of atoms from elementary particles form a huge but dead state of energy in the real atmosphere.
However, these forms of unavailable energy are not taken into account in this study.
It is assumed that the main gases, N${}_2$, O${}_2$ and H${}_2$O, that compose the moist atmosphere only interact with the kinetic and potential energies via the standard thermal enthalpies defined by (\ref{def_abs_h}).

\begin{figure}[t]
\centering
\includegraphics[width=0.8\linewidth,angle=0,clip=true]{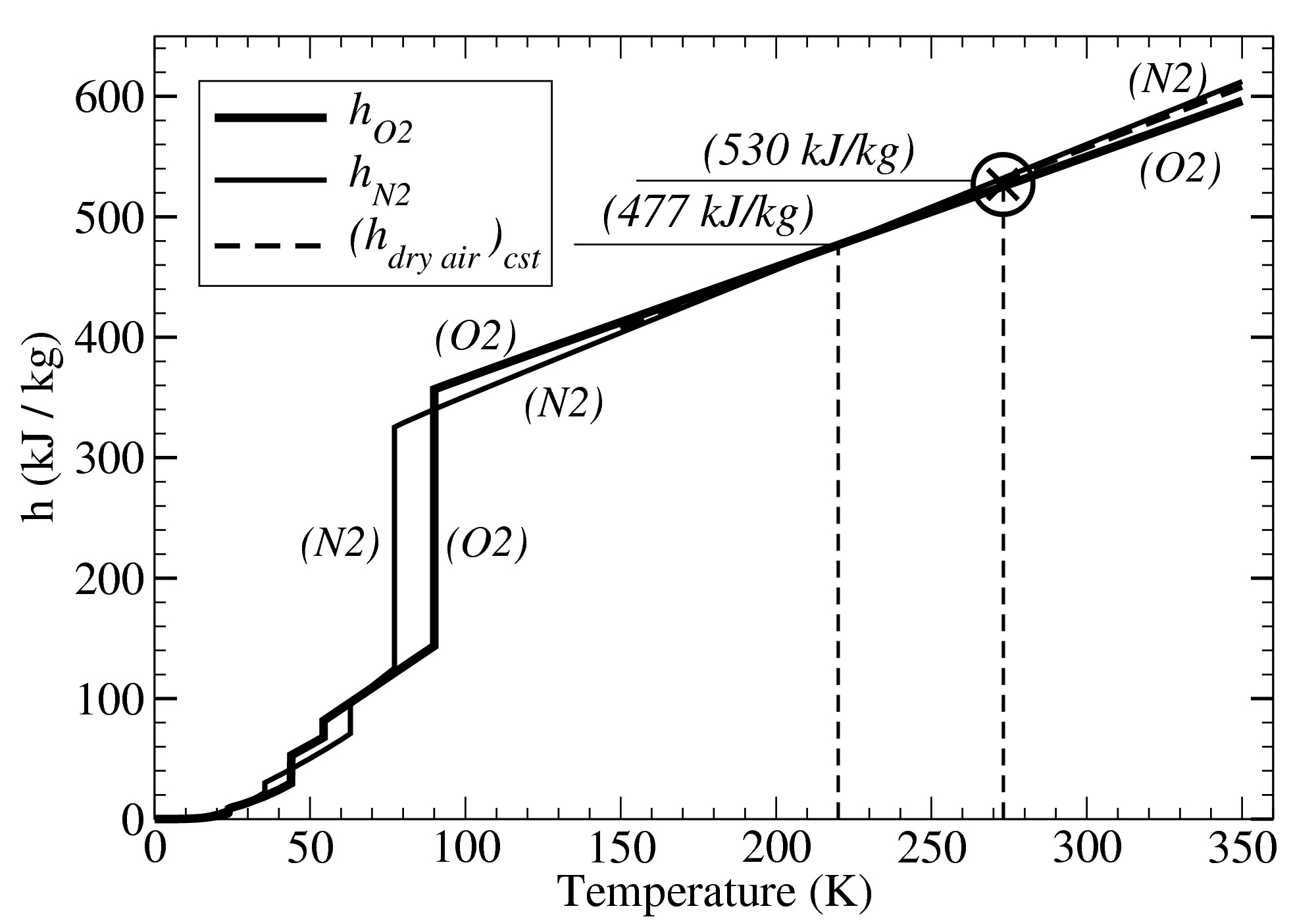}
\caption{
{\it \small
Enthalpies for O${}_2$ (bold solid line) and N${}_2$ (thin solid line) as a function of the absolute temperature, with units of kJ~kg${}^{-1}$.
The dry-air curve (dashed line) represents the dry-air enthalpy ($h_d$), computed and plotted from $100$ to $320$~K with the  ``constant specific heat capacity'' assumption ($c_{pd}=1004.7$~J~K${}^{-1}$~kg${}^{-1}$).
}
\label{fig_All_Enthalpies2}}
\end{figure}

\begin{figure}[t]
\centering
\includegraphics[width=0.8\linewidth,angle=0,clip=true]{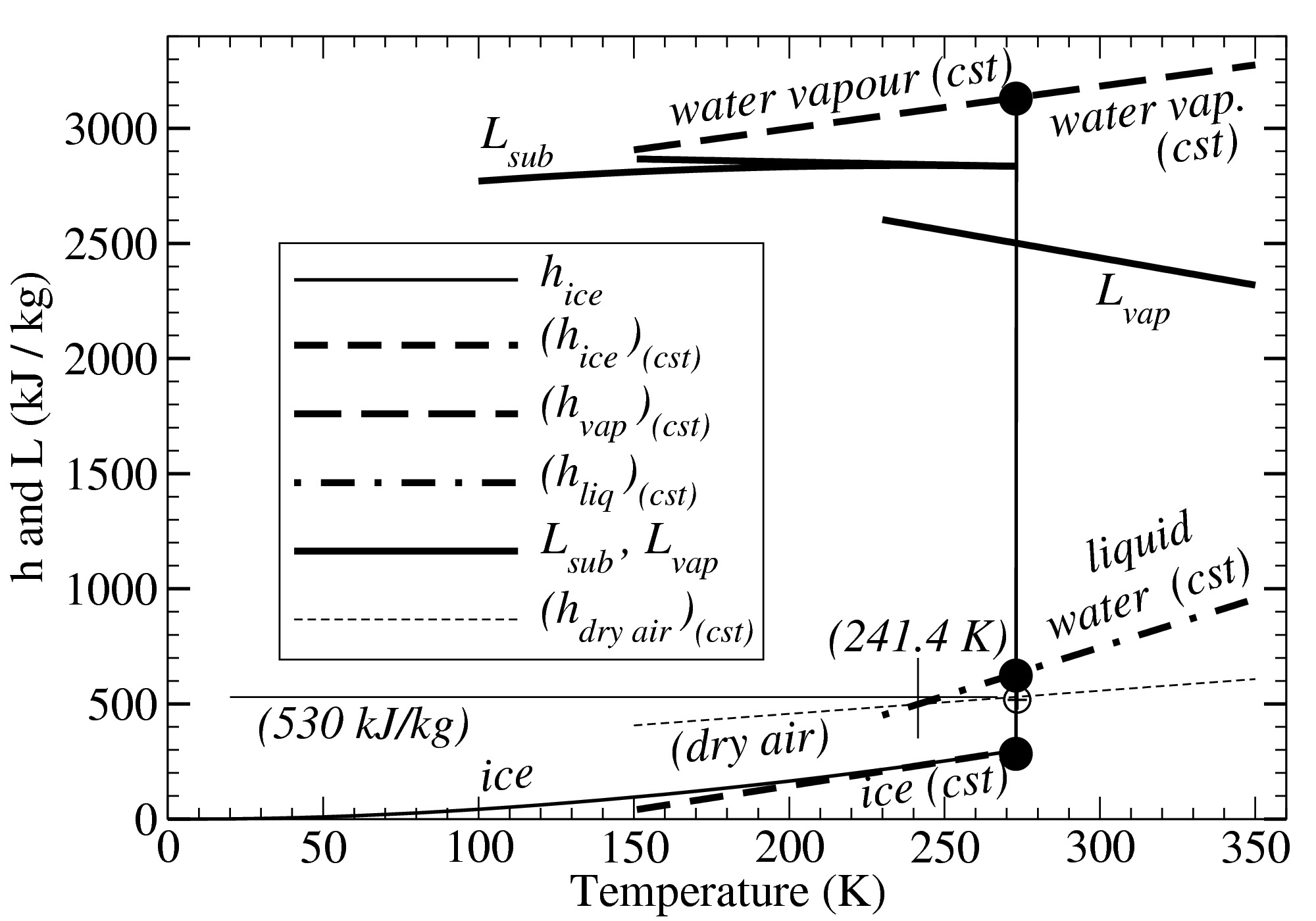}
\caption{
{\it \small
Enthalpies for dry-air and water species as a function of the absolute temperature, with units of kJ~kg${}^{-1}$.
The enthalpy of ice-Ih is plotted in the range of temperature from $0$ to $273.15$~K (thin solid line).
The three dashed lines represent the ice, liquid-water and water-vapour enthalpies computed with  ``constant specific heat capacity'' assumptions ($c_i=2106$, $c_l=4218$ and  $c_{pv}=1846.1$~J~K${}^{-1}$~kg${}^{-1}$).
The three bold dots represent the standard values $h^0_i \approx 298$, $h^0_l \approx 632$ and  $h^0_v \approx 3133$~J~kg${}^{-1}$.
Latent heat curves are described in the text.
}
\label{fig_All_Enthalpies1}}
\end{figure}

The integral of $c_p(T)$ in the thermal enthalpy formulation (\ref{def_abs_h}) can be computed by using the specific heats described in the Tables~\ref{Table_cp_O2}  to \ref{Table_H2O_ice} for  O${}_2$,  N${}_2$ and ice-Ih (see Appendix~B).
The latent heats are indicated in Figures~\ref{fig_Cp_O2}, \ref{fig_Cp_N2} and \ref{fig_All_Enthalpies1}.
The remaining problem is thus to determine the value of the thermal enthalpy at $0$~K.

The standard entropies used  in H87 and M11 are based on the Third Law.
It is assumed that a kind of Third Law can apply to the thermal enthalpy for all substances with no chemical reaction, i.e. zero-enthalpy for $T=0$~K.

This means that the thermal enthalpies of  N${}_2$, O${}_2$ and H${}_2$O are equal to zero for the most stable solid states at $0$~K (thermally dead states).
Otherwise, if any indeterminacy existed for the values of thermal enthalpies at $0$~K, except a possible identical global shift valid for all species, the thermal enthalpy of moist air could not be defined locally and would not have a clear physical meaning for all values of $T_0$ in (\ref{def_abs_h}).

The curves for the thermal enthalpies of dry-air species are shown in Figure~\ref{fig_All_Enthalpies2}.
They were computed with the values of $c_p(T)$ and the latent heats given in Appendix~B for O${}_2$ and N${}_2$ (Tables~\ref{Table_cp_O2} and \ref{Table_cp_N2}).
The thermal enthalpies of O${}_2$ and N${}_2$ have about the same value of $477$~kJ~kg${}^{-1}$ at $T=220$~K, with about the same value for dry air (composed at 99\% of O${}_2$ plus N${}_2$).
A thin horizontal line has been added to represent this common value $477$~J~kg${}^{-1}$.

According to Figures~\ref{fig_Cp_O2} and \ref{fig_Cp_N2}, the variation with $T$ of specific heat for O${}_2$ and N${}_2$ is small  above $150$~K.
This corresponds to constant values for $c_{pd}$ and therefore the dry-air thermal enthalpy can be approximated by
\begin{align}
  h_d (T)  & \approx \: 477  \mbox{~kJ} \mbox{~kg}^{-1}
  \: + \:c_{pd} \: ( \: T - 220 \mbox{~K} )
  \: \label{def_hd_term_220}
\end{align}
for $T>150$~K, leading to a standard value of $h^0_d $ of $530$~kJ~kg${}^{-1}$ at $273.15$~K represented by the ``crossed circle'' symbol on Figure~\ref{fig_All_Enthalpies2}.

The curves for the thermal enthalpies of water species are depicted in Figure~\ref{fig_All_Enthalpies1}.
They were computed with the values of $c_p(T)$ plotted in Figure~\ref{fig_Cp_Ice} and the specific heats given in Appendix~C for phase Ih of ice (Table~\ref{Table_H2O_ice}), plus the values of  latent heats of fusion and vaporization $L^0_{fus}$ and $L^0_{vap}$ given in Appendix~A.
The latent heats of sublimation ($L_{sub} \equiv h_v-h_i$) and vaporization ($L_{vap} \equiv h_v-h_l$) are depicted by solid bold lines.
The maximum at $T \approx 240$~K for $L_{sub}$ corresponds to a known thermodynamic property.

The same dry-air dashed line as in Figure~\ref{fig_All_Enthalpies2} is shown in Figure~\ref{fig_All_Enthalpies1} from $150$ to $320$~K, with the same ``crossed circle'' symbol at $T=273.15$~K.
The thin horizontal line represents the same constant value of $530$~J~kg${}^{-1}$ as plotted in Figure~\ref{fig_All_Enthalpies2}, in order to facilitate the comparison of the vertical scales in Figures~\ref{fig_All_Enthalpies1} and \ref{fig_All_Enthalpies2}.

Finally, the values of the standard thermal enthalpies for dry-air and water species are given by
\begin{align}
   h^0_d  & \;  =  \;  h_d\:(T_0)  \; \approx \: 530  \mbox{~kJ} \mbox{~kg}^{-1}
   \: , \label{def_h0d} \\
   h^0_v  & \;  = \;  h_v\:(T_0)  \; \approx \: 3133  \mbox{~kJ} \mbox{~kg}^{-1}
   \: , \label{def_h0v} \\
   h^0_l   & \;  = \;  h_l\:(T_0)  \; \approx \: 632  \mbox{~kJ} \mbox{~kg}^{-1}
   \: , \label{def_h0l} \\
   h^0_i   & \;  = \;  h_i\:(T_0)  \; \approx \: 298  \mbox{~kJ} \mbox{~kg}^{-1}
   \: . \label{def_h0i}
\end{align}
The standard values $h^0_d$ and $h^0_v$ lead to   ${\Upsilon\!}_0(T_0)\approx 9.5$  in (\ref{def_Upsilon}), $h_{ref} \approx 256 $~kJ~kg${}^{-1}$  in (\ref{def_Href0}) and $T_{\Upsilon} \approx 2362$~K  in (\ref{def_Tr_Upsilon_cste}).

The term $T_{\Upsilon}/ T \approx 10$  is close to $L_{vap}/(c_{pd}\:T) \approx 9$ and it is 
about ten times larger than $\lambda \approx 0.84$.
This means that $c_{pd} \: T_{\Upsilon} \: q_t$ is a dominant term in the last term of (\ref{def_h_bis}), which is thus almost equal to $L_{vap}\:q_t$.
Thus the bracketed term in (\ref{def_Th}) cannot be neglected, because it is of the same order of magnitude as the other moist terms depending on $q_t$, $q_l$ or $q_i$ in (\ref{def_T_l}).

If the specific heat for dry air, ice, liquid water and water vapour are assumed to be constant with $T$ above $150$~K, the thermal enthalpies can be approximated by
\begin{align}
  h_d (T)  & \approx \: h^0_d
  \: + \:c_{pd} \: ( \: T - T_0 \: )
  \: , \label{def_hd_term} \\
  h_v (T)  & \approx \: h^0_v
  \: + \:c_{pv} \: ( \: T - T_0 \: )
  \: , \label{def_hv_term} \\
  h_l (T)  & \approx \: h^0_l
  \: + \:c_{l} \: ( \: T - T_0 \: )
  \: , \label{def_hl_term} \\
  h_i (T)  & \approx \: h^0_i
  \: + \:c_{i} \: ( \: T - T_0 \: )
  \: . \label{def_hi_term} 
\end{align}
Reference values are computed with $T=T_r$ in (\ref{def_hd_term})-(\ref{def_hi_term}).

The important result is that hypotheses of  linear laws ($h_k \approx c_{pk} T$) for enthalpies of dry-air or water species are clearly invalidated since the straight lines $h_k(T)$ intersect at $0$~K  in Figure~\ref{fig_All_Enthalpies1}, at the non-zero values $h^0_k -  c_{pk} \: T_0$.

It is worth noting that the same dataset used to compute the standard values of enthalpies can be used to compute the standard values of entropies for O${}_2$,  N${}_2$ and H${}_2$O.
It is important to check this so as to demonstrate the good level of accuracy of the dataset and method described in this section.
The results (not shown) indicated that the standard entropies are close to the values published in HH87 and used in M11: $0.5$\% deviation for N${}_2$ vapour,  $1.3$\% for  O${}_2$ vapour and $0.2$\%  for ice, including a residual entropy of $189$~J~K${}^{-1}$~kg${}^{-1}$ for ice-Ih at $0$~K due to proton disorder and to the remaining randomness of hydrogen bonds at $0$~K (Pauling, 1935, Nagle, 1966).

\section{\bf Results.} 
\label{section_results}

\subsection{Impacts of a coincidence and of Trouton's rule.} 
\label{subsection_Trouton}

It is assumed in many studies (B82, E94 or A10) that $h_d(T_0) \approx h_l(T_0)$ at $T_0=0^{\circ}$~C and that this common value for the enthalpies is equal to zero.
The accuracies of these two approximations are analyzed in this section.

The special temperature for which $h_d(T) = h_l(T)$ can be computed by setting (\ref{def_hd_term}) and (\ref{def_hl_term}) equal.
The result $241.4$~K $=-31.75^{\circ}$~C is precisely the temperature for which the curves for dry-air and liquid-water enthalpies intersect in Figure~\ref{fig_All_Enthalpies1}.
However, the difference $h_d(T) - h_l(T)$ is positive and becomes progressively larger for increasing  temperatures and in particular for temperatures that are positive on the Celsius scale.

The first hypothesis $h_d(T_0) \approx h_l(T_0)$ is thus an approximation based on a coincidence.
It is not a fundamental physical property and it is not valid for all temperatures.
This means that it could be worthwhile to avoid this approximation and to compute the moist-air enthalpy by (\ref{def_h_bis}) and with the more accurate observed values for $h_d(T)$ and $h_l(T)$ given by (\ref{def_hd_term}) and (\ref{def_hl_term}).

The second hypothesis is that both $h_d(T_0)$ and $h_l(T_0)$ could be equal to zero, or that they could be small terms in comparison with the latent heat $L_{vap}(T_0)$.
A large vertical step is indeed observed at $T_0$ in  Figure~\ref{fig_All_Enthalpies1} between the water vapour and both liquid-water and dry-air enthalpies.
This means that $L^0_{vap}$ is the dominant term in this diagram.

The reason why the latent heat of vaporization for H${}_2$O is much larger than those for N${}_2$ or O${}_2$  ($2501$ versus $200$ and $213$~kJ~kg${}^{-1}$) is due to Trouton's rule (Wisniak, 2001).
This rule states that almost all substances follow the general result $L_{vap}/\:T \approx 88$~J~mol${}^{-1}$~K${}^{-1}$ at their boiling temperature and at normal pressure of $1000$~hPa.
The dominant feature for $L^0_{vap}$ is thus a consequence of the boiling temperature of H${}_2$O (373.15~K), which is about $4.4$ times greater than those for N${}_2$ (77.4~K) and  O${}_2$ (90~K).

The consequence of the coincidence $h^0_d \approx h^0_l$ and of Trouton's rule  is that ${c}_{pd} \: T_{\Upsilon} \approx {c}_{pd}\: T_0 \: {\Upsilon\!}_0  \approx  L^0_{vap} $ is valid to within $5$~\%.
The last term of (\ref{def_h_bis}) can thus be approximated by $L_{vap}\, q_t$, leading to the approximate formula for the thermal enthalpy of moist air
\begin{align}
 h  & \approx\: h_{ref}  
\: + \: 
{c}_{pd} \: T
 \: + \:  L_{vap}\:q_v  
 \: - \:  L_{fus}\:q_i
  \: . \label{def_h_ter}
\end{align}
Unlike (\ref{def_h_bis}), this approximate formula is not symmetric if $L_{vap}\:q_l$ is replaced by $L_{sub}\:q_i$, since it is the term $L_{fus}\:q_i$ that is involved in (\ref{def_h_ter}).
This means that the internal liquid-ice symmetry of the enthalpy equation (\ref{def_h_bis}) may not be observed in  approximate versions of $h$.

The results presented in this section show the extent to which the thermodynamic properties of water are special and different from the properties of dry air.
The coincidence $h^0_d \approx h^0_l$ and Trouton's rule associated with the high boiling temperature of H${}_2$O are empirical results.
Since it is possible to avoid these approximations, and since they can lead to systematic errors, theoretical investigation dealing with moist-air thermal enthalpy should be based on the more general formula (\ref{def_h_bis}). Otherwise, some physical properties might be poorly represented or misinterpreted.

\subsection{Analytic comparisons between enthalpy and two MSE quantities.} 
\label{subsection_comp_h_MSE}

The structures of $h+\phi$, with $h$ given by (\ref{def_h_bis}) or by the approximation (\ref{def_h_ter}), are clearly similar to the MSE quantities recalled in Section~\ref{subsection_MSE}, except for the constant term $h_{ref}$, which has no physical meaning since it corresponds to a global shift in the moist enthalpy units.
It will be shown in the following sections that $h+\phi$ is especially close to MSE$_d$ and MSE$_m$ given by (\ref{def_MSEd}) and (\ref{def_MSEm}).

Accordingly, in this section, comparisons are made between the analytical formulations of the thermal enthalpy $h-h_{ref}$ computed with (\ref{def_h_bis}) and the quantities TMSE$_d$ and TMSE$_m$ given by (\ref{def_MSEd}) and (\ref{def_MSEm}).
The relative accuracies of the approximation of $h$ by TMSE$_d$ or TMSE$_m$ are evaluated by the terms $X_d = [ \, \mbox{TMSE}_d - (h-h_{ref}) \, ]/(c_{pd}\:T)$ and $X_m = [ \, \mbox{TMSE}_m - (h-h_{ref}) \, ]/(c_{pd}\:T)$.
After some algebra, they can be written as
\begin{align}
 X_d & \:= \: 
       \left(  
              \frac{L_{vap}}{{c}_{pd}\: T}  
              \:  - \: 
              \frac{T_{\Upsilon}}{T}
              \:  - \: 
               \lambda
     \right)  q_t
   \:+ \: \left(   
      \frac{L_{fus}}{{c}_{pd}\: T} 
       \right)   \:  q_i
  \: , \label{def_deltah3} \\
 X_m & \:= \:
       \left(
              \frac{L_{vap}}{{c}_{pd}\: T}  
              \:  - \: 
              \frac{T_{\Upsilon}}{T}
     \right)  q_t
   \: + \:
       \left(  \frac{c_l-{c}_{pv}}{{c}_{pd}} 
     \right)  q_l
    \:+ \:
       \left(   \frac{c_i-{c}_{pv}}{{c}_{pd}} 
                + \frac{L_{fus}}{{c}_{pd}\: T} 
     \right)  q_i
  \label{def_deltah2} \: .
\!\!
 \end{align}
Since the specific contents $q_t$, $q_l$ and  $q_i$ are small terms (less than $0.03$~kg~kg${}^{-1}$), $X_h$ and $X_h$ are less than $10$~\% if absolute values of the  terms in parentheses in  (\ref{def_deltah3}) and (\ref{def_deltah2}) are lower than $3$.

The  last two terms in parentheses in (\ref{def_deltah2}) can  be approximated by $2.4$ for $q_l$ and $1.5$ for $q_i$.
The first term in parentheses in (\ref{def_deltah2}) can be evaluated from the results $T_{\Upsilon} \approx 2362$~K and $L_{vap}(T_0)/{c}_{pd}\approx 2490$, leading to a positive value of about $0.5$ for the first quantity in factor $q_t$.
Similarly,  the  first term in parentheses in (\ref{def_deltah3}) is about $-0.4$ at $T_0=273.15$~K, and the second term in parentheses in (\ref{def_deltah3})  is equal to $1.7$.
If $q_t$ is replaced by $q_v+q_l+q_i$, $X_m$ and $X_d$  can be approximated at $T_0$ by
\begin{align}
   X_m \, (T_0) & \approx \: 0.5\:q_v +3\:q_l+2\:q_i
  \: , \label{def_Xm_approx} \\
   X_d  \, (T_0) & \approx \: 
      - 0.4\:q_v-0.4\:q_l+0.9\:q_i
  \: . \label{def_Xd_approx}
 \end{align}
$X_m$ is thus a positive difference indicating that the larger the water species contents $q_v$ and $q_l$ are, the more TMSE$_m$ overestimates the moist thermal enthalpy $h-h_{ref}$.
Except for very high and unrealistic values of $q_i$,  $X_d$ is negative and TMSE$_d$ underestimates  $h-h_{ref}$.
The specific ice content $q_i$ further increases the overestimation of TMSE$_m$, while it decreases the underestimation of TMSE$_d$.

The approximations of $(h-h_{ref})$ by TMSE$_m$  or TMSE$_d$ can thus be considered as accurate for under-saturated conditions ($q_l=q_i=0$), with $X_m$ and $|X_d|$ lower than $1$\% for $q_v$ lower than $20$~g~kg${}^{-1}$.
The differences become larger within clouds, and in particular for TMSE$_m$, with  $X_m$ increasing by about $0.3$\%  for each  $1$~g~kg${}^{-1}$ of liquid-water content.

Possible systematic differences of about $1$\% correspond to differences in $T_h$ and $T$  of about $3$~K.
This can be of some importance for the accurate determination of the moist thermal enthalpy within clouds or moist regions.
Such large differences may modify the analysis of the impact of drying or moistening processes on $h$ and on the local temperature.
This is another justification for the use of the complete formula (\ref{def_h_bis}) when evaluating $h$, with no approximation for the second line depending on $q_t$.

\subsection{Thermal enthalpy diagrams.} 
\label{subsection_s_h_diagrams}

\begin{figure}[hbt]
\centering
\includegraphics[width=0.8\linewidth,angle=0,clip=true]{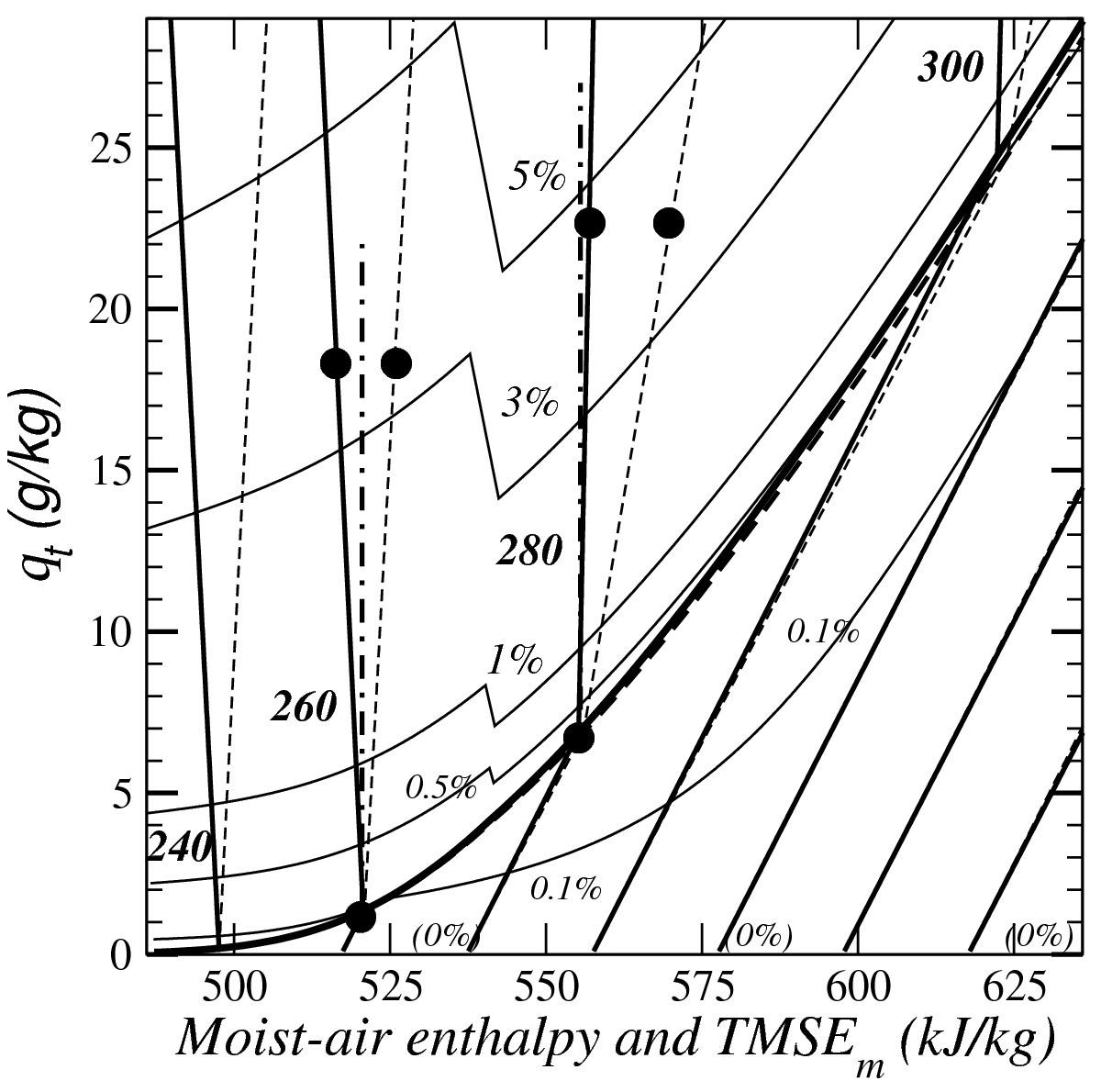}
\caption{\it \small
Enthalpy diagram at the constant pressure $p_1=900$~hPa.
The total specific water content is plotted against the moist enthalpy $h$ or against $h_{ref}+\:$TMSE$_m$, with $h_{ref}=h^0_d=530$~kJ~kg${}^{-1}$.
The saturation curves are plotted in the centre of the diagram (bold solid lines for $h$ and dashed lines for TMSE$_m$).
The saturated regions are located above the saturation curves.
Isotherms are labelled every $20$ K, with bold solid lines for $h$ and dashed lines for TMSE$_m$.
Other elements are described in the text.
\label{fig_h_qt_diagrams}}
\end{figure}

The possibility of computing specific moist values for the thermal enthalpy $h$ offers the opportunity to plot the enthalpy diagram shown in Figure~\ref{fig_h_qt_diagrams}, where $q_t$ is plotted as a function of $h$.
This diagram is the enthalpy counterpart of the entropy diagram ($q_t$, $s$) given in Marquet and Geleyn (2013).

In this enthalpy diagram, the moist thermal enthalpy $h$ given by (\ref{def_h_bis}) is compared with the quantity $h_{ref}+\mbox{TMSE}_m$, where TMSE$_m = c_p \: T + L_{vap}\:q_v$ is the thermal part of  (\ref{def_MSEm}).
The use of TMSE$_m$ instead of $h$ generates an error $X_m$ measured by (\ref{def_deltah2}) and represented by the thin solid lines  (values $0, 0.1, 0.5, 1, 3$ and $5$~\%).
Positive values of $X_m$ correspond to dashed lines located to the right of the solid lines, or equivalently to TMSE$_m\:>h$.
The water-vapour saturation pressure is computed with the liquid-water formulation if $T>0$~C and with the ice formulation otherwise.
This creates zigzag features for $X_m$ in the saturated regions and around $0$~C (about $535$~kJ~kg${}^{-1}$).

The isotherms are represented on Figure~\ref{fig_h_qt_diagrams} for either $h$ (solid lines) or $h_{ref}+\:$TMSE$_m$ (dashed lines) as abscissae (from $240$ to $300$~K by steps of $20$~K).
The two sets of isotherms are almost superimposed in the non-saturated domain.
This means that the approximation $h \approx h_{ref}+\mbox{TMSE}_m$  is accurate in the non-saturated region, where $X_m$ increases gradually from $0$~\% for the dry-air case to less than $0.1$~\% to $0.5$~\%, depending on the temperature.
It is, however, worth noting that a value of $0.2$\% for $X_m$ corresponds to large differences in $T$ of about $0.6$~K.

\begin{figure}[t]
\centering
(a)\includegraphics[width=0.43\linewidth,angle=0,clip=true]{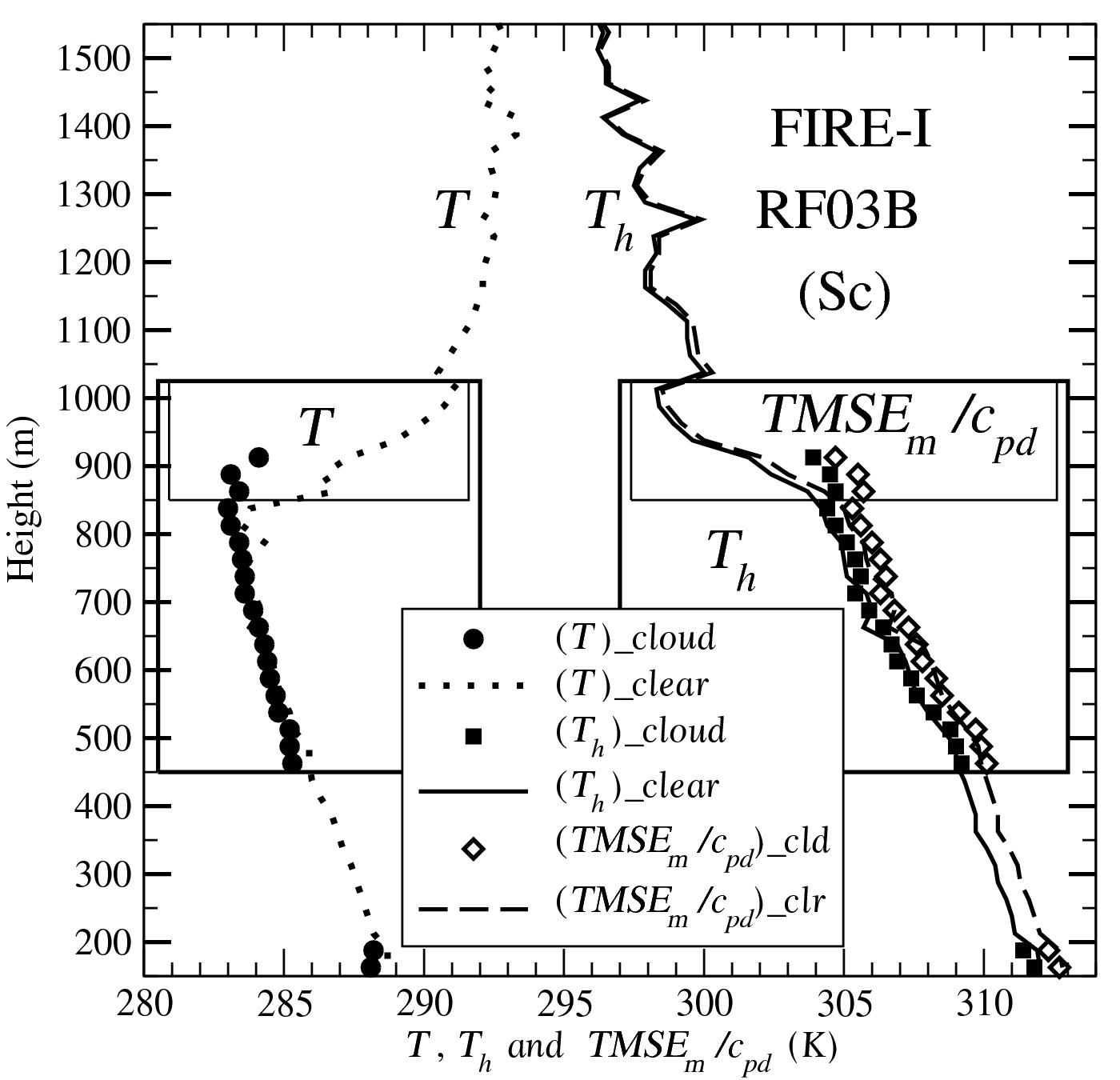}
(b)\includegraphics[width=0.27\linewidth,angle=0,clip=true]{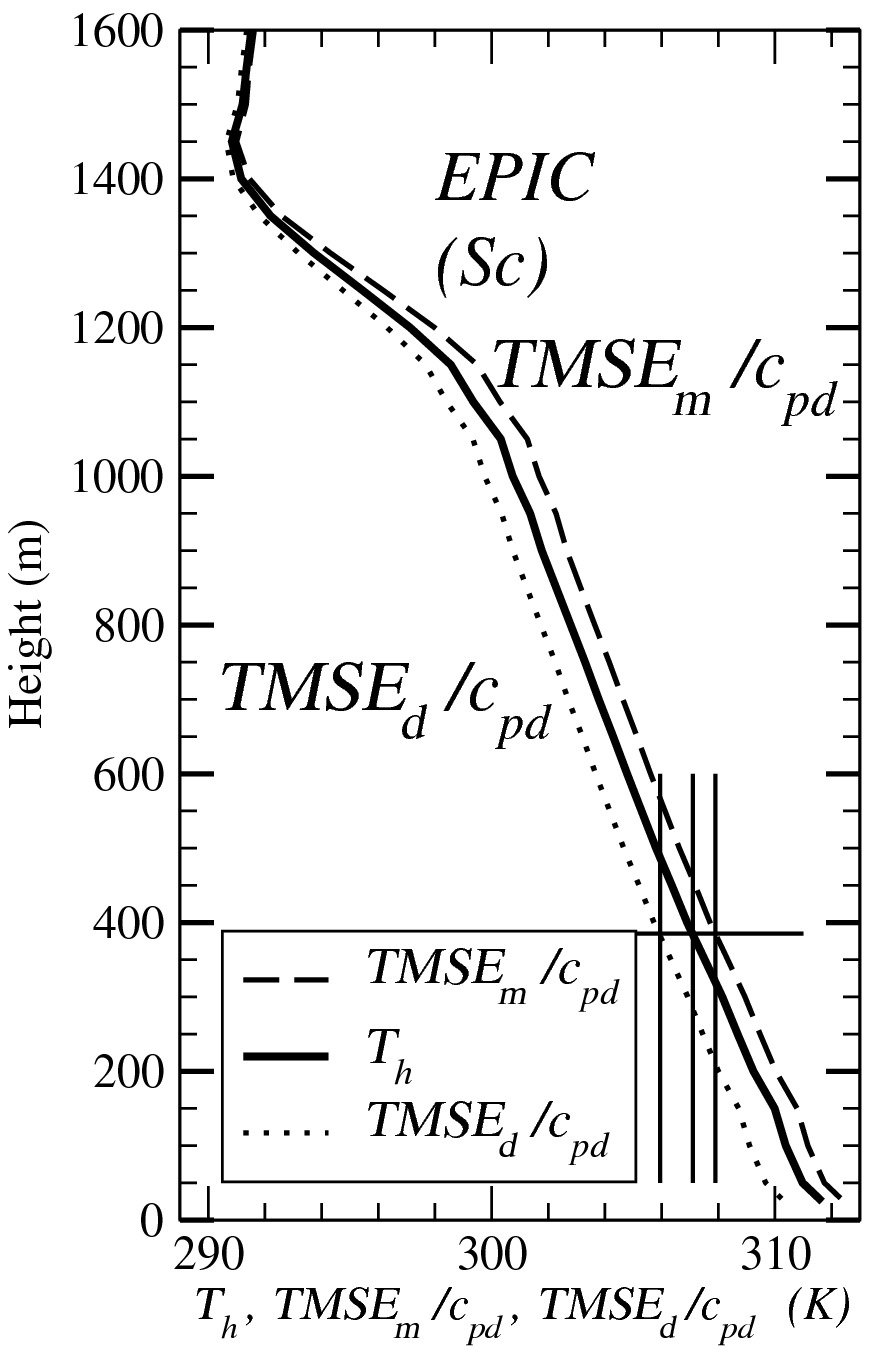}
(c)\includegraphics[width=0.27\linewidth,angle=0,clip=true]{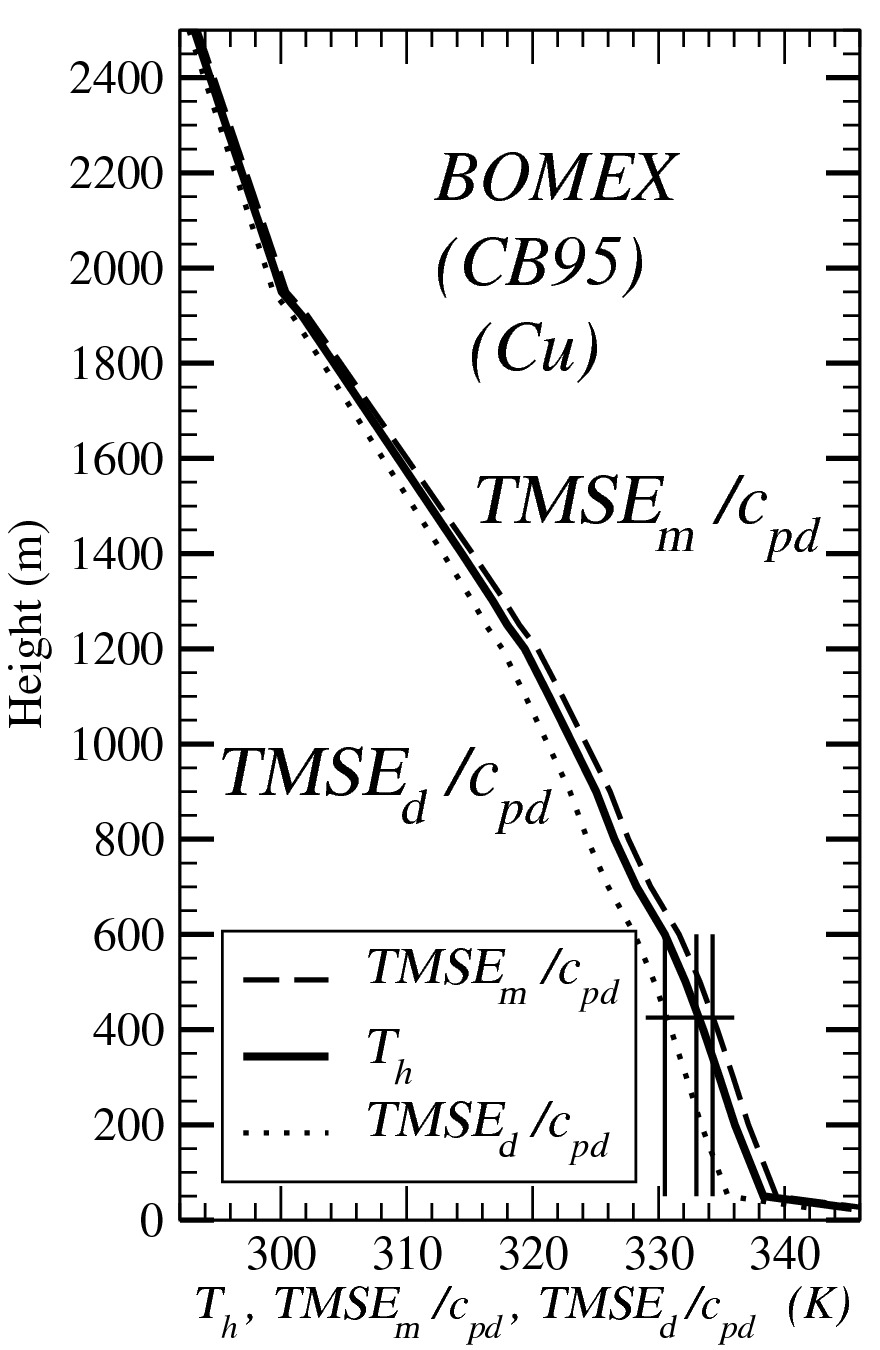}
\caption{
{\it \small
(a) The stratocumulus FIRE-I RF03B (2 July 1987) data flight.
The temperature $T$ is depicted as a dotted line for clear-air conditions and as dark circles for in-cloud regions.
The enthalpy temperature $T_h$ and TMSE$_m\,/c_{pd}$ are represented by a solid line and dark squares for $T_h$ (clear-air and in-cloud regions) and by a bold dashed line and open diamonds for TMSE$_m\,/c_{pd}$ (clear-air and in-cloud regions).
(b) The EPIC stratocumulus vertical profile and (c) the BOMEX shallow cumulus vertical profile, where the profiles of  TMSE$_d\,/c_{pd}$ are plotted.
Thin vertical solid lines highlight differences between $T_h$, TMSE$_m\,/c_{pd}$ and TMSE$_d\,/c_{pd}$.
}
\label{Fig_Hm_MSE_Sc}}
\end{figure}

According to the formula (\ref{def_Xm_approx}), the approximations are greater in the saturated region (above the saturation curves), reaching $1$~\% for $q_l=3$~g~kg${}^{-1}$ or $q_i=5$~g~kg${}^{-1}$.
Larger differences between moist thermal enthalpy and TMSE$_m$ values  are indeed observed in the saturated region of the enthalpy Figure~\ref{fig_h_qt_diagrams}.

The black spots represent moistening processes associated with large isothermal increases of $q_t$ which originate at the saturation line at $260$ and $280$~K.
These processes correspond to a decrease in $h$ in the ice region and to almost constant values in the liquid-water domain.
Conversely, TMSE$_m$ exhibits a clear increase with $q_t$ in the saturated domains and for the whole range of temperatures.

Although increases of more than $15$~g~kg${}^{-1}$ for $q_t$ are large and unrealistic, this is only for the sake of clarity and the same kind of changes in $h$ or TMSE$_m$ are observed for smaller and more realistic increases of $q_t$.
These differences suggest that TMSE$_m$ does not give an accurate measurement of the thermal enthalpy for saturated moist air, because they correspond to different physical properties.
Such differences may be observed within stratocumulus or cumulus clouds, implying possible systematic biases for the impacts of entrainment and detrainment processes on the budget of moist enthalpy (and thus on the local temperature).

\subsection{Evaluations of $T_h$ and two TMSE quantities for stratocumulus and cumulus profiles.} 
\label{subsection_h_TMSE_profiles}

\begin{figure}[t]
\centering
\includegraphics[width=0.49\linewidth,angle=0,clip=true]{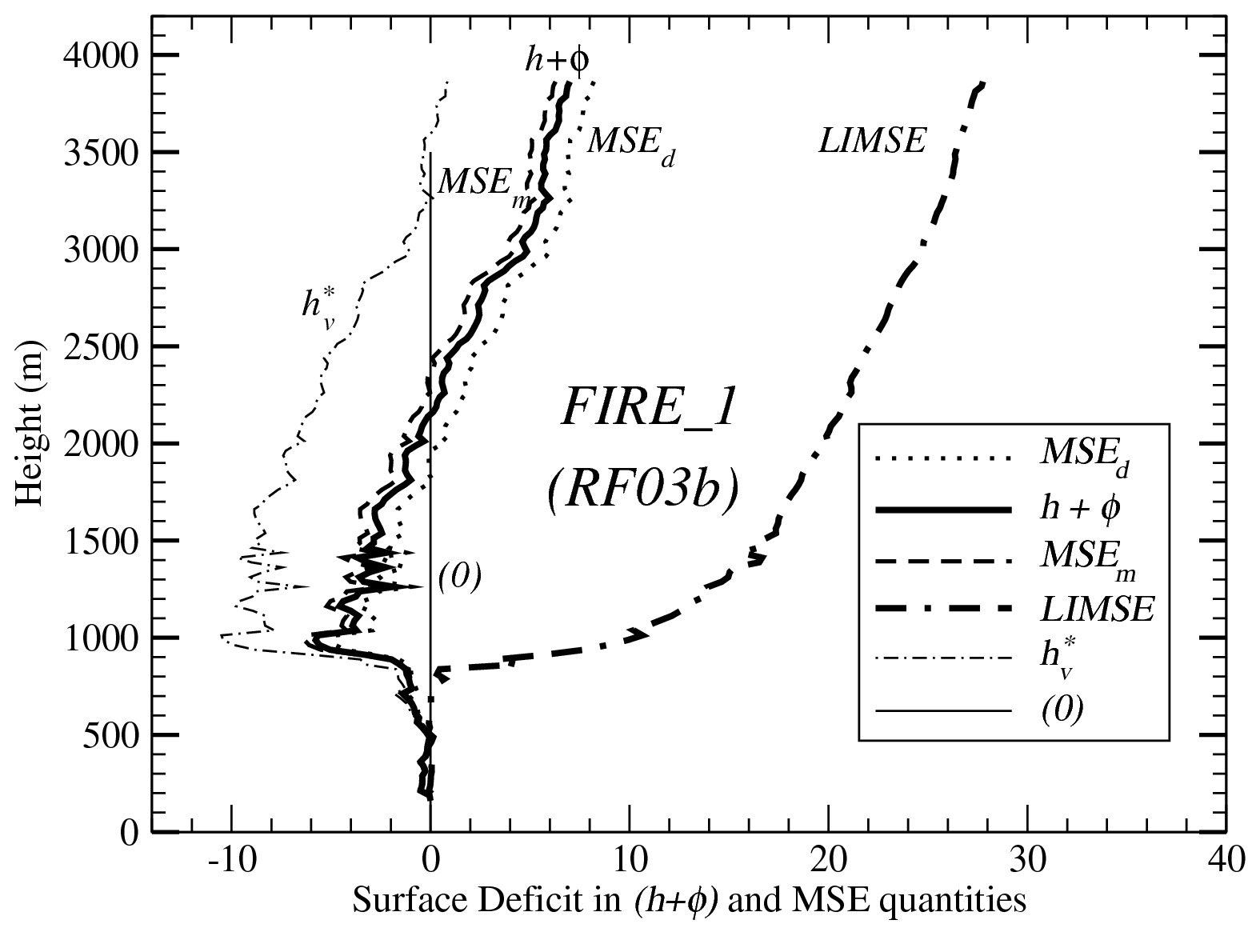}
\includegraphics[width=0.49\linewidth,angle=0,clip=true]{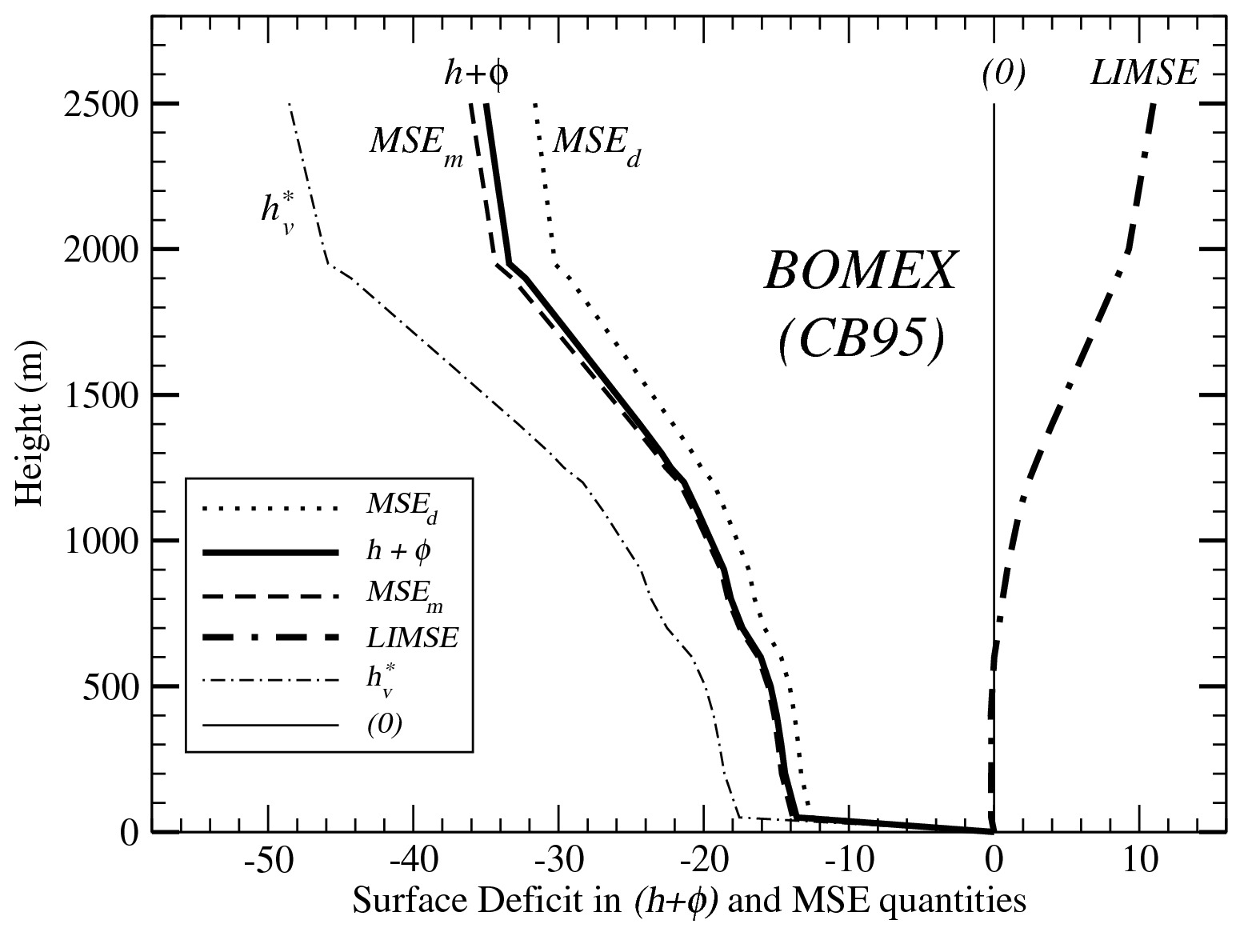} \\
(a) \hspace*{8cm}(b)\\
\caption{
{\it \small
Vertical profiles of surface deficit of $h+\phi$ and of several MSE quantities defined in (\ref{def_MSEd})-(\ref{def_MSEm})
  for (a) the FIRE-I (RF03b) stratocumulus and (b) the BOMEX shallow cumulus.
Units are in kJ~kg${}^{-1}$.}
\label{Fig_MSE_deficit}}
\end{figure}

Vertical profiles of $T_h$, TMSE$_m$ and TMSE$_d$  are compared  in Figures~\ref{Fig_Hm_MSE_Sc}(a)-(c) for two stratocumulus cases (FIRE-I and EPIC) and one shallow cumulus case (BOMEX).
They correspond to the datasets given in Appendix~C.
Cloud liquid-water contents are small, but the values of $q_v$ in the PBL are large enough to allow clear evaluations of impacts of water vapour.
For the sake of readability, vertical profiles of TMSE$_d$ are not  plotted for FIRE-I in (a).

A large positive top-of-PBL jump in $T$ is associated with large negative jumps in both $T_h$ and TMSE$_m\,/c_{pd}$ in (a).
Moreover, in-cloud and clear-air values are different in the entrainment region represented by the thin solid line boxes.
These results are different from the important properties observed in M11 with the same dataset but with moist entropy, where it is shown that in-cloud and clear-air values of $\theta_s$ are almost equal and almost constant in the whole PBL, including in the entrainment region at the top.
It can be concluded that the specific moist enthalpy represented by $T_h$ is not conserved (nor well-mixed) by the moist turbulence within the PBL of FIRE-I stratocumulus, contrary to what happens for the specific moist entropy.

The approximation of (\ref{def_Xm_approx}) suggests that TMSE$_m\,/c_{pd}$ is  warmer than the enthalpy temperature $T_h$, due to the term $0.5\:q_v$, whereas (\ref{def_Xd_approx}) suggests that TMSE$_d\,/c_{pd}$ is  cooler, due to the term $-0.4\:q_v$.
This is observed for all profiles (a)-(c), where TMSE$_m$ overestimates $T_h$ (and thus the moist enthalpy) by about $1$~K in the warm and moist lower troposphere, whereas TMSE$_d$ underestimates $T_h$  by about $2$~K  for EPIC and BOMEX.

According to (\ref{def_deltah3}) to (\ref{def_Xd_approx}), the three formulations for $h$, TMSE and TMSE$_d$ are the same in the dry-air limit ($q_v=q_l=q_i=0$).
This is confirmed by the vertical profiles depicted on Figures~\ref{Fig_Hm_MSE_Sc}(a)-(c): the curves converge above the top of the PBL, where $q_v$ is small.

In order to investigate the comparison between enthalpy and MSE quantities differently, vertical profiles of deficit in surface values are shown in Figure~\ref{Fig_MSE_deficit}(a) for the FIRE-I stratocumulus case and in (b) for the BOMEX shallow cumulus case.
This kind of surface-deficit chart is commonly used in studies of convective processes.

It is shown that MSE$_m$, $h+\phi$ and MSE$_d$ remain close to each other at each level and for the two cases, with the generalized enthalpy located  between the others and with MSE$_m$ being a better approximation for $h+\phi$.
This confirms the results observed in Figures~\ref{Fig_Hm_MSE_Sc}(a) and (c).

In Figure~\ref{Fig_MSE_deficit}(a)-(b), LIMSE and $h^{\star}_v$ are more different from $h+\phi$ than MSE$_m$ and MSE$_d$.
This means that the thermal part of LIMSE and $h^{\star}_v$ cannot represent the moist-air thermal enthalpy $h$ accurately.

Large jumps in all variables are observed  close to the surface for BOMEX, except for LIMSE, due to the impact of large values of surface  specific humidity that are not taken into account in LIMSE.
The impact of surface values are not observed for FIRE-I, since the in-flight measurements were taken at altitude.

The conclusion of this section is that MSE$_m$ is probably the best candidate for approximating $h+\phi$.
However, the fact that systematic differences exist between $h$ and MSE quantities in the moist lower PBL only, and not in the dry air above, may have significant physical implications if the purpose is to accurately analyse moist enthalpy budgets or differential budgets, or to understand the convective processes (entrainment and detrainment), or to validate the long-term budgets for NWP models and GCMs by comparing them with climatology or reanalyses.

\subsection{Wet-bulb temperatures and psychrometric equations.} 
\label{subsection_Tw_psychro}

\begin{figure}[hbt]
\centering
\includegraphics[width=0.90\linewidth,angle=0,clip=true]{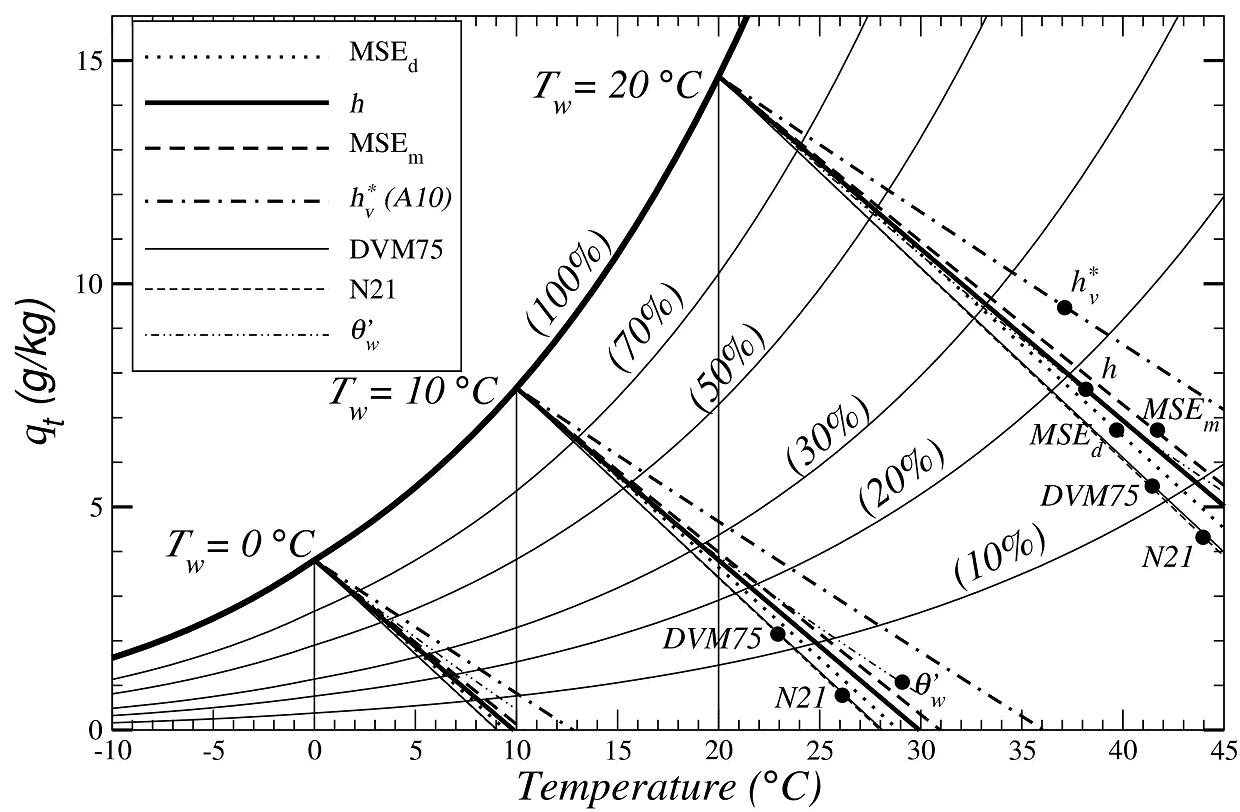}
\caption{\it \small
Psychrometric diagram for $1000$~hPa and for $\phi=0$.
Vertical isotherms and lines of constant $T_w$ (thin solid lines) are  for $T_w = 0^{\circ}$C, $10^{\circ}$C and $20^{\circ}$C in a non-saturation region (where $q_l=q_i=0$).
The two versions derived from DVM75 and N21 can be compared using lines of constant $\theta'_w$.
Bold lines for constant enthalpy ($h$) or MSE quantities (MSE$_d$, MSE$_m$ and $h^{\star}_v$) decrease with $T$.
Lines of constant  relative humidity  curve upwards (from $70$\% to $10$\%), the saturation curve corresponding to $100$\%.
\label{fig_T_qt_h_psychro}}
\end{figure}

The differences between $h+\phi$ and MSE quantities observed in Figures~\ref{fig_h_qt_diagrams} to \ref{Fig_MSE_deficit} can be interpreted in a different way.
If  enthalpy is replaced in abscissa by temperature, the enthalpy of Figure~\ref{fig_h_qt_diagrams} is transformed into the psychrometric chart plotted on Figure~\ref{fig_T_qt_h_psychro}, where the isotherms are vertical.

The aim of psychrometric charts is to determine and plot the lines of constant wet-bulb temperature $T_w$.
It is assumed in N21 or DVM75 that $T_w$ is the temperature attained by a mass of moist air brought to saturation by water evaporating into it, with a latent heat continuously supplied by the wet bulb.
It is thus an isenthalpic process at constant pressure for the whole system of the moist air plus the wet bulb, with the mass of dry air assumed to be constant.
The WMO (2008) psychrometric equation is derived in DVM75. 
It can be written as
\begin{align}
(c_{pd} + c_{pv} \: r_v )\: (T - T_w)
&  \: = \: 
L_{vap}(T_w) \: ( r_{sw} - r_v )
  \: . \label{def_psychro_DVM75} 
\end{align}
The wet-bulb temperature may be defined differently.
The left hand side is approximated by $c_{pd} \: (T - T_w)$ in N21.
The definition (6.67) or (6.78) published in Bohren and Albrecht (1998) corresponds to $c_{pv}\: r_v$ being replaced by $c_l\: r_{sw}$.
Psychrometric equations are derived in E94 (4.6.4) and A10 (5.41) by assuming conservation of $h^{\star}_v/q_d$ expressed per unit mass of dry air.
They correspond to $c_{pv} \: r_v$ being replaced by $c_{pv} \: r_l + (c_{pv} - c_l)\: r_v$.
All these terms depending on mixing ratios are expected to be small in comparison with $c_{pd}$ in (\ref{def_psychro_DVM75}).

The wet-bulb temperature may be derived from the Second Law and the conservation of the pseudo-adiabatic potential temperature $\theta'_w$. It is the temperature attained by a parcel of fluid  brought adiabatically to saturation by upward displacement and at a lower pressure, then carried adiabatically back to the original pressure, with water assumed to be continuously supplied to maintain saturation.


Psychrometric lines of equal $T_w$ and lines of constant $\theta'_w$ are plotted on Figure~\ref{fig_T_qt_h_psychro}.
The figure shows that the versions of N21 and DVM75 are very close to each other.
This is a confirmation  that $c_{pd} \: (T - T_w)$ is the leading-order term in (\ref{def_psychro_DVM75}).
Larger differences are observed with lines of constant $\theta'_w$, in particular for $RH<50$~\%.
The wet-bulb temperatures computed from the methods described in N21 or DVM75 and the one using $\theta'_w$ are thus different.

It is  common practice to plot  isenthalpic lines on the same psychrometric charts.
These lines of constant specific enthalpy expressed per unit mass of dry or moist air correspond to open systems, since values of $q_t=q_v$ vary from $0$ to the saturating value $q_{sw}$.
We thus have an opportunity to compare the properties of $h$ given by (\ref{def_h_bis}) with those from MSE formulations, because MSE quantities are derived for closed systems only and with arbitrary assumptions like $h^0_d=h^0_l$ at $0^{\circ}$~C, whereas the definition of $h$ is  valid for both closed and open systems \textit{a priori} and is derived from a Third Law.
This study is the continuation of the comparison of $h$ and TMSE$_m$ in the Figure~\ref{fig_h_qt_diagrams}.

All isenthalpic lines are oriented in a downward direction on Figure~\ref{fig_T_qt_h_psychro}.
They are almost parallel to psychrometric lines of constant values of $T_w$, but do not coincide with them.
This probably invalidates the possibility that psychrometric lines might be computed as constant values of $h$, although additional measurements are required to determine which curve is relevant for low RH.

Moreover, systematic differences exist on Figure~\ref{fig_T_qt_h_psychro}: the curves of constant $h$ are located between the curves of constant MSE$_d$  and MSE$_m$, whereas the curves of constant $h^{\star}_v$ are more different and are located above the others.
The closed system assumption and the use of arbitrary definitions for the reference enthalpies explain the differences observed between the curves plotted with $h$ and with other MSE quantities.

\section{\bf Discussion and conclusion.} 
\label{section_conclusion}

In the same way that moist entropy is expressed in M11 in terms of $\, c_{pd} \, \ln(\theta_s)$ up to a constant entropy reference value, moist-air thermal enthalpy is expressed in terms of $\, c_{pd} \, T_h$ up to a constant enthalpy reference value.
The potential temperature $\theta_s$ and the enthalpy temperature $T_h$ are thus equivalent to the specific moist-air entropy ($s$) and thermal enthalpy ($h$).

Thermal enthalpies are generated by variations of $c_p(T)$ with $T$ corresponding to progressive excitations of the translational, rotational and vibrational states of the molecules, and by possible changes of phase represented by latent heats (with negligible impact of changes of pressure).

The original feature is that both $T_h$ and $\theta_s$ depend on standard values of enthalpies and entropies, via the terms $\Upsilon_r$ and $\Lambda_r$, respectively.
In this paper, standard values are determined with an error of a few percent for the cryogenic properties of the dry-air and water-components of moist air.
Different choices for these standard values would generate other values for $T_h$ and $\theta_s$.
The use of cryogenic values for the specific heats of solid and liquid phases, in addition to latent heats when changes of phases occur, eliminate the issue of referring to the perfect gas law when the temperature is too low.

The important feature in the definition of $h$ and $s$ is that the specific contents of the moist-air components multiply several constant terms which depend on the reference values $T_r$ or $p_r$.
This  leads naturally to additional varying terms depending on $q_t$.
It is for this reason that the standard values $h^0_d$, $h^0_v$, $s^0_d$ and $s^0_v$ must determined for varying $q_t$.
This explains why the Third Law versions for $h$ and $s$ are different from the ones obtained in DVM75, E94 or P10, where only quantities per unit of dry-air are considered and where hypotheses like $h^0_d = 0$,  $h^0_l = 0$, $h^0_v = 0$ or $h^0_d=h^0_l$ are made at a temperature different from $0$~K.
This suggests that the Third Law of thermodynamics may have an important and somewhat surprising physical meaning when moist-air thermodynamics is considered.

The constant volume or constant dry-air point of view may be useful for laboratory experiments, but it is not relevant to isolate the dry-air part from a moving open moist-air parcel with multiple components.
The specific value approach is more suitable for applications to a barycentric view and to equations of compressible fluids.
This may explain some of the differences between the present Third Law results and previous ones.

It has been shown that the enthalpy temperature is expressed in terms of the quantity  $T_{\Upsilon}$, which depends on the difference between the standard values of dry-air and water-vapour enthalpies (in the same way as the moist entropy depends, via $\Lambda_r$,  on the difference between the standard values of dry-air and of water-vapour entropies).

It should be noted that $T_{\Upsilon}$ is a constant term independent of the reference temperature $T_r$.
This means that the same property observed by entropy in M11 is also valid for the thermal enthalpy, i.e. they are both independent of the choice made for the reference temperature and pressure $T_r$ and $p_r$.
Since they are well-defined local quantities, it becomes possible to use specific values of thermal enthalpy or entropy to compute local budgets, integral or mean values of $h$ and $s$, either in space or time.

A more surprising result concerns the assumption that generalized enthalpy might be roughly represented by MSE$_m$ or MSE$_d$.
It is shown that it is serendipitous since:  
i) the numerical values of standard enthalpies of dry air and liquid water are close to each other; and
ii) $L^0_{vap}$ is a large dominant term of $T_{\Upsilon}$.

It is shown in M11 that no such coincidence exists for the entropy, thus explaining the large impact of the new term $\Lambda_r$ introduced for the entropy computations.
However, even if MSE$_m$ or MSE$_d$ might be considered as relevant approximations of $h+\phi$ (at least for non-saturated conditions), they do not follow the same internal liquid-ice symmetry as in the case of the specific moist-air enthalpy formulation.

The enthalpy diagram illustrates the possibility of representing all the thermodynamic properties of moist air as a function of $h$, $p$ and $q_t$ alone, as in Pauluis and Schumacher (2010) who suggest using the set ($s$, $p$, $q_t$) to compute any of the basic variables ($T$, $p$, $q_v$, $q_l$, $q_i$).
It is thus possible to use the First- or Second-Law set of variables ($h$, $p$, $q_t$) or ($s$, $p$, $q_t$) as prognostic variables.
This is true for LES models where each grid cell represents a homogeneous parcel of moist air (as long as the pressure is known, in spite of non-hydrostatic processes).

This result is no longer valid for a parcel subject to sub-grid variability.
All properties of a parcel in NWP models or GCMs are understood as weighted sums of properties observed for a given fraction of unsaturated air plus the other part from saturated air (with weighting factors depending on cloud fractions).
However, the method of representing moist-air properties by ($s$, $p$, $q_t$) or by ($h$, $p$, $q_t$) may be used even in NWP models or GCMs, provided that: it is i) applied separately for each of the unsaturated and saturated fractions of moist air, and ii) the values of cloud fraction can be determined by some other processes.

It thus becomes possible to compute and study the tendencies of $h$ or $s$ generated by various processes (dynamics, radiation, convection, turbulence) and to compare the results with reanalysis outputs.
An alternative method could consist of evaluating the finite differences $h(t+dt)-h(t)$ at each point, or $s(t+dt)-s(t)$, with the specific thermal enthalpy or entropy computed with zero values at $0$~K.

Another application concerns pure isenthalpic processes, as plotted on psychrometric charts.
It is shown in this paper that, depending on how the specific enthalpy is defined, the isenthalpic lines are different.
Since there is only one kind of isenthalpic process in the real world, it might be valuable to compare the new thermal enthalpy versions involving real processes and/or observations.

The differences between MSE quantities and $h+\phi$ are analysed for several cumulus and stratocumulus vertical profiles.
MSE$_m$ is greater, and  MSE$_d$ smaller, than the generalized enthalpy $h+\phi$ in moist conditions.
These systematic differences of more than $1$\% may have significant physical implications for the accurate determination of the moist thermal enthalpy within clouds.
They may modify results from analyses of the impact of drying or moistening processes on $h$, and therefore on local temperature.
This confirms the striking results obtained when referring to the enthalpy diagrams, and suggests that the formulation of thermal enthalpy (\ref{def_h_bis}) should be used to compute moist enthalpy or energy budgets accurately.
This may be important for the validation of NWP models and/or GCMs, when the budget of thermal enthalpy simulated by these models is compared with those obtained from climatology or reanalyses. 

All of the above suggests new possible applications for the Third Law formulation of entropy and thermal enthalpy.
It may be used to study dry-air entrainment and moist-air detrainment occurring along cloud edges, as well as the conservative properties used when parameterizing  shallow or deep-convection.
For the latter phenomena, MSE quantities should be replaced by the specific generalized enthalpy $h+\phi$.



\vspace{5mm}
\noindent{\large\bf Acknowledgements}
\vspace{2mm}

The author is most grateful to Jean-Fran\c{c}ois Geleyn and Maarten Ambaum for stimulating discussions.
He would like to thank the anonymous referees for their constructive comments, which helped to improve the manuscript.
Thanks to S. Becker and Y. Tourre for the improvements of the written English.
S. R. de Roode and Q. Wang kindly provided the validation data from NASA Flights during the FIRE I experiment.


\vspace{4mm}
\noindent
{\bf Appendix A. List of symbols and acronyms.}
             \label{appendixSymbol}
\renewcommand{\theequation}{A.\arabic{equation}}
  \renewcommand{\thefigure}{A.\arabic{figure}}
   \renewcommand{\thetable}{A.\arabic{table}}
      \setcounter{equation}{0}
        \setcounter{figure}{0}
         \setcounter{table}{0}
\vspace{1mm}
\hrule

\begin{tabbing}
 ------------ \= -------------------------------------- \= \kill
 $c_{pd}$ \> specific heat of dry air   \>($1004.7$~J~K${}^{-1}$~kg${}^{-1}$) \\
 $c_{pv}$ \> spec. heat of water vapour \>($1846.1$~J~K${}^{-1}$~kg${}^{-1}$) \\
 $c_{l}$  \> spec. heat of liquid water \>($4218$~J~K${}^{-1}$~kg${}^{-1}$) \\
 $c_{i}$  \> spec. heat of ice          \>($2106$~J~K${}^{-1}$~kg${}^{-1}$) \\
 $c_p$ \> specific heat at constant pressure for moist air 
          ($c_p = \: q_d \: c_{pd} + q_v \: c_{pv} + q_l \: c_l + q_i \: c_i$) \\
 $\delta$ \> $=R_v/R_d-1 \approx 0.608$ \\
 $\eta$   \> $=1+\delta =R_v/R_d \approx 1.608$ \\
 $\varepsilon$ \> $=1/\eta=R_d/R_v \approx 0.622$ \\
 $\kappa$ \> $=R_d/c_{pd}\approx 0.2857$ \\
 $\gamma$ \> $= \eta \: \kappa \ = R_v/c_{pd} \approx 0.46$ \\
 $\lambda$ \> $= c_{pv}/c_{pd}-1 \approx 0.8375$ \\
 $e$       \> water-vapour partial pressure \\
 $e_{sw}(T)$ \> partial saturating pressure over liquid water \\
 $e_r$      \> water-vapour reference partial pressure:  
               $\: e_r = e_{ws}(T_r=T_0) \approx 6.11$~hPa \\
 $e_i$      \> internal energy: $e_i = h - p/\rho = h - R\:T$\\
 $g$        \> magnitude of Earth's gravity: $9.8065$~m${}^{2}$~s${}^{-2}$\\
 $h$        \> specific enthalpy \\
 $(h_{d})_r$  \> reference enthalpy of dry air at $T_r$\\
 $(h_{v})_r$  \> reference enthalpy of water vapour  at $T_r$\\
 $h^0_d$ \> standard specific enthalpy of dry air at $T_0$ 
                        \hspace{10mm} ($530$~kJ~kg${}^{-1}$) \\
 $h^0_v$ \> standard specific enthalpy of water vapour at $T_0$  
                                     ($3133$~kJ~kg${}^{-1}$) \\
 $h^0_l$ \> standard specific enthalpy of liquid water at $T_0$ 
                        \hspace{1mm} ($632$~kJ~kg${}^{-1}$) \\
 $h^0_i$ \> standard specific enthalpy of ice water at $T_0$ 
                        \hspace{6mm} ($298$~kJ~kg${}^{-1}$) \\
 $h^{\star}_v$ \> a specific moist static energy (A10) \\
 ${\Lambda}_r$ \> $= [ (s_{v})_r - (s_{d})_r ] / c_{pd} \approx 5.87$ \\
 $L_{vap}$    \> $=h_v-h_l$: latent heat of vaporization \\
 $L^0_{vap}$  \> $= 2.501$~$10^{6}$~J~kg${}^{-1}$ at $T_0$\\
 $L_{fus} $   \> $=h_l-h_i$: latent heat of fusion \\
 $L^0_{fus} $ \> $= 0.334$~$10^{6}$~J~kg${}^{-1}$ at $T_0$ \\
 $L_{sub}$    \> $=h_v-h_i$: latent heat of sublimation \\
 $L^0_{sub}$  \> $= 2.835$~$10^{6}$~J~kg${}^{-1}$ at $T_0$ \\
 $m$  \> a mass of moist air \\
 $p$      \> $=p_d + e$: local value of pressure \\
 $p_r$  \> $=(p_d)_r + e_r$: reference pressure ($p_r=p_0$)\\
 $p_d$     \> local dry-air partial pressure \\
 $(p_d)_r$ \> reference dry-air partial pressure ($\equiv p_r-e_r$)\\
 $p_0$     \> $=1000$~hPa: conventional pressure \\
 $q_{d}$   \> $={\rho}_d / {\rho}$: specific   dry air content\\
 $q_{v}$   \> $={\rho}_v / {\rho}$: specific   water vapour content\\
 $q_{l}$   \> $={\rho}_l / {\rho}$: specific   liquid water content\\
 $q_{i}$   \> $={\rho}_i / {\rho}$: specific   ice content content\\
 $q_t  $   \> $= q_v+q_l+q_i$: total specific   water content \\
 $q_s$     \> specific  for saturating water vapour content \\
 $r_{v}$   \> $=q_{v}/q_{d}$: mixing ratio for water vapour \\
 $r_{l}$   \> $=q_{l}/q_{d}$: mixing ratio for liquid water \\
 $r_{i}$   \> $=q_{i}/q_{d}$: mixing ratio for ice \\
 $r_{r}$   \> reference mixing ratio for water species: $\eta\:r_{r} \equiv e_r / (p_d)_r$
             and $r_{r} \approx 3.82$~g~kg${}^{-1}$ \\
 $r_s$     \> mixing ratio for saturating water vapour \\
 $r_{t}$   \> $=q_{t}/q_{d}$: mixing ratio for total water \\
 ${\rho}_d$   \> specific mass of dry air  \\
 ${\rho}_v$   \> specific mass of  water vapour \\
 ${\rho}_l$   \> specific mass of  liquid water \\
 ${\rho}_i$   \> specific mass of  ice \\
 ${\rho}$   \> specific mass of  moist air 
             $({\rho} = {\rho}_d+{\rho}_v+{\rho}_l+{\rho}_i)$  \\
 $RH$   \> relative humidity ($100 \times e/e_{sw}$ for liquid water) \\
 $R_v$   \> water-vapour gas constant --\= ($461.52$~J~K${}^{-1}$~kg${}^{-1}$) \kill
 $R_d$   \> dry-air gas constant     \> ($287.06$~J~K${}^{-1}$~kg${}^{-1}$) \\
 $R_v$   \> water-vapour gas constant \> ($461.53$~J~K${}^{-1}$~kg${}^{-1}$) \\
 $R$     \> $ = q_d \: R_d + q_v \: R_v$: gas constant for moist air \\
 $s$       \> specific entropy \\
 $(s_{d})_r$  \>  reference values for the entropy of dry air, \\
 $(s_{v})_r$  \> reference values for the entropy of water vapour, \\
 $s^0_d$   \> standard specific entropy of dry air 
              at $T_0$ and $p_0$ : $6775$~J~K${}^{-1}$~kg${}^{-1}$ \\
 $s^0_v$   \> standard specific entropy of water vapour
              at $T_0$ and $p_0$ : $10320$~J~K${}^{-1}$~kg${}^{-1}$ \\
 $s^0_l$   \> standard specific entropy of liquid water
              at $T_0$ and $p_0$ : $3517$~J~K${}^{-1}$~kg${}^{-1}$ \\
 $s^0_i$   \> standard specific entropy of solid water
              at $T_0$ and $p_0$ : $2296$~J~K${}^{-1}$~kg${}^{-1}$ \\
 $S_e, S_l$  \>   entropies in P10 ($S_a$ is their weighted sum)\\
 $T$       \> local temperature \\
 $T_r$   \> reference temperature ($T_r\equiv T_0$) \\
 $T_w$   \> isenthalpic wet-bulb temperature  \\
 $T_{il}$   \> ice-liquid water temperature \\
 $T_{0}$   \> zero Celsius temperature ($273.15$~K) \\
 $T_{\Upsilon}$    \> a constant temperature ($2362$~K) \\
 $\theta$         \> $ = T\:(p_0/p)^{\kappa}$: potential temperature\\
 ${\theta}_{e}$   \> equivalent potential temperature\\
 ${\theta}_{l}$   \> liquid-water potential temperature \\
 ${\theta}'_w$   \> pseudo-adiabatic wet-bulb potential temperature \\
 ${\theta}_{s}$   \> moist entropy potential temperature (M11) \\
 ${\Upsilon\!}_r$ \> $= [ (h_v)_r - (h_d)_r ] / (c_{pd}\:T_r)$, ${\Upsilon\!}_0(T_0) \approx 9.5$\\
 $\boldmath{v}_k$, $\boldmath{v}$ \> individual and barycentric mean velocities  \\
$\boldmath{J}_k$  \> individual barycentric diffusion flux  \\
 $w$  \> vertical component of the velocity  \\
$\boldmath{\nabla} $ \>   3D-gradient operator  \\
${\mu}$ \>  Gibbs' function ($h-T\:s$)   \\
$b_k$ \>  $=(h_k)_r - c_{pk}\:T_r$  for species $k$ (B82) \\
$d_e$, $d_i$ \> external and internal changes    \\
$\dot{Q}$ \>  diabatic heating rate   \\
$\dot{q}_{v}$, $\dot{q}_{l}$ \>  rate of change of $q_v$  and $q_l$  \\
$\dot{S}_{irr}$ \> irreversible source of entropy    \\
$\phi$ \>  gravitational potential energy ($=g\: z + \phi_0$ )    \\
GCM \> General Circulation Model \\
 NWP \> Numerical Weather Prediction \\
 MSE \> Moist Static Energy \\
 LES \> Large Eddy Simulation \\
 TMSE \> Thermal Moist Static Energy
\end{tabbing}

\vspace{4mm}
\noindent
{\bf Appendix~B. Standard values of enthalpies.}
             \label{Thermo_values}
\renewcommand{\theequation}{B.\arabic{equation}}
  \renewcommand{\thefigure}{B.\arabic{figure}}
   \renewcommand{\thetable}{B.\arabic{table}}
      \setcounter{equation}{0}
        \setcounter{figure}{0}
         \setcounter{table}{0}
\vspace{2mm}

\begin{table}
\caption{Values of the specific heat for O${}_2$ are given as a function of the absolute temperature.
The first table corresponds to the solid-$\alpha$ form.
Units are K for $T$ and J~K${}^{-1}$~mol${}^{-1}$ for $c_p$, to be divided by $0.032$~kg~mol${}^{-1}$ to obtain units of J~K${}^{-1}$~kg${}^{-1}$.
Data were obtained up to $20$~K from Table-I of FH69, with some interpolation performed from their Figures~1 and 2 for the range $21$ to $23.84$~K.
The second table corresponds to the solid-$\beta$ form.
Data were obtained from Table I of FH69 for the range $30$ to $43.78$~K, with interpolation performed from their Figures~1 and 2 for the range $23.855$ to $28$~K.
The third table corresponds to the solid-$\gamma$  form ($43.78$ to $54.4$~K) and the fourth table to the liquid form ($54.4$ to $90$~K).
Data were obtained from FH69, with linear interpolation from their Figure~1.
The fifth table corresponds to the vapour form  at $1013.25$~hPa (above $90$~K) and with $c_p$ directly expressed in J~K${}^{-1}$~kg${}^{-1}$.
Data were obtained from Jacobsen \textit{et al.} (1997, Table~5.79).
\vspace*{2mm}
\label{Table_cp_O2}}
\centering
$c_p(T)$  for O${}_2$ (solid-$\alpha$) -  Unit of J~K${}^{-1}$~mol${}^{-1}$
\begin{tabular}{||c|c||c|c||c|c||}
\hline 
\hline 
    $T$  & $c_p$  & $T$ & $c_p$      & $T$ & $c_p$ \\ 
\hline 
    $0$ & $0$     & $12$ & $4$     & $23$ & $23$ \\ 
    $2$ & $0.01$  & $13$ & $5$     & $23.35$ & $30$ \\ 
    $3$ & $0.05$  & $14$ & $6$     & $23.52$ & $40$ \\ 
    $4$ & $0.12$  & $15$ & $7.1$   & $23.6$ & $50$ \\ 
    $5$ & $0.24$  & $16$ & $8.3$   & $23.68$ & $70$ \\ 
    $6$ & $0.43$  & $17$ & $9.6$   & $23.71$ & $100$ \\ 
    $7$ & $0.72$  & $18$ & $11.1$  & $23.76$ & $200$ \\ 
    $8$ & $1.12$  & $19$ & $12.5$  & $23.80$ & $500$ \\ 
   $9$ & $1.65$  & $20$ & $14$    & $23.82$ & $1000$ \\ 
    $10$ & $2.30$  & $21$ & $16$   & $23.84$ & $2000$ \\ 
    $11$ & $3.10$  & $22$ & $18.2$ & $ $ & $ $\\ 
\hline
\hline 
\end{tabular}
\\\vspace*{1mm}
$c_p(T)$ for O${}_2$ (solid-$\beta$)  -  Unit of J~K${}^{-1}$~mol${}^{-1}$
\begin{tabular}{||c|c||c|c||c|c||}
\hline 
\hline 
    $T$  & $c_p$  & $T$ & $c_p$      & $T$ & $c_p$ \\ 
\hline
    $23.855$ & $2000$  & $24.03$ & $50$  & $28$ & $26$ \\ 
    $23.860$ & $1000$  & $24.10$ & $40$  & $30$ & $28$ \\ 
    $23.865$ & $500$   & $24.32$ & $30$  & $35$ & $34$ \\ 
    $23.87$ & $200$    & $25$ & $23.2$   & $40$ & $41$ \\ 
    $23.88$ & $100$     & $26$ & $23.7$   & $43.78$ & $45.4$ \\ 
    $23.91$ & $70$     & $27$ & $24.7$   & $ $ & $ $ \\ 
\hline 
\hline
\end{tabular}
\\\vspace*{1mm}
$c_p(T)$ for O${}_2$ (solid-$\gamma$)  -  Unit of J~K${}^{-1}$~mol${}^{-1}$
\begin{tabular}{||c|c||c|c||c|c||}
\hline 
\hline 
    $T$  & $c_p$  & $T$ & $c_p$      & $T$ & $c_p$ \\ 
\hline 
    $43.78$ & $45.4$  & $50$ & $46$  & $54.4$ & $46.5$ \\ 
\hline 
\hline
\end{tabular}
\\\vspace*{1mm}
$c_p(T)$ for O${}_2$ (liquid)  -  Unit of J~K${}^{-1}$~mol${}^{-1}$
\begin{tabular}{||c|c||c|c||c|c||}
\hline 
\hline 
    $T$  & $c_p$  & $T$ & $c_p$      & $T$ & $c_p$ \\ 
\hline 
    $54.4$ & $55$  & $70$ & $55.66$  & $90$ & $56.5$ \\ 
\hline 
\hline
\end{tabular}
\\\vspace*{1mm}
$c_p(T)$ for O${}_2$ (vapour)  -  Unit of J~K${}^{-1}$~kg${}^{-1}$
\begin{tabular}{||c|c||c|c||c|c||}
\hline 
\hline 
    $T$  & $c_p$  & $T$ & $c_p$      & $T$ & $c_p$ \\ 
\hline
\hline 
    $ 90$ & $970.5$  & $135$ & $923.1$  & $240$ & $914.5$ \\ 
    $ 95$ & $941.3$  & $140$ & $921.8$  & $250$ & $915.0$ \\ 
    $100$ & $935.2$  & $145$ & $920.7$  & $260$ & $915.6$ \\ 
    $105$ & $933.2$  & $150$ & $919.6$  & $270$ & $916.4$ \\ 
    $110$ & $931.6$  & $170$ & $916.7$  & $280$ & $917.4$ \\ 
    $115$ & $929.8$  & $190$ & $915.1$  & $290$ & $918.5$ \\ 
    $120$ & $928.0$  & $210$ & $914.3$  & $300$ & $919.9$ \\ 
    $125$ & $926.2$  & $230$ & $914.3$  & $ $ & $$ \\ 
    $130$ & $924.6$  & $235$ & $914.4$  & $ $ & $$ \\ 
\hline 
\hline
\end{tabular}
\end{table}

The values of the specific heat of oxygen are given in Table~\ref{Table_cp_O2} for the $0$ to $300$~K range of temperature.
The resulting variation of $c_p(T)$ is depicted in Figure~\ref{fig_Cp_O2}.

\begin{table}
\caption{Values of the specific heat for N${}_2$ are given as a function of the absolute temperature.
The first table corresponds to the solid-$\alpha$ form.
Data were obtained up to $35.6$~K from Table-14.4 of MF97.
Units are K for $T$ and J~K${}^{-1}$~mol${}^{-1}$ for $c_p$, to be divided by $0.028016$~kg~mol${}^{-1}$ to obtain units of J~K${}^{-1}$~kg${}^{-1}$.
The second table corresponds to the solid-$\beta$ form.
Data were obtained up to $56$~K directly from Table-14.4 of MF97.
The data in the range $56$ to $63$~K were obtained from the measured values depicted on Figure~1 of Kudryavtsev and Nemchenko (2001), with a change in the slope of the curve $c_p(T)$.
The third and fourth tables correspond to the liquid and the vapour forms  (at $1013.25$~hPa) and with $c_p$ directly expressed in  J~K${}^{-1}$~kg${}^{-1}$.
Data were obtained from Jacobsen \textit{et al.} (1997, Table~5.73).
\label{Table_cp_N2}}
\vspace*{2mm}
\centering
$c_p(T)$ for N${}_2$ (solid-$\alpha$)  -  Unit of J~K${}^{-1}$~mol${}^{-1}$
\begin{tabular}{||c|c||c|c||c|c||}
\hline 
\hline 
    $T$  & $c_p$  & $T$ & $c_p$      & $T$ & $c_p$ \\ 
\hline 
    $0$ & $0$      & $20$ & $19.9$   & $30$   & $31.16$ \\ 
    $2$ & $0.03$   & $21$ & $21.37$  & $31$   & $35.80$ \\ 
    $4$ & $0.24$   & $22$ & $22.86$  & $32$   & $37.64$ \\ 
    $6$ & $0.91$   & $23$ & $24.24$  & $33$   & $39.77$ \\ 
    $8$ & $2.43$   & $24$ & $25.67$  & $33.5$ & $40.97$ \\ 
   $10$ & $4.83$   & $25$ & $27.05$  & $34$   & $42.16$ \\ 
   $12$ & $7.64$   & $26$ & $28.45$  & $34.5$ & $43.38$ \\ 
   $14$ & $10.66$  & $27$ & $29.82$  & $35$   & $44.63$ \\ 
   $16$ & $13.61$  & $28$ & $31.19$  & $35.3$ & $45.53$ \\ 
   $18$ & $16.64$  & $29$ & $32.64$  & $35.6$ & $45.91$ \\ 
\hline
\hline 
\end{tabular}
\\\vspace*{1mm}
$c_p(T)$ for N${}_2$ (solid-$\beta$)  -  Unit of J~K${}^{-1}$~mol${}^{-1}$
\begin{tabular}{||c|c||c|c||c|c||}
\hline 
\hline 
    $T$  & $c_p$  & $T$ & $c_p$      & $T$ & $c_p$ \\ 
\hline 
   $35.6$ & $36.08$  & $44$ & $39.27$  & $56$ & $$43.5 \\ 
   $36$ & $36.26$  & $46$ & $40.03$  & $58$ & $44.5$ \\ 
   $38$ & $37.02$  & $48$ & $40.78$  & $60$ & $45.5$ \\ 
   $39$ & $37.39$  & $50$ & $41.53$  & $62$ & $46.5$ \\ 
   $40$ & $37.76$  & $52$ & $42.29$  & $63.1$ & $47.0$ \\ 
   $42$ & $38.52$  & $54$ & $43.04$  & $ $ & $ $ \\ 
\hline
\hline 
\end{tabular}
\\\vspace*{1mm}
$c_p(T)$ for N${}_2$ (liquid)  -  Unit of J~K${}^{-1}$~kg${}^{-1}$
\begin{tabular}{||c|c||c|c||c|c||}
\hline 
\hline 
    $T$  & $c_p$  & $T$ & $c_p$      & $T$ & $c_p$ \\ 
\hline
\hline 
    $63.1$ & $2019$  & $70$ & $2015$  & $77.4$ & $2042$ \\ 
\hline 
\hline
\end{tabular}
\\\vspace*{1mm}
$c_p(T)$ for N${}_2$ (vapour)  -  Unit of J~K${}^{-1}$~kg${}^{-1}$
\begin{tabular}{||c|c||c|c||c|c||}
\hline 
\hline 
    $T$  & $c_p$  & $T$ & $c_p$      & $T$ & $c_p$ \\ 
\hline
\hline 
    $77.4$ & $1340$  & $120$ & $1056$  & $200$ & $1043$ \\
    $80$   & $1191$  & $140$ & $1050$  & $250$ & $1042$ \\
    $90$   & $1081$  & $160$ & $1047$  & $300$ & $1041$ \\
    $100$  & $1067$  & $180$ & $1045$  & $ $ & $ $ \\
\hline 
\hline
\end{tabular}
\end{table}

According to Fagerstroem and Hollis Hallet (1969, hereafter referred to as FH69), the solid $\alpha$-$\beta$ transition occurs at about $23.85$~K, with no latent heat associated with it.
The solid $\beta$-$\gamma$ transition occurs at about $43.78$~K with a latent heat of $23.2$~kJ~kg${}^{-1}$.
The latent heat of melting occurring at the triple point ($54.4$~K) is equal to $13.9$~kJ~kg${}^{-1}$.
The latent heat of vaporization occurring at $90$~K is equal to $213$~kJ~kg${}^{-1}$.

The specific heat for the liquid phase of O${}_2$ is about $2$~J~K${}^{-1}$~mol${}^{-1}$ lower in Jacobsen \textit{et al.} (1997), giving an indication on the level of accuracy ($\approx 4$~\%) in the measurements of properties of cryogenic substances.

The values of the specific heat of nitrogen are given in Table~\ref{Table_cp_N2} for the $0$ to $300$~K range of temperature.
The resulting variation of $c_p(T)$ is depicted in  Figure~\ref{fig_Cp_N2}.

According to Manzhelii and Freiman (1997, hereafter referred to as MF97), the solid $\alpha$-$\beta$ transition occurs at about $35.6$~K and the  associated latent heat is equal to $8.2$~kJ~kg${}^{-1}$.
This latent heat is equal to $7.7$~kJ~kg${}^{-1}$ in Lipi\'{n}ski \textit{et al.} (2007), with a  discrepancy in the values illustrating the errors ($\approx 6 $~\%) made in measurements of properties of cryogenic substances.
The latent heat of melting at the triple point ($63.1$~K) is equal to $25.7$~kJ~kg${}^{-1}$ in MF97.
The latent heat of vaporization at $77.4$~K is equal to $200$~kJ~kg${}^{-1}$.

The values of the specific heat of ice-Ih are given in Table~\ref{Table_H2O_ice} for the range of temperatures from $0$ to $273$~K.
The resulting variation of $c_p(T)$ is depicted in Figure~\ref{fig_Cp_Ice}.

As for O${}_2$ and N${}_2$, the thermal enthalpy of the water species are obtained from (\ref{def_abs_h}) by integrating the values of $c_p(T)$ shown in Figure~\ref{fig_Cp_Ice} and by summing all the corresponding latent heats, which leads to a standard thermal enthalpy for ice at $T_0$ of $h_l (T_0) \approx 298$~kJ~kg${}^{-1}$.
The latent heat of fusion and sublimation at temperature $T_0$ are $334$~kJ~kg${}^{-1}$ and $2835$~kJ~kg${}^{-1}$, respectively, leading to the standard values $h_l (T_0) \approx 632$~kJ~kg${}^{-1}$ and $h_v (T_0) \approx 3133$~kJ~kg${}^{-1}$.

\begin{table}
\caption{Values of the specific heat for the solid (ice-Ih) form of H${}_2$O as a function of the absolute temperature.
Data were obtained up to $273$~K from Table-13 of Feistel and  Wagner (2006) and at a pressure of $1013.25$~hPa.
Units are K for $T$ and J~K${}^{-1}$~kg${}^{-1}$ for $c_p$.
\label{Table_H2O_ice}}
\vspace*{2mm}
\centering
$c_p(T)$ for H${}_2$O (ice-Ih)  -  Unit of J~K${}^{-1}$~kg${}^{-1}$
\begin{tabular}{||c|c||c|c||c|c||}
\hline
\hline 
    $T$  & $c_p$  & $T$ & $c_p$      & $T$ & $c_p$ \\ 
\hline 
 $ 0$ &  $  0   $  &   $100$ & $ 874.14$ &  $200$ & $1568.35$ \\ 
 $10$ &  $ 14.80$  &   $110$ & $ 949.38$ &  $210$ & $1638.86$ \\ 
 $20$ &  $111.43$  &   $120$ & $1021.30$ &  $220$ & $1710.03$ \\ 
 $30$ &  $230.66$  &   $130$ & $1090.80$ &  $230$ & $1781.79$ \\ 
 $40$ &  $337.89$  &   $140$ & $1158.82$ &  $240$ & $1854.08$ \\ 
 $50$ &  $437.49$  &   $150$ & $1226.18$ &  $250$ & $1926.83$ \\ 
 $60$ &  $532.56$  &   $160$ & $1293.51$ &  $260$ & $1999.98$ \\ 
 $70$ &  $623.92$  &   $170$ & $1361.21$ &  $270$ & $2073.48$ \\ 
 $80$ &  $711.48$  &   $180$ & $1429.53$ &  $273$ & $2095.59$ \\ 
 $90$ &  $794.93$  &   $190$ & $1498.57$ &  $ $   & $ $ \\ 
\hline
\hline 
\end{tabular}
\end{table}

From (\ref{def_hi_term}) the standard value of thermal enthalpy for ice  ($298$~kJ~kg${}^{-1}$) is in good agreement with the value suggested in Feistel and Wagner (2006), Table~14, from which $h_i(T_0)-h_i(0)\approx 298.35$~kJ~kg${}^{-1}$.

Conversely, Bannon, 2005 (hereafter referred to as B05) used the Thermochemical Tables of Chase, 1998 (hereafter referred to as C98) to evaluate dry-air and water-vapour enthalpies as $273.5$ and $503.1$~kJ~kg${}^{-1}$, respectively.
The large differences with the corresponding results derived in this section ($530$ and $3133$~kJ~kg${}^{-1}$) can be explained by the choice of  different definitions of the (thermal) enthalpy in C98, where only the integral of the vapour value of $c_p$ from $0$ to $T$ are computed.
The contribution of the latent heats (solid-solid, solid-liquid and liquid-vapour), or the integral of the liquid or solid values of $c_p$ at low temperature (including  Debye's law) are not taken into account in C98.

For the water vapour, if the latent heat of sublimation ($2835$~kJ~kg${}^{-1}$)  is roughly added to the value $503.1$~kJ~kg${}^{-1}$ deduced from C98, the result $3338$ is close to, but greater than, the value of $h^0_v=3133$~kJ~kg${}^{-1}$ given in (\ref{def_h0v}).
But the  integral of the ice values of $c_p$, with $c_p$ varying from $0$ to $2106$~J~K${}^{-1}$~kg${}^{-1}$ as shown in Figure~\ref{fig_Cp_Ice}, must be smaller than if the integral is computed with the almost constant water-vapour value $1846.1$~J~K${}^{-1}$~kg${}^{-1}$.
This must explain why the value $3338$ used in C98 and B05 is greater than the value of $3133$ computed in this Appendix: it corresponds to the integral of almost constant vapour (perfect gas?) values of $c_p$.

The sum of the latent heats described in Figures~\ref{fig_Cp_O2} and \ref{fig_Cp_N2} are equal to $250$ for O${}_2$ and $234$~kJ~kg${}^{-1}$  and N${}_2$.
If the corresponding dry-air average  value $237$ is added to the value $273.5$ found in B05, the result  $511$~kJ~kg${}^{-1}$ is close to, but smaller than, the value $h^0_d=530$~kJ~kg${}^{-1}$ given in (\ref{def_h0d}).
This difference can be explained by the integral of the solid and liquid values of $c_p$ which are greater than the vapour values extrapolated toward $0$~K, as shown in Figures~\ref{fig_Cp_O2} and \ref{fig_Cp_N2}.

\vspace{4mm}
\noindent
{\bf Appendix~C. Stratocumulus and Cumulus datasets.}
             \label{Sc_Cu_datasets}
\renewcommand{\theequation}{C.\arabic{equation}}
  \renewcommand{\thefigure}{C.\arabic{figure}}
      \setcounter{equation}{0}
        \setcounter{figure}{0}
\vspace{2mm}

Numerical values used for plotting the Figures~\ref{Fig_Hm_MSE_Sc}~(a)-(c) and \ref{Fig_MSE_deficit}~(a)-(b) are described in this Appendix.

The same FIRE-I (RF03B) dataset as used in M11 is given in Tables~\ref{Table_FIRE_I_cloud} and  \ref{Table_FIRE_I_clear} for in-cloud and clear-air values, respectively.
Values are defined every $25$~m up to $3862.5$~m in the Figures.
Only a few selected levels are kept  above $1562.5$~m in  Table~\ref{Table_FIRE_I_clear}.
Liquid water contents reach a maximum ($0.163$~g~kg${}^{-1}$) close to $900$m.
\begin{table}
\caption{The dataset for FIRE-I (in-cloud).
\label{Table_FIRE_I_cloud}}
\vspace*{2mm}
\centering
\begin{tabular}{||c|c|c|c|c||}
\hline
\hline 
    $Z$~(m)  & $p$~(hPa)  & $T$~(K) & $q_v$~(g/kg) &  $q_l$~(g/kg) \\ 
\hline 
 $   912.5 $ & $ 915.5 $ & $284.13 $ & $  7.58 $ & $  0.089 $ \\
 $   887.5 $ & $ 917.5 $ & $283.13 $ & $  8.21 $ & $  0.163 $ \\
 $   862.5 $ & $ 920.1 $ & $283.37 $ & $  8.20 $ & $  0.140 $ \\
 $   837.5 $ & $ 922.9 $ & $283.01 $ & $  8.20 $ & $  0.153 $ \\
 $   812.5 $ & $ 926.0 $ & $283.15 $ & $  8.27 $ & $  0.132 $ \\
 $   787.5 $ & $ 928.8 $ & $283.39 $ & $  8.36 $ & $  0.094 $ \\
 $   762.5 $ & $ 931.7 $ & $283.48 $ & $  8.41 $ & $  0.078 $ \\
 $   737.5 $ & $ 933.7 $ & $283.63 $ & $  8.46 $ & $  0.066 $ \\
 $   712.5 $ & $ 936.9 $ & $283.64 $ & $  8.38 $ & $  0.054 $ \\
 $   687.5 $ & $ 939.7 $ & $283.86 $ & $  8.48 $ & $  0.050 $ \\
 $   662.5 $ & $ 942.2 $ & $284.08 $ & $  8.57 $ & $  0.044 $ \\
 $   637.5 $ & $ 944.6 $ & $284.26 $ & $  8.62 $ & $  0.034 $ \\
 $   612.5 $ & $ 947.3 $ & $284.37 $ & $  8.66 $ & $  0.030 $ \\
 $   587.5 $ & $ 950.0 $ & $284.49 $ & $  8.80 $ & $  0.028 $ \\
 $   562.5 $ & $ 952.8 $ & $284.66 $ & $  8.84 $ & $  0.019 $ \\
 $   537.5 $ & $ 955.3 $ & $284.84 $ & $  8.97 $ & $  0.023 $ \\
 $   512.5 $ & $ 958.7 $ & $285.20 $ & $  9.07 $ & $  0.017 $ \\
 $   487.5 $ & $ 961.2 $ & $285.20 $ & $  9.13 $ & $  0.044 $ \\
 $   462.5 $ & $ 962.7 $ & $285.30 $ & $  9.18 $ & $  0.037 $ \\
\hline
\hline 
\end{tabular}
\end{table}
\begin{table}
\caption{The dataset for FIRE-I (clear-air).
\label{Table_FIRE_I_clear}}
\vspace*{2mm}
\centering
\begin{tabular}{||c|c|c|c||}
\hline
\hline 
    $Z$~(m)  & $p$~(hPa)  & $T$~(K) & $q_v$~(g/kg) \\ 
\hline 
 $  3862.5 $ & $ 645.7 $ & $280.37 $ & $  1.03 $  \\
 $  3562.5 $ & $ 668.7 $ & $282.08 $ & $  1.04 $  \\
 $  3062.5 $ & $ 710.3 $ & $285.41 $ & $  1.27 $  \\
 $  2562.5 $ & $ 753.8 $ & $287.86 $ & $  1.17 $  \\
 $  2062.5 $ & $ 799.7 $ & $290.38 $ & $  1.12 $  \\
 $  1562.5 $ & $ 847.7 $ & $292.55 $ & $  1.36 $  \\
 $  1537.5 $ & $ 850.3 $ & $292.81 $ & $  1.39 $  \\
 $  1512.5 $ & $ 852.4 $ & $292.51 $ & $  1.42 $  \\
 $  1487.5 $ & $ 855.1 $ & $292.23 $ & $  1.63 $  \\
 $  1462.5 $ & $ 858.0 $ & $292.64 $ & $  1.47 $  \\
 $  1437.5 $ & $ 860.8 $ & $292.04 $ & $  2.17 $  \\
 $  1412.5 $ & $ 862.8 $ & $293.32 $ & $  1.19 $  \\
 $  1387.5 $ & $ 864.5 $ & $293.34 $ & $  1.43 $  \\
 $  1362.5 $ & $ 867.6 $ & $292.38 $ & $  2.27 $  \\
 $  1337.5 $ & $ 870.1 $ & $292.40 $ & $  2.05 $  \\
 $  1312.5 $ & $ 872.8 $ & $292.57 $ & $  1.88 $  \\
 $  1287.5 $ & $ 875.4 $ & $292.49 $ & $  2.08 $  \\
 $  1262.5 $ & $ 878.2 $ & $292.08 $ & $  2.92 $  \\
 $  1237.5 $ & $ 880.4 $ & $292.45 $ & $  2.20 $  \\
 $  1212.5 $ & $ 883.1 $ & $292.35 $ & $  2.27 $  \\
 $  1187.5 $ & $ 885.8 $ & $292.09 $ & $  2.22 $  \\
 $  1162.5 $ & $ 888.3 $ & $292.06 $ & $  2.25 $  \\
 $  1137.5 $ & $ 891.0 $ & $291.89 $ & $  2.63 $  \\
 $  1112.5 $ & $ 893.7 $ & $291.69 $ & $  2.94 $  \\
 $  1087.5 $ & $ 896.4 $ & $291.34 $ & $  3.10 $  \\
 $  1062.5 $ & $ 899.0 $ & $290.94 $ & $  3.31 $  \\
 $  1037.5 $ & $ 901.9 $ & $290.53 $ & $  3.64 $  \\
 $  1012.5 $ & $ 904.3 $ & $291.20 $ & $  2.73 $  \\
\hline
\hline 
\end{tabular}
\begin{tabular}{||c|c|c|c||}
\hline
\hline 
    $Z$~(m)  & $p$~(hPa)  & $T$~(K) & $q_v$~(g/kg) \\ 
\hline 
 $   987.5 $ & $ 906.5 $ & $290.66 $ & $  2.97 $  \\
 $   962.5 $ & $ 909.4 $ & $289.88 $ & $  3.44 $  \\
 $   937.5 $ & $ 912.4 $ & $289.01 $ & $  4.08 $  \\
 $   912.5 $ & $ 914.9 $ & $287.32 $ & $  5.48 $  \\
 $   887.5 $ & $ 917.6 $ & $286.39 $ & $  6.17 $  \\
 $   862.5 $ & $ 920.3 $ & $286.83 $ & $  6.49 $  \\
 $   837.5 $ & $ 923.4 $ & $283.76 $ & $  7.87 $  \\
 $   787.5 $ & $ 929.9 $ & $284.25 $ & $  7.96 $  \\
 $   762.5 $ & $ 931.3 $ & $283.83 $ & $  8.13 $  \\
 $   612.5 $ & $ 947.6 $ & $284.64 $ & $  8.68 $  \\
 $   587.5 $ & $ 950.1 $ & $284.75 $ & $  8.72 $  \\
 $   562.5 $ & $ 953.0 $ & $284.91 $ & $  8.72 $  \\
 $   537.5 $ & $ 956.0 $ & $285.20 $ & $  8.78 $  \\
 $   512.5 $ & $ 958.7 $ & $285.28 $ & $  8.94 $  \\
 $   487.5 $ & $ 962.1 $ & $285.86 $ & $  8.88 $  \\
 $   462.5 $ & $ 963.3 $ & $285.85 $ & $  8.91 $  \\
 $   437.5 $ & $ 966.7 $ & $285.96 $ & $  8.96 $  \\
 $   412.5 $ & $ 969.2 $ & $286.35 $ & $  8.89 $  \\
 $   387.5 $ & $ 972.6 $ & $286.66 $ & $  8.84 $  \\
 $   362.5 $ & $ 975.4 $ & $286.90 $ & $  8.75 $  \\
 $   337.5 $ & $ 978.4 $ & $287.13 $ & $  8.82 $  \\
 $   312.5 $ & $ 981.3 $ & $287.39 $ & $  8.85 $  \\
 $   287.5 $ & $ 984.1 $ & $287.58 $ & $  8.80 $  \\
 $   262.5 $ & $ 986.9 $ & $287.80 $ & $  8.82 $  \\
 $   237.5 $ & $ 989.8 $ & $288.03 $ & $  8.81 $  \\
 $   212.5 $ & $ 992.7 $ & $288.20 $ & $  8.81 $  \\
 $   187.5 $ & $ 996.3 $ & $288.72 $ & $  8.90 $  \\
 $   162.5 $ & $ 997.3 $ & $288.67 $ & $  8.97 $  \\
\hline
\hline 
\end{tabular}
\end{table}

The EPIC dataset is given in Table~\ref{Table_EPIC}.
It  corresponds to the LES profiles of ($\theta$, $q_v$, $q_l$) given in Bretherton \textit{et al.} (2004) for a 6-day mean sounding in October 2001.
Liquid water contents reach a maximum of $0.283$~g~kg${}^{-1}$ at $1200$m.
\begin{table}
\caption{The dataset for EPIC.
\label{Table_EPIC}}
\vspace*{2mm}
\centering
\begin{tabular}{||c|c|c|c|c||}
\hline
\hline 
    $Z$~(m)  & $p$~(hPa)  & $T$~(K) & $q_v$~(g/kg)     &  $q_l$~(g/kg) \\ 
\hline 
$1600$ & $827.6  $ & $288.66    $ & $1.10    $ & $0.0 $\\
$1550$ & $832.5  $ & $288.48    $ & $1.10    $ & $0.004 $\\
$1500$ & $837.4  $ & $288.21    $ & $1.15    $ & $0.033 $\\
$1450$ & $842.4  $ & $287.70    $ & $1.20    $ & $0.075$ \\
$1400$ & $847.4  $ & $286.71    $ & $1.70    $ & $0.175$ \\
$1350$ & $852.4  $ & $285.66    $ & $2.50    $ & $0.200$ \\
$1300$ & $857.5  $ & $283.95    $ & $3.75    $ & $0.262 $\\
$1250$ & $862.5  $ & $282.41    $ & $5.00    $ & $0.267$ \\
$1200$ & $867.7  $ & $281.50    $ & $6.00    $ & $0.283$ \\
$1150$ & $872.8  $ & $280.87    $ & $6.80    $ & $0.250$ \\
$1100$ & $878.0  $ & $280.62    $ & $7.20    $ & $0.204 $\\
$1050$ & $883.2  $ & $280.57    $ & $7.60    $ & $0.150$ \\
$1000$ & $888.4  $ & $280.60    $ & $7.75    $ & $0.104$ \\
$950$ & $893.7  $ & $280.84    $ & $7.90    $ & $0.067$ \\
$900$ & $899.0  $ & $281.12    $ & $7.95    $ & $0.031 $\\
$850$ & $904.3  $ & $281.50    $ & $8.00    $ & $0.008$ \\
$800$ & $909.7  $ & $281.88    $ & $8.05    $ & $0.003 $\\
$750$ & $915.1  $ & $282.25    $ & $8.10    $ & $0.001 $\\
$700$ & $920.5  $ & $282.68    $ & $8.12    $ & $0.0$ \\
$650$ & $926.0  $ & $283.11    $ & $8.15    $ & $0.0 $\\
$600$ & $931.5  $ & $283.54    $ & $8.17    $ & $0.0 $\\
$550$ & $937.0  $ & $283.97    $ & $8.20    $ & $0.0$ \\
$500$ & $942.6  $ & $284.40    $ & $8.23    $ & $0.0 $\\
$450$ & $948.2  $ & $284.86    $ & $8.27    $ & $0.0 $\\
$400$ & $953.8  $ & $285.32    $ & $8.30    $ & $0.0 $\\
$350$ & $959.4  $ & $285.80    $ & $8.35    $ & $0.0 $\\
$300$ & $965.1  $ & $286.28    $ & $8.40    $ & $0.0 $\\
$250$ & $970.9  $ & $286.77    $ & $8.42    $ & $0.0 $\\
$200$ & $976.6  $ & $287.25    $ & $8.45    $ & $0.0 $\\
$150$ & $982.4  $ & $287.74    $ & $8.55    $ & $0.0 $\\
$100$ & $988.2  $ & $288.13    $ & $8.55    $ & $0.0 $\\
$50$ & $994.1   $ & $288.71    $ & $8.50    $ & $0.0 $\\
$25$ & $997.0   $ & $289.00    $ & $8.70    $ & $0.0$\\
$ 0$ & $1000.0  $ & $288.80    $ & $9.10    $ & $0.0 $\\
\hline
\hline 
\end{tabular}
\end{table}

The BOMEX dataset is given in Table~\ref{Table_BOMEX}.
It corresponds to the LES profiles depicted in Cuijpers and Bechtold (1995) for ($\theta_l$, $q_t$).
Liquid water content is less than $0.01$~g~kg${}^{-1}$ for this shallow cumulus (Siebesma \textit{et al.}, 2003) and is set to $0$ for this application.
\begin{table}
\caption{The dataset for BOMEX.
\label{Table_BOMEX}}
\vspace*{2mm}
\centering
\begin{tabular}{||c|c|c|c||}
\hline
\hline 
    $Z$~(m)  & $p$~(hPa)  & $T$~(K) & $q_v$~(g/kg) \\ 
\hline 
 $2500$ & $ 751.4 $ & $ 285.32 $ & $  3.00 $  \\
 $2400$ & $ 760.0 $ & $ 285.96 $ & $  3.24 $  \\ 
 $2300$ & $ 768.8 $ & $ 286.60 $ & $  3.48 $  \\ 
 $2200$ & $ 777.6 $ & $ 287.24 $ & $  3.72 $  \\ 
 $2100$ & $ 786.6 $ & $ 287.88 $ & $  3.96 $  \\ 
 $2000$ & $ 795.6 $ & $ 288.52 $ & $  4.20 $  \\ 
 $1950$ & $ 800.2 $ & $ 288.62 $ & $  4.40 $  \\ 
 $1900$ & $ 804.7 $ & $ 288.67 $ & $  5.00 $  \\ 
 $1800$ & $ 814.0 $ & $ 288.76 $ & $  5.94 $  \\ 
 $1700$ & $ 823.4 $ & $ 288.86 $ & $  6.88 $  \\ 
 $1600$ & $ 832.8 $ & $ 288.95 $ & $  7.82 $  \\ 
 $1500$ & $ 842.4 $ & $ 289.04 $ & $  8.76 $  \\ 
 $1400$ & $ 852.1 $ & $ 289.12 $ & $  9.70 $  \\ 
 $1300$ & $ 861.9 $ & $ 289.25 $ & $  10.60 $  \\ 
 $1250$ & $ 866.8 $ & $ 289.34 $ & $  11.00 $  \\ 
 $1200$ & $ 871.8 $ & $ 289.43 $ & $  11.50 $  \\ 
 $1100$ & $ 881.8 $ & $ 289.80 $ & $  12.10 $  \\ 
 $1000$ & $ 892.0 $ & $ 290.36 $ & $  12.60 $  \\ 
  $900$ & $ 902.2 $ & $ 290.92 $ & $  13.10 $  \\ 
  $800$ & $ 912.6 $ & $ 291.58 $ & $  13.40 $  \\ 
  $700$ & $ 923.1 $ & $ 292.29$ & $  13.80 $  \\ 
  $600$ & $ 933.7 $ & $ 292.95$ & $  14.40 $  \\ 
  $500$ & $ 944.4 $ & $ 293.81$ & $  14.70 $  \\ 
  $400$ & $ 955.3 $ & $ 294.67$ & $  14.90 $  \\ 
  $300$ & $ 966.3 $ & $ 295.69$ & $  15.00 $  \\ 
  $200$ & $ 977.4 $ & $ 296.66$ & $  15.10 $  \\ 
  $100$ & $ 988.6 $ & $ 297.63$ & $  15.30 $  \\ 
  $ 50$ & $ 994.3 $ & $ 298.11$ & $  15.40 $  \\ 
  $ 25$ & $ 997.1 $ & $ 298.46$ & $  17.95 $  \\ 
   $ 0$ & $1000.0 $ & $ 298.80$ & $  20.50 $  \\ 
\hline
\hline 
\end{tabular}
\end{table}




\newpage

\vspace{5mm}
\noindent{\large\bf References}
\vspace{2mm}

\noindent{$\bullet$ Betts AK.} {1973 (B73)}.
{Non-precipitating cumulus convection and its parameterization.
{\it Q. J. R. Meteorol. Soc.}
{\bf 99} (419):
178--196.}

\noindent{$\bullet$ Ambaum MHP.} {2010}.
{Thermal physics of the atmosphere.}
Advancing weather and climate science.
Wiley-Blackwell.
John Willey and sons.
Chichester (A10).

\noindent{$\bullet$ Arakawa A, Schubert WH.} {1974}.
{Interaction of a cumulus cloud ensemble with the 
 large-scale environment, Part~I.
{\it J. Atmos. Sci.}
{\bf 31} (3):
674--701 (AS73).}

\noindent{$\bullet$ Bannon PR.} {2005}.
{Eulerian available energetics in moist atmosphere.
{\it J. Atmos. Sci.}
{\bf 62} (12):
4238--4252 (B05).}

\noindent{$\bullet$ Bechtold P, Bazile E, 
      Guichard F, Mascart P, Richard E.} {2001}.
{A mass-flux convection scheme for regional and global models.
{\it Q. J. R. Meteorol. Soc.}
{\bf 127} (573):
869--886.}

\noindent{$\bullet$ Betts AK.} {1973}.
{Non-precipitating cumulus convection and its parameterization.
{\it Q. J. R. Meteorol. Soc.}
{\bf 99} (419):
178--196 (B73).}

\noindent{$\bullet$ Betts AK.} {1974}.
{Further coments on ``A comparison of the 
equivalent potential temperature and the 
static energy''.
{\it J. Atmos. Sci.}
{\bf 31} (6):
1713--1715 (B74).}

\noindent{$\bullet$ Betts AK.} {1975}.
{Parametric interpretation of Trade-Wind cumulus budget studies.
{\it J. Atmos. Sci.}
{\bf 32} (10):
1934--1945 (B75).}

\noindent{$\bullet$ Bretherton CS, Uttal T, Fairall CW, Yuter SE, 
Weller RA, Baumgardner D, Comstock K, Wood R, 
Raga GB.} {2004}.
{The EPIC 2001 Stratocumulus study.
{\it Bull. Amer. Meteor. Soc.}
{\bf 85,} (7):
967--977.}

\noindent{$\bullet$ Bretherton CS, Blossey PN., Khairoutdinov M.} {2005}.
{An energy-balance analysis of deep convective 
self-aggregation above uniform SST.
{\it J. Atmos. Sci.}
{\bf 62} (12):
4273-4292.}

\noindent{$\bullet$ Bohren CF, Albrecht BA.} {1998}.
{Atmospheric thermodynamics.}
Pp.1--402.
Oxford University Press.

\noindent{$\bullet$ Bougeault Ph.} {1985}.
{A simple parameterization of the large-scale effects of cumulus convection.
{\it Mon. Wea. Rev.}
{\bf 113} (12):
2108--2121.}

\noindent{$\bullet$ Businger JA.} {1982}.
{The fluxes of specific enthalpy, sensible heat and latent heat near the Earth's surface.
{\it J. Atmos. Sci.}
{\bf 39} (8):
1889-1892 (B82).}

\noindent{$\bullet$ Catry B, Geleyn JF, Tudor M, B\'enard P, 
      Troj\'{a}kov\'{a} A.} {2007}.
{Flux-conservative thermodynamic equations in a mass-weighted
framework.
{\it Tellus A.}
{\bf 59,} (1):
71--79.}

\noindent{$\bullet$ Chase MW Jr.} {1998}.
{Journal of Physics and Chemical Reference Data.
Monograph No.9.
NIST-JANAF Thermochemical Tables.  
4th ed.
{\it
American Chemical Society
and American Institute of Physics. }
Vol.1, Pp. 1--957. Vol.2, Pp. 959-1951 (C98).}

\noindent{$\bullet$ Cuijpers JWM, Bechtold P.} {1995}.
{A simple parameterization of cloud water related 
 variables for use in boundary layer models.
{\it J. Atmos. Sci.}
{\bf 52} (13):
2486--2490.}

\noindent{$\bullet$ De Groot SR, Mazur P.} {1962}.
{Non-equilibrium Thermodynamics. 
{\it North-Holland Publishing Company}. Amsterdam}

\noindent{$\bullet$ Derbyshire SH, Beau I, Bechtold P, Grandpeix J-Y, 
      Piriou J-M, Redelsperger J-L, Soares PMM.} {2004}.
{Sensitivity of moist convection to environmental humidity.
{\it Q. J. R. Meteorol. Soc.}
{\bf 130} (604):
178--196.}

\noindent{$\bullet$ Dufour L, Van Mieghem J.} {1975}.
{Thermodynamique de l'atmosph\`ere.}
Institut Royal M\'et\'eorologique de Belgique.
Bruxelle (DVM75).

\noindent{$\bullet$ Emanuel KA.} {1994}.
{Atmospheric convection.}
Pp.1--580.
Oxford University Press: New York and Oxford (E94).

\noindent{$\bullet$ Emanuel KA.} {2004}.
{Tropical cycle energetics and structure.}
Chapter 8 in ``Atmospheric Turbulence and 
mesoscale meteorology. 
Scientific Research Inspired by Doug Lilly.''
Edited by E.E. Fedorovich, R. Rotuno and B. Stevens.
p:165-192.
Cambridge University Press.

\noindent{$\bullet$ Fagerstroem CH., Hollis Hallet AC.} {1969}.
{The specific heat of solid oxygen.
{\it Journal of low temperature Physics.}
{\bf 1} (1).
3--12 (FH69).}

\noindent{$\bullet$ Feistel R., Wagner W.} {2006}.
{A new equation of state for H${}_2$O ice Ih.
{\it J. Phys. Chem. Ref. Data}
{\bf 35} (2).
1021--1047.}

\noindent{$\bullet$ Fuehrer PL, Friehe CA.} {2002}.
{Flux corrections revisited.
{\it Boundary-Layer Meteorol.}
{\bf 102} (3):
415-457.}

\noindent{$\bullet$ Gerard L, Piriou J-F, Brozkova R, Geleyn J-F, Banciu D.} {2009}.
{Cloud and precipitation parameterezization in a meso-gamma-scale operational weather prediction model.
{\it Mon. Wea. Rev.}
{\bf 137} (11):
3960--3977.}

\noindent{$\bullet$ Glansdorff P, Prigogine I.} {1971}.
{Structure stabilit\'e et fluctuations.
{\it Masson Ed}. Paris}
(Also available in English: Thermodynamic theory of 
structure, stability and fluctuations, Wiley-Interscience).

\noindent{$\bullet$ Iribarne JV,. Godson WL.} {1973}.
{Atmospheric thermodynamics.
{\it Geophysics and astrophysics monographs. 
     D. Reidel Pub. Company}. 
 Dordrecht-Holland and Boston-U.S.A.}

\noindent{$\bullet$ Hauf T, H\"{o}ller H.} {1987}.
{Entropy and potential temperature.
{\it J. Atmos. Sci.}
{\bf 44} (20):
2887--2901 (HH87).}

\noindent{$\bullet$ Jacobsen RT, Penoncello SG, Lemmon EW.} {1997}.
{Thermodynamic properties of Cryogenic fluids.
Pp.1--312.
The international cryogenics monograph series.
Springer, New-York.}

\noindent{$\bullet$ Khairoutdinov MF, Randall DA.} {2003}.
{Cloud resolving modeling of the ARM summer 1997 IOP: 
 model formulation, results, uncertainties, and sensitivities.
{\it J. Atmos. Sci.}
{\bf 60} (4):
607--625.}

\noindent{$\bullet$ Kudryavtsev IN, Nemchenko KE.} {2001}.
{Lattice dynamics and heat capacity of solid nitrogen.}
Proceeding of the 10th international conference 
on phonon scattering in condensed matter.
August 12-17, 2001.
Dartmouth, USA.

\noindent{$\bullet$ Lipi\'{n}ski L, Kowal A, Szmyrka-Grzebyk A, 
 Manuszkiewocz H, Steur PPM., Pavese F.} {2007}.
{The $\alpha$-$\beta$ transition of Nitrogen.
{\it Int. J. Thermophys.}
{\bf 28}.
1904--1912.}

\noindent{$\bullet$ Madden RA, Robitaille FE.} {1970}.
{A comparison of the equivalent potential temperature 
 and the static energy.
{\it J. Atmos. Sci.}
{\bf 27} (2):
327--329.}

\noindent{$\bullet$ Manzhelii VG, Freiman YA.} {1997}.
{Physics of Cryocrystals.}
Pp.1--691.
Springer, New-York.

\noindent{$\bullet$ Marquet P.} {1993}.
{Exergy in meteorology: definition and properties
of moist available enthalpy.
{\it Q. J. R. Meteorol. Soc.}
{\bf 119} (511):
567--590.} 

\noindent{$\bullet$ Marquet P.} {2011}.
{Definition of a moist entropic potential temperature. 
Application to FIRE-I data flights.
{\it Q. J. R. Meteorol. Soc.}
{\bf 137} (656):
768--791 (M11).
\url{http://arxiv.org/abs/1401.1097}.
{\tt arXiv:1401.1097 [ao-ph]}}

\noindent{$\bullet$ Marquet P, Geleyn J-F.} {2013}.
{On a general definition of the squared Brunt-V\"{a}is\"{a}l\"{a} 
 frequency associated with the specific 
 moist entropy potential temperature.
{\it Q. J. R. Meteorol. Soc.}
{\bf 139} (670) :
85--100.
\url{http://arxiv.org/abs/1401.2379}.
{\tt arXiv:1401.2379 [ao-ph]}}

\noindent{$\bullet$ Marquet P.} {2013}.
{On the definition of a moist-air potential vorticity.
{\it Q. J. R. Meteorol. Soc.}
Accepted in April, 2013. Early view stage\\
\url{http://arxiv.org/abs/1401.2006}.
{\tt arXiv:1401.2006 [ao-ph]}}

\noindent{$\bullet$ Nagle JF.} {1966}.
{Lattice statistics of Hydrogen bonded crystals.
I. The residual entropy of Ice.
{\it J. Math. Phys.}
{\bf 7} (8):
1484--1491.}

\noindent{$\bullet$ Normand CWB.} {1921}.
{Wet bulb temperatures and the thermodynamics of the air.
{\it Indian Metl. Memoirs.}
{\bf 23}. Part 1:
1--22 (N21).}

\noindent{$\bullet$ Pauling L.} {1935}.
{The structure and entropy of ice and
of other crystals with some randomness
of atomic arrangement.
{\it J. Am. Chem. Soc.}
{\bf 57} (12):
2680--2684.}

\noindent{$\bullet$ Pauluis O., Czaja A, Korty R.} {2010}.
{The global atmospheric circulation in moist 
isentropic coordinates.
{\it J. Climate.}
{\bf 23}.
3077--3093 (P10).}

\noindent{$\bullet$ Pauluis O., Schumacher J.} {2010}.
{Idealized moist Rayleigh-B\'enard convection with piecewise
linear equation of state.
{\it Commun. math. Sci.}
{\bf 8} (1):
295--319.}

\noindent{$\bullet$ Peterson TC., Willett KM., Thorne PW.} {2011}.
{Observed changes in surface atmospheric energy over land.
{\it Geophys. Res. Lett.}
{\bf 38} (L16707):
1--6.}

\noindent{$\bullet$ Richardson LF.} {1922}.
{Weather prediction by numerical process.
{\it Cambridge University Press} (R22).}

\noindent{$\bullet$ Saunders PM.} {1957}.
{The thermodynamics of saturated air:
 a contribution to the classical theory.
{\it Q. J. R. Meteorol. Soc.}
{\bf 83} (357):
342--350.}

\noindent{$\bullet$ Siebesma AP, Bretherton CS, Brown A, Chlond A, Cuxart J,
Duynkerke PG, Jiang H, Khairoutdinov M, Lewellen D, Moeng CH,
S\'{a}nchez  E, Stevens B, Stevens DE.} {2003}.
{A large eddy simulation intercomparison study of shallow cumulus convection.
{\it J. Atmos. Sci.}
{\bf 60} (10):
1870--1891.}

\noindent{$\bullet$ Tripoli GJ, Cotton WR.} {1981}.
{The use of ice-liquid water potential temperature as 
a thermodynamic variable in deep atmospheric models.
{\it Mon. Weather Rev.}
{\bf 109,} (5) :
1094--1102.}

\noindent{$\bullet$ Wisniak J.} {2001}.
{Frederick Thomas Trouton: the man, the rule, and the ratio.
{\it Chem. Educator.}
{\bf 6} (1):
55--61.}

\noindent{$\bullet$ WMO-No.8.} {2008}.
{Guide to meteorological instruments and methods of observation.
7th edition.
{\it World Meteorological Organization}.}

\noindent{$\bullet$ Zdunkowski W, Bott A.} {2004}.
{Thermodynamics of the atmosphere.
  A course in theoretical meteorology.
{\it Cambridge University Press}.}

\end{document}